\documentclass[pre,longbibliography,twocolumn]{revtex4-1}
\usepackage[main=english,spanish]{babel}
\usepackage[utf8]{inputenc}
\usepackage{makecell}
\usepackage{natbib}
\usepackage{amssymb,amsmath}
\usepackage{epsfig}
\usepackage{bm}
\usepackage{color}
\usepackage{appendix}
\usepackage{comment}
\usepackage{booktabs}
\usepackage{xurl}
\usepackage{shellesc}

\usepackage{graphicx}
\usepackage[normalem]{ulem}
\usepackage{soul}

\usepackage{color}
\newcommand{\vicente}[1]{{\color{black} #1}}
\newcommand{\david}[1]{{\color{black} #1}}
\newcommand{\rgg}[1]{{\color{black} #1}}

\newcommand\beq{\begin{equation}}
\newcommand\eeq{\end{equation}}
\newcommand\beqa{\begin{eqnarray}}
\newcommand\eeqa{\end{eqnarray}}
\newcommand{\dd}{\text{d}}

\newcommand{\al}{\alpha}

\begin{document}

\title{Dynamic properties \vicente{in a collisional model} of a confined quasi--two--dimensional granular fluid driven by a stochastic bath with friction}

\author{David Gonz\'alez M\'endez}
\email{Electronic address: dgonzalezm@unex.es}
\affiliation{Departamento de F\'{\i}sica, Universidad de Extremadura, E-06071 Badajoz, Spain}

\author{Rub\'en G\'omez Gonz\'alez}
\email{Electronic address: ruben@unex.es}
\affiliation{Departamento de Did\'actica de las Ciencias Experimentales y las Matem\'aticas, Universidad de Extremadura, E-10003 C\'aceres, Spain}

\author{Vicente Garz\'{o}}
\thanks{Electronic address: vicenteg@unex.es\\
URL: \url{https://fisteor.cms.unex.es/investigadores/vicente-garzo-puertos/}}
\affiliation{Departamento de F\'{\i}sica and Instituto de Computaci\'on Cient\'{\i}fica Avanzada (ICCAEx), Universidad de Extremadura, E-06071 Badajoz, Spain}

\begin{abstract}

This paper investigates the dynamic properties of a confined quasi--two--dimensional granular fluid at moderate densities, modeled within the framework of the Enskog kinetic equation. \david{The confinement is not treated explicitly at the geometrical level; instead}, it is incorporated \vicente{in an effective way} through the so–called $\Delta$--model (\vicente{a collisional model that includes energy injection through modified collision rules).  The referred model is extended in this paper} to account for the influence of an interstitial gas via a viscous drag force and a stochastic Langevin--like term. By applying the Chapman--Enskog method, the Navier--Stokes transport coefficients and the cooling rate are derived analytically considering the leading terms in a Sonine polynomial expansion. The study focuses on steady--state conditions and examines how the combined effects of inelastic collisions and external driving influence transport properties such as the viscosity and the thermal conductivity. Theoretical predictions for the steady temperature and the kurtosis are validated against direct simulation Monte Carlo (DSMC) results, showing excellent agreement. The findings reveal that the external driving significantly alters the transport coefficients compared to dry (no gas phase) granular systems, challenging previous assumptions that neglected these effects. Additionally, a linear stability analysis demonstrates that the homogeneous steady state is stable across the explored parameter space.

\end{abstract}


\date{\today}
\maketitle

\section{Introduction}
\label{sec1}

In recent years, the study of the transport properties of confined granular systems has received considerable attention (see, for example, Refs.\ \cite{OU98,LCG99,PMEU04,CMS12,CMS15,GS18,CMSSGS19}). In an experimental setup that reproduces this situation, the particles are confined in a box in which the vertical $z$--direction is slightly larger than the diameter of a particle. Due to their collisions with the vibrating bottom plate, energy is injected into the vertical degrees of freedom of grains. This kinetic energy is subsequently transferred to the horizontal degrees of freedom of the granular particles via inter--particle collisions. Since the vertical dynamics is usually faster than the horizontal one, it is natural to start with an effective two--dimensional description, in which the effects of confinement and vibration are encoded in modified collision rules for horizontal velocities.

Needless to say, describing confined granular systems with kinetic theory tools is quite intricate, primarily due to confinement's restrictions on the inelastic Boltzmann/Enskog collision operator. Thus, although recent advances have been made using this approach (see, for example, Refs.\ \cite{MBGM22,MGB22,MPGM23}), for simplicity's sake, it is common in the granular literature to adopt a coarse--grained model in which the influence of confinement on grain dynamics is considered via a collisional model. The so--called $\Delta$--model, proposed years ago by Brito \emph{et al.} \cite{BRS13}, was developed with this objective in mind. Apart from the normal coefficient of restitution, which characterizes the inelasticity of collisions, the $\Delta$--model considers an additional velocity increment of fixed magnitude ($\Delta>0$) along the normal collision direction. Physically, this increment represents the transfer of kinetic energy from a particle's vertical degrees of freedom to its horizontal ones. The addition of the $\Delta$ term, which adds that amount of velocity, serves as a thermostat that balances the collisional dissipation coming from the normal coefficient of restitution. More specific details on the $\Delta$--model can be found in the review \cite{BSG26}.

In the context of kinetic theory, the $\Delta$--model has been used to study the dynamic properties of monocomponent confined granular gases. These results are particularly relevant to the low--density regime, for which the inelastic Boltzmann equation is used as a starting point \cite{BGMB13,BMGB14,SRB14,BBMG15,BBGM16}. These results have then been extended to moderate densities by considering the inelastic version of the Enskog kinetic equation \cite{GBS18,GBS20}. Additionally, several studies have determined the transport properties in confined granular mixtures \cite{BSG20,GBS21,GBS24,GGBS24,GMG26}. Apart from results in kinetic theory, the $\Delta$--model has also been employed in recent years in the study of systems with long--range interactions \cite{JMV16}, absorbing phase transitions in driven granular systems \cite{MPSTSF24,MPSF25}, the formation of quasi--long--range ordered phases \cite{PMFBRSF24,MP24}, the non--equilibrium coexistence between a fluid and a crystal of granular hard disks \cite{MPSF25a}, and the study of hyperuniformity \cite{MCh25}. These examples demonstrate the relevance of the $\Delta$--model in capturing the trends observed in confined granular systems.

\begin{figure}[t]
\begin{center}
\begin{tabular}{lr}
\resizebox{8.5cm}{!}{\includegraphics{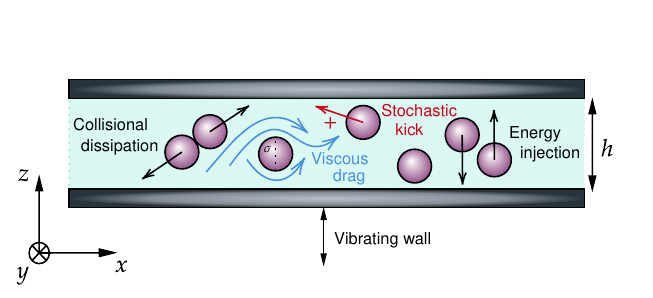}}
\end{tabular}
\end{center}
\caption{Schematic illustration of the confined quasi--two-dimensional system. Vertical vibration is imposed on the box to inject external energy into the system. During nonplanar collisions between grains, the vertical energy gained from the vibration of the plate is transferred to the $xy$ components of the velocities of grains. This kinetic energy is then dissipated and redistributed among the particles' horizontal degrees of freedom due to their collisions. Additionally, the interstitial medium (e.g., a fluid or gas) is modeled effectively as a viscous drag plus a stochastic kick, which introduces further dissipation and fluctuation in the horizontal dynamics. It is important to note that, although our study aims to capture the phenomenology of confined systems, we are modeling an unconfined two--dimensional system where collisions are described by the $\Delta$--model.
\label{fig0}}
\end{figure}

Nevertheless, although granular matter in nature is surrounded by an interstitial fluid, such as air, most previous theoretical and computational studies of the $\Delta$--model have neglected the impact of the gas phase on the dynamics of solid particles. As discussed in previous papers (see, for example, Ref.\ \cite{GGG19a}), the kinetic description of granular suspensions (i.e., an ensemble of solid particles immersed in a molecular viscous gas) is quite complex since a complete description of the system requires knowledge of the velocity distribution functions of each phase. Thus, although progress has recently been made using this approach (see Refs.\ \cite{GG22,GChG24,GG25}), a common model for describing gas--solid flows considers a kinetic equation for solid particles, in which the interstitial gas' effect on them is through an effective external (nonconservative) force \cite{KH01}.  This external solid--fluid force consists of two terms \cite{GTSH12}: (i) a viscous drag force (involving a drift or friction coefficient $\gamma$) that mimics the friction of grains on the surrounding viscous gas, and (ii) a stochastic Langevin--like term (involving the bath temperature $T_\text{b}$) accounting for the energy gained by the grains due to their interactions with the faster particles of the interstitial gas. This effective suspension model has been employed by many researchers to assess the effect of the gas phase on the solid particles \cite{TK95,SMTK96,KS99,WZLH09,GTSH12,H13,WGZS14,SA17,HTG17,GGG19a,SA20,THSG20,GGKG20,PG26}. A schematic illustration of the studied system is provided in Fig.\ \ref{fig0}.

To the best of our knowledge, the only study in the context of the $\Delta$--model that partially accounted for the effect of the surrounding gas on grains has been recently reported by Maire \emph{et al.} \cite{MPSF25}. In this paper, the authors introduce a viscous drag force $\gamma$ to model the friction of grains on the walls during the free flight between collisions. However, since the authors assume that corrections to the transport coefficients due to the presence of either $\Delta$ or $\gamma$ are very small over a wide range of values, they use the conventional expressions of the Navier--Stokes transport coefficients for \emph{elastic} hard disks to analyze different problems.
This raises the question of whether, and if so to what extent, the transport coefficient expressions employed in Ref.\  \cite{MPSF25} may change due to the combined effect of $\Delta$ and $\gamma$.

In this paper, we address the above question by determining the Navier--Stokes transport coefficients of a confined, quasi--two--dimensional, monocomponent granular fluid modeled by the $\Delta$--model and driven by a stochastic bath with friction. We perform the analysis within the framework of the Enskog kinetic equation; therefore, the results apply to moderate densities (let us say for instance, solid volume fraction $\phi \lesssim 0.25$ for hard spheres). As in Ref.\ \cite{GGG19a}, the starting point is the suspension model constituted by the viscous drag force plus the stochastic Langevin term. The transport coefficients are obtained by solving the Enskog equation by means of the application of the Chapman--Enskog method \cite{CC70}, adapted to dissipative dynamics. As usual \cite{GBS18}, the transport coefficients are given in terms of the solutions of a set of coupled linear integral equations. These equations are approximately solved by considering the leading terms in a Sonine polynomial expansion. To achieve analytical forms for these coefficients, we consider the relevant state of confined granular suspensions with \emph{steady} temperature. Our results clearly show that, in general, the dependence of the Navier--Stokes transport coefficients on the coefficient of restitution differs significantly from that previously obtained in the absence of the gas phase \cite{GBS18}. This conclusion calls into question the assumption made in Ref.\  \cite{MPSF25} where the transport coefficients are independent of the friction coefficient $\gamma$ and the $\Delta$ parameter.

The plan of the paper is as follows. In Sec.\ \ref{sec2}, the $\Delta$--model and the Enskog kinetic equation for granular suspensions are introduced. The balance equations for the densities of mass, momentum, and energy are also derived in this section and the kinetic and collisional contributions to the fluxes are given in terms of the one--particle velocity distribution function $f(\mathbf{r}, \mathbf{v};t)$. Section \ref{sec3} deals with the homogeneous steady state (HSS) where (approximate) theoretical expressions for the (steady) temperature $\theta_\text{s}$ and the kurtosis $a_{2,\text{s}}$ are obtained as functions of the coefficient of restitution $\al$, the (reduced) Delta parameter $\Delta_\text{b}^*=\Delta/\sqrt{2T_\text{b}/m}$ ($m$ being the mass of a solid particle), the (reduced) background temperature $T_\text{b}^*=T_\text{b}/(m\sigma^2 \gamma^2_\text{St})$ ($\sigma$ being the diameter of the particle, and $\gamma_\text{St}$ the Stokes drag coefficient, with $\gamma_\text{St}\propto \eta_\text{g}$, where $\eta_\text{g}$ is the gas viscosity), and the solid volume fraction $\phi$. These theoretical results are compared with computer simulation results obtained by numerically solving the Enskog equation by the direct simulation Monte Carlo (DSMC) method \cite{B94}. The comparison shows an excellent agreement between theory and DSMC results. Once the distribution function of the HSS is well characterized, we take the time--dependent homogeneous distribution $f^{(0)}(\mathbf{r}, \mathbf{v};t)$ as the reference state and solve the Enskog equation by the Chapman--Enskog expansion \cite{CC70} around this distribution. The explicit expressions of the Navier--Stokes transport coefficients and the cooling rate are displayed in Sec.\ \ref{sec5} for steady state conditions. The dependence of the transport coefficients and the cooling rate on the parameter space of the system is illustrated in Sec.\ \ref{sec6} for several systems. The results clearly show that the influence of the gas phase on them is in general quite important. As an application, a linear stability analysis of the HSS is carried out in Sec.\ \ref{sec7}; an exhaustive analysis of the dependence of the eigenvalues on the parameter space shows that the HSS is linearly stable. The paper is closed in Sec.\ \ref{sec8} with a brief discussion on the results reported here.

\section{Enskog kinetic equation for confined granular suspensions}
\label{sec2}

\subsection{Collision rules in the $\Delta$--model}

We consider a set of solid particles of diameter $\sigma$ and mass $m$ immersed in a viscous gas. Although the physical system is primarily intended to model quasi--two--dimensional configurations, for the sake of generality we carry out the analysis in an arbitrary spatial dimension $d$, where $d=2$ corresponds to hard disks and $d=3$ to hard spheres. Collisions between grains are inelastic and are characterized by a (positive) constant coefficient of normal restitution $\al \leq 1$, where $\al=1$ corresponds to elastic collisions (ordinary gases). In the context of the so--called $\Delta$--model \cite{BRS13}, the relationship between the pre--collisional velocities $(\mathbf{v}_1, \mathbf{v}_2)$ and the post--collisional velocities $(\mathbf{v}_1',\mathbf{v}_2')$ reads
\beq
\label{1.1}
\mathbf{v}_1'=\mathbf{v}_1-\frac{1}{2}\left(1+\alpha\right)(\widehat{{\boldsymbol {\sigma }}}\cdot \mathbf{g})\widehat{{\boldsymbol {\sigma }}}-\Delta \widehat{{\boldsymbol {\sigma }}},
\eeq
\beq
\label{1.1.0}
{\bf v}_{2}'=\mathbf{v}_{2}+\frac{1}{2}\left(1+\alpha\right)
(\widehat{{\boldsymbol {\sigma}}}\cdot \mathbf{g})
\widehat{\boldsymbol {\sigma}}+\Delta \widehat{{\boldsymbol {\sigma}}},
\eeq
where $\mathbf{g}=\mathbf{v}_1-\mathbf{v}_2$ is the relative velocity and $\widehat{{\boldsymbol {\sigma}}}$ is the unit collision vector joining
the centers of the two colliding spheres (pointing from particle 1 to particle 2). Particles are approaching if $\widehat{{\boldsymbol {\sigma}}}\cdot \mathbf{g}>0$. In Eqs.\  \eqref{1.1}--\eqref{1.1.0}, $\Delta$ is an extra (positive) velocity added to the relative motion. This extra velocity points outward in the normal direction $\widehat{\boldsymbol {\sigma}}$, as required by the conservation of angular momentum \cite{L04}. The relative velocity after collision $\mathbf{g}'=\mathbf{v}_1'-\mathbf{v}_2'$ can be easily obtained from Eqs.\  \eqref{1.1} and \eqref{1.1.0} as
\beq
\label{1.2}
\mathbf{g}'=\mathbf{g}-(1+\al)(\widehat{{\boldsymbol {\sigma}}}\cdot \mathbf{g})
\widehat{\boldsymbol {\sigma}}-2\Delta \widehat{{\boldsymbol {\sigma }}},
\eeq
so that
\beq
\label{1.3}
(\widehat{{\boldsymbol {\sigma}}}\cdot \mathbf{g}')=-\al (\widehat{{\boldsymbol {\sigma}}}\cdot \mathbf{g})-2\Delta. \vspace{2.5mm}
\eeq

With the set of collision rules \eqref{1.1}--\eqref{1.1.0}, momentum is conserved but energy is not. The change in kinetic energy upon collision is
\beqa
\label{1.4}
\Delta E&\equiv& \frac{m}{2}\left(v_1'^{2}+v_2'^{2}-v_1^2-v_2^2\right)\nonumber\\
&=&m\left[\Delta^2+\al \Delta (\widehat{{\boldsymbol {\sigma}}}\cdot \mathbf{g})-\frac{1-\al^2}{4}(\widehat{{\boldsymbol {\sigma}}}\cdot \mathbf{g})^2\right].
\eeqa
The right--hand side of Eq.\ \eqref{1.4} vanishes for elastic collisions ($\al=1$) and $\Delta=0$. Moreover, it appears that energy can be gained or lost in a collision depending on whether $\widehat{{\boldsymbol {\sigma}}}\cdot \mathbf{g}$ is smaller than or larger than $2\Delta /(1-\al)$.

For practical purposes, it is also convenient to consider the restituting collision $\left(\mathbf{v}_1'',\mathbf{v}_2''\right)\to \left(\mathbf{v}_1,\mathbf{v}_2\right)$ with the same collision vector $\widehat{{\boldsymbol {\sigma }}}$. The restituting velocities are given by
\beq
\label{1.5}
\mathbf{v}_1''=\mathbf{v}_1-\frac{1}{2}\left(1+\alpha^{-1}\right)(\widehat{{\boldsymbol {\sigma }}}\cdot \mathbf{g})\widehat{{\boldsymbol {\sigma }}}-\alpha^{-1}\Delta \widehat{{\boldsymbol {\sigma }}},
\eeq
\beq
\label{1.5.0}
\mathbf{v}_2''=\mathbf{v}_2+\frac{1}{2}\left(1+\alpha^{-1}\right)(\widehat{{\boldsymbol {\sigma }}}\cdot \mathbf{g})\widehat{{\boldsymbol {\sigma }}}+\alpha^{-1}\Delta \widehat{{\boldsymbol {\sigma }}}.
\eeq
Equations \eqref{1.5}--\eqref{1.5.0} lead to the result
\beq
\label{1.5.1}
(\widehat{{\boldsymbol {\sigma}}}\cdot \mathbf{g}'')=-\al^{-1} (\widehat{{\boldsymbol {\sigma}}}\cdot \mathbf{g})-2\Delta \al^{-1},
\eeq
where $\mathbf{g}''=\mathbf{v}_1''-\mathbf{v}_2''$. Moreover, the volume transformation in velocity space for a direct collision is $\dd\mathbf{v}_1' \dd \mathbf{v}_2'=\al\, \dd\mathbf{v}_1
\dd\mathbf{v}_2$, and for the restituting collision is
$\dd \mathbf{v}_1'' \dd \mathbf{v}_2''=\al^{-1} \dd \mathbf{v}_1 \dd \mathbf{v}_2$.

\subsection{Enskog kinetic equation for a confined granular fluid driven by a stochastic bath with friction}

At a kinetic level, all the relevant information on the state of the granular fluid is provided by the one--particle velocity distribution function of solid particles  $f(\mathbf{r}, \mathbf{v}; t)$. At moderate densities, the distribution $f$ obeys the Enskog kinetic equation
\beq
\label{1.6}
\frac{\partial f}{\partial t}+\mathbf{v}\cdot \nabla f+\mathcal{F} f=J_\text{E}[\mathbf{r},\mathbf{v}|f,f],
\eeq
where the Enskog collision operator in the $\Delta$--model is given by \cite{GBS18}
\begin{widetext}
\begin{align}
\label{1.7}
J_\text{E}[\mathbf{r},\mathbf{v}_1|f,f]\equiv \sigma^{d-1}\int \dd{\bf v}_{2}\int \dd \widehat{\boldsymbol{\sigma}}\;
\Theta (-\widehat{{\boldsymbol {\sigma }}}\cdot {\bf g}-2\Delta)
(-\widehat{\boldsymbol {\sigma }}\cdot {\bf g}-2\Delta)
\al^{-2}\chi(\mathbf{r},\mathbf{r}+\boldsymbol{\sigma}) f(\mathbf{r},\mathbf{v}_1'',t)
f(\mathbf{r}+\boldsymbol{\sigma},\mathbf{v}_2'',t)\nonumber\\
-\sigma^{d-1}\int\ \dd{\bf v}_{2}\int \dd\widehat{\boldsymbol{\sigma}}
\;\Theta (\widehat{{\boldsymbol {\sigma }}}\cdot {\bf g})
(\widehat{\boldsymbol {\sigma }}\cdot {\bf g})
\chi(\mathbf{r},\mathbf{r}+\boldsymbol{\sigma}) f(\mathbf{r},\mathbf{v}_1,t)
f(\mathbf{r}+\boldsymbol{\sigma},\mathbf{v}_2,t).
\end{align}
\end{widetext}
Here, $\chi[{\bf r},{\bf r}\pm\boldsymbol{\sigma}|n(t)] $ is the equilibrium pair correlation function at contact as a functional of the nonequilibrium density field $n({\bf r}, t)$ defined by
\begin{equation}
\label{1.9}
n({\bf r}, t)=\int \dd{\bf v} \;f({\bf r},{\bf v},t).
\end{equation}

As mentioned in the Introduction, we assume that the confined solid particles are surrounded by an interstitial viscous gas. As usual \cite{TK95,SMTK96,GTSH12,GGG19a}, for the sake of simplicity, a coarse--grained approach is considered and the effect of the viscous gas on solid particles is accounted for through an effective fluid--solid interaction force. This force is represented by the operator $\mathcal{F}$ in Eq.\ \eqref{1.6}. For low Reynolds numbers, it is assumed that the external force $\mathbf{F}$ acting on solid particles is composed of two independent terms. One term corresponds to a viscous drag force $\mathbf{F}^{\text{drag}}$ proportional to the (instantaneous) velocity of particle $\mathbf{v}$. This term takes into account the friction of grains on the viscous gas. Since the model attempts to mimic gas--solid flows, the drag force is defined in terms of the relative velocity $\mathbf{v}-\mathbf{U}_\text{g}$ where $\mathbf{U}_\text{g}$ is the (known) mean flow velocity of the surrounding molecular gas. Thus, the drag force $\mathbf{F}^{\text{drag}}=-m\gamma \left(\mathbf{v}-\mathbf{U}_\text{g}\right)$ is represented in the Enskog equation \eqref{1.6} by the term
\beq
\label{1.10}
\mathcal{F}^{\text{drag}}f\to -\gamma \frac{\partial}{\partial \mathbf{v}}\cdot\left(\mathbf{v}-\mathbf{U}_\text{g}\right)f.
\eeq
Here, $\gamma$ is the drag or friction coefficient. The second term in the total force $\mathbf{F}$ corresponds to a stochastic force intended to mimic the effect of random collisions with the molecules of the background fluid. This stochastic force $\mathbf{F}^{\text{st}}$ has the form of a Gaussian white noise with the properties
\beq
\label{1.11}
\langle \mathbf{F}_i^{\text{st}}(t) \rangle =\mathbf{0}, \quad
\langle \mathbf{F}_i^{\text{st}}(t) \mathbf{F}_j^{\text{st}}(t') \rangle = 2 m^2 \gamma T_\text{b} \mathsf{I} \delta_{ij}\delta(t-t'),
\eeq
where $\mathsf{I}$ is the identity tensor and $i$ and $j$ refer to two different particles. Here, $T_\text{b}$ can be interpreted as the temperature of the background (or bath) fluid. The amplitude of the stochastic (Langevin) force is chosen to satisfy the fluctuation–dissipation theorem in the elastic limit. In the Enskog kinetic Eq.\ \eqref{1.6}, the stochastic external force is represented by a Fokker--Planck operator of the form \cite{GTSH12}
\beq
\label{1.12}
\mathcal{F}^{\text{st}}f\to -\frac{\gamma T_\text{b}}{m}\frac{\partial^2 f}{\partial v^2}.
\eeq
As in previous works \cite{GGG19a}, we henceforth assume that $\gamma$ is a scalar quantity which, in the dilute limit, reduces to the Stokes value $\gamma_\text{St}$ (which is proportional to the viscosity of the surrounding gas $\eta_\text{g}$). In the case of hard spheres ($d=3$) and for very dilute suspensions, the corresponding drift coefficient is
\beq
\label{1.12.1}
\gamma\equiv \gamma_{\text{St}}=\frac{3\pi \sigma \eta_\text{g}}{m}.
\eeq
However, for moderately dense suspensions, hydrodynamic interactions and lubrication effects may modify the isolated--particle drag. In this regime, for low Reynolds number, these effects can be (phenomenologically) accounted for by writing $\gamma$ as
\beq
\label{1.13.1}
\gamma=\gamma_{\text{St}} R(\phi),
\eeq
where $R(\phi)$ is a function of the solid volume fraction
\beq
\label{1.13}
\phi=\frac{\pi^{d/2}}{2^{d-1}d\Gamma \left(\frac{d}{2}\right)}n\sigma^d.
\eeq
In the three--dimensional case ($d=3$), several expressions for the density dependence of the drag coefficient have been proposed from numerical simulations of low--Reynolds--number flow through arrays of spheres \cite{VdHBK05,Beetstra07,YinSundaresan09}. These studies provide explicit closures for $R(\phi)$ in suspensions of spherical particles. To the best of our knowledge, however, no comparable expression is available in the granular literature for a two--dimensional setting ($d=2$). In the present work, we therefore keep $R(\phi)$ arbitrary. This allows us to retain a general formulation, so that the results derived below are independent of the particular choice of the function $R(\phi)$.



Therefore, according to Eqs.\ \eqref{1.11} and \eqref{1.12}, the Enskog equation \eqref{1.6} can be written as
\beq
\label{1.14}
\frac{\partial f}{\partial t}+\mathbf{v}\cdot \nabla f-\gamma\Delta \mathbf{U}\cdot \frac{\partial f}{\partial \mathbf{v}}-\gamma\frac{\partial}{\partial \mathbf{v}}\cdot \mathbf{V} f-\gamma \frac{T_{\text{b}}}{m}\frac{\partial^2 f}{\partial v^2}=J_\text{E}[f,f],
\eeq
where $\Delta \mathbf{U}=\mathbf{U}-\mathbf{U}_\text{g}$ and $\mathbf{V}=\mathbf{v}-\mathbf{U}$ is the peculiar velocity. Here,
\beq
\label{1.15}
\mathbf{U}(\mathbf{r},t)=\frac{1}{n(\mathbf{r},t)}\int \dd\mathbf{v}\; \mathbf{v} f(\mathbf{r},\mathbf{v},t)
\eeq
is the mean flow velocity of the solid particles. Another relevant hydrodynamic field is the \emph{granular} temperature $T(\mathbf{r},t)$ defined as
\beq
\label{1.16}
T(\mathbf{r},t)=\frac{m}{d n(\mathbf{r},t)} \int \dd\mathbf{v}\; V^2 f(\mathbf{r},\mathbf{v},t).
\eeq

An important property of the integrals involving the Enskog collision operator is \cite{GBS18}
\beqa
\label{1.17}
I_\psi&\equiv& \int\; \dd\mathbf{v}_1\; \psi(\mathbf{v}_1) J_\text{E}[\mathbf{r},\mathbf{v}_1|f,f]\nonumber\\
&=&\sigma^{d-1}\int  \dd\mathbf{v}_1\int\ \dd{\bf v}_{2}\int \dd\widehat{\boldsymbol{\sigma}}\,
\Theta (\widehat{{\boldsymbol {\sigma }}}\cdot {\bf g})(\widehat{\boldsymbol {\sigma }}\cdot {\bf g})
\chi(\mathbf{r},\mathbf{r}+\boldsymbol{\sigma})\nonumber\\ 
& & \times f(\mathbf{r},\mathbf{v}_1,t)
f(\mathbf{r}+\boldsymbol{\sigma},\mathbf{v}_2,t)
\left[\psi(\mathbf{v}_1')-\psi(\mathbf{v}_1)\right],
\eeqa
where $\mathbf{v}_1'$ is defined by Eq.\ \eqref{1.1}. The identity \eqref{1.17} is the same as for the conventional inelastic hard sphere (IHS) model \cite{BP04,G19}. An immediate consequence of the relation \eqref{1.17} is that the balance equations associated with the density of mass, momentum and energy are formally equivalent to those obtained from the IHS model \cite{GGG19a}. They are (respectively) given by
\begin{equation}
\text{D}_{t}n+n\nabla \cdot {\bf U}=0\;, \label{1.18}
\end{equation}
\begin{equation}
\text{D}_{t}{\bf U}=-\rho ^{-1}\nabla \cdot \mathsf{P}-\gamma \Delta \mathbf{U}\;,
\label{1.19}
\end{equation}
\begin{equation}
\text{D}_{t}T+\frac{2}{dn} \left( \nabla \cdot {\bf q}+\mathsf{P}:\nabla {\bf U}\right) =2\gamma \left(T_\text{b}-T\right)-\zeta \,T.
\label{1.20}
\end{equation}
In the above equations, $\text{D}_{t}=\partial_{t}+{\bf U}\cdot \nabla$ is the material derivative and $\rho=m n$ is the mass density. The pressure tensor ${\sf P}({\bf r},t)$ and
the heat flux ${\bf q}({\bf r},t)$ have both {\em kinetic} and {\em collisional transfer} contributions, i.e., ${\sf P}={\sf P}_\text{k}+{\sf P}_\text{c}$ and ${\bf q}={\bf q}_\text{k}+{\bf q}_\text{c}$. Their kinetic contributions
are given as usual as
\begin{align}
\label{1.21}
\mathsf{P}_\text{k}({\bf r}, t)&=\int \dd{\bf v} \;m{\bf V}{\bf V}f({\bf r},{\bf v},t),\\
\label{1.22}
{\bf q}_\text{k}({\bf r}, t)&=\int \dd{\bf v} \:\frac{m}{2}V^2{\bf V}f({\bf r},{\bf v},t).
\end{align}
The collisional transfer contributions are \cite{GBS18}:
\beqa
\label{1.22.1}
\mathsf{P}_\text{c}&=&\frac{1+\alpha}{4}m \sigma^{d}
\int \dd\mathbf{v}_{1}\int \dd\mathbf{v}_{2}\int
\dd\widehat{\boldsymbol {\sigma }}\;\Theta (\widehat{\boldsymbol
{\sigma }}\cdot
\mathbf{g})(\widehat{\boldsymbol {\sigma}}\cdot \mathbf{g})
\nonumber\\
& & \times 
\widehat{\boldsymbol {\sigma}}\widehat{\boldsymbol {\sigma }}\left[(\widehat{\boldsymbol {\sigma}}\cdot \mathbf{g})+\frac{2\Delta}{1+\al}\right]\nonumber\\
& & \times 
\int_{0}^{1} \dd \lambda\; f_2\left(\mathbf{r}-\lambda{\boldsymbol{\sigma}},\mathbf{v}_{1},\mathbf{r}+(1-\lambda){\boldsymbol{\sigma}},\mathbf{v}_2,t\right),
\nonumber\\
\eeqa
\beqa
\label{1.23}
{\bf q}_\text{c}&=&\frac{1+\alpha}{4}m \sigma^{d}
\int \dd\mathbf{v}_{1}\int \dd\mathbf{v}_{2}\int
\dd\widehat{\boldsymbol {\sigma}}\;\Theta (\widehat{\boldsymbol{\sigma}}\cdot
\mathbf{g})(\widehat{\boldsymbol {\sigma}}\cdot \mathbf{g})^{2}
\nonumber\\
& & \times 
(\widehat{\boldsymbol {\sigma}}\cdot {\bf G})
\widehat{\boldsymbol {\sigma}}\int_{0}^{1} \dd \lambda f_2\left(\mathbf{r}-
\lambda{\boldsymbol{\sigma}},\mathbf{v}_1,\mathbf{r}+(1-\lambda)
{\boldsymbol {\sigma}},\mathbf{v}_{2},t\right)
\nonumber\\
& & 
-\Delta \frac{m \sigma^d}{4}
\int \dd\mathbf{v}_{1}\int \dd\mathbf{v}_{2}\int
\dd\widehat{\boldsymbol {\sigma}}\;\Theta (\widehat{\boldsymbol{\sigma}}\cdot
\mathbf{g})(\widehat{\boldsymbol {\sigma}}\cdot \mathbf{g})\widehat{\boldsymbol {\sigma}} 
\nonumber\\
& & \times 
\left[\Delta +\al (\widehat{\boldsymbol {\sigma}}\cdot \mathbf{g})-2 (\widehat{\boldsymbol {\sigma}}\cdot \mathbf{G})\right]
\nonumber\\
& & 
 \times
\int_{0}^{1} \dd \lambda\; f_2\left(\mathbf{r}-
\lambda {\boldsymbol{\sigma}},\mathbf{v}_1,\mathbf{r}+(1-\lambda)
{\boldsymbol {\sigma}},\mathbf{v}_{2},t\right).
\eeqa
Here, ${\bf G}=\frac{1}{2}({\bf V}_1+{\bf V}_2)$ is the velocity of the center of mass and 
$f_2$ is defined as
\begin{equation}
\label{1.24}
f_2({\bf r}_1, {\bf v}_1, \mathbf{r}_2, {\bf v}_2, t)\equiv \chi({\bf r}_1, {\bf r}_2)f({\bf r}_1, {\bf v}_1, t) f({\bf r}_2, {\bf v}_2, t).
\end{equation}
Finally, the cooling rate is given by
\beqa
\label{1.25}
\zeta&=&-\frac{m}{d n T}\sigma^{d-1}
\int \dd\mathbf{v}_{1}\int \dd\mathbf{v}_{2}\int \dd\widehat{\boldsymbol {\sigma}}
\;\Theta (\widehat{\boldsymbol {\sigma}}\cdot
\mathbf{g})(\widehat{ \boldsymbol {\sigma}}\cdot
\mathbf{g})\nonumber\\
& & \times \left[\Delta^2+\al \Delta (\widehat{{\boldsymbol {\sigma}}}\cdot \mathbf{g})-\frac{1-\al^2}{4}(\widehat{{\boldsymbol {\sigma}}}\cdot \mathbf{g})^2\right]\nonumber\\
& & \times f_2(\mathbf{r},\mathbf{v}_{1}, \mathbf{r}+\boldsymbol {\sigma
}, \mathbf{v}_{2},t).
\eeqa

Note that the applicability of the suspension model employed here is restricted to low Reynolds numbers and moderate densities. A relevant point is that the form of the Enskog collision operator $J_\text{E}[\mathbf{r},\mathbf{v}_1|f,f]$ for the $\Delta$--model defined in Eq.\ \eqref{1.7} is assumed to be the same as for dry granular fluids (namely, in the absence of interstitial gas). Thus, the effect of the interstitial solvent gas is absent in the collision dynamics of grains. As it has been discussed in previous papers \cite{KH01,K90,TK95,WKL03}, the above assumption holds only when the mean--free time between collisions is assumed to be much less than the time needed by the environmental gas to significantly affect the binary collisions between solid particles. This assumption is reliable when the impact of the gas phase on the motion of solid particles is weak (for instance, in the case of solid particles immersed in air) but fails for instance in the case of liquid flows. In this situation one has to take into account the presence of the interstitial fluid in the collision process.

\section{Homogeneous state}
\label{sec3}

Before considering spatial perturbations, we first analyze the homogeneous state, which is of intrinsic interest and provides the reference state for the Chapman--Enskog--like expansion \cite{CC70}. For homogeneous time--dependent states, the density $n$ and the temperature $T$ are spatially uniform, and with an appropriate selection of the frame of reference, the mean flow velocities vanish ($\mathbf{U}=\mathbf{U}_\text{g}=\mathbf{0}$). Consequently, the Enskog equation \eqref{1.14} becomes
\beq
\label{2.1}
\frac{\partial f}{\partial t}-\gamma\frac{\partial}{\partial \mathbf{v}}\cdot \mathbf{v} f-\gamma \frac{T_{\text{b}}}{m}\frac{\partial^2 f}{\partial v^2}=J_\text{E}^{\text{H}}[\mathbf{V}|f,f],
\eeq
where the Enskog collision operator for homogeneous states is
\beqa
\label{2.2}
J_\text{E}^{\text{H}}[f,f]&=&\sigma^{d-1}\chi \int \dd{\bf v}_{2}\int \dd \widehat{\boldsymbol{\sigma}}
\;\Theta (-\widehat{{\boldsymbol {\sigma }}}\cdot {\bf g}-2\Delta)\nonumber\\
& & \times 
(-\widehat{\boldsymbol {\sigma }}\cdot {\bf g}-2\Delta)
\al^{-2} f(\mathbf{v}_1'',t)f(\mathbf{v}_2'',t)-\sigma^{d-1}\chi\nonumber\\
& & \times
 \int \dd{\bf v}_{2}\int \dd\widehat{\boldsymbol{\sigma}}
\;\Theta (\widehat{{\boldsymbol {\sigma }}}\cdot {\bf g})
(\widehat{\boldsymbol {\sigma }}\cdot {\bf g})
f(\mathbf{v}_1,t)f(\mathbf{v}_2,t).\nonumber\\
\eeqa
Here, $\chi$ is the pair correlation function evaluated at the (homogeneous) density $n$. The collision operator \eqref{2.2} can be recognized as the Boltzmann operator for inelastic collisions multiplied by the factor $\chi$. For homogeneous time--dependent states, the balance equation \eqref{1.20} for the granular temperature $T$ reads
\beq
\label{2.3}
\frac{\partial T}{\partial t} =2\gamma\left(T_{\text{b}}-T\right)-\zeta T.
\eeq
As usual, for times longer than the mean free time, the system has completely forgotten its initial conditions and so, one expects that the system achieves a hydrodynamic regime. In this regime,  the distribution $f$ qualifies as a \emph{normal} distribution in the sense that $f$ depends on time only through its dependence on the temperature $T$. This means that
\beq
\label{2.3.1}
\frac{\partial f}{\partial t}=\left(\frac{\partial f}{\partial T}\right)\left(\frac{\partial T}{\partial t}\right)=\big[2\gamma\left(\theta^{-1}-1\right)-\zeta\big]T\frac{\partial f}{\partial T},
\eeq
and equation \eqref{2.1} becomes
\beq
\label{2.4}
\bigg[2\gamma\left(\theta^{-1}-1\right)-\zeta\bigg]T\frac{\partial f}{\partial T}-\gamma\frac{\partial}{\partial\mathbf{v}}\cdot\mathbf{v}f-\frac{\gamma T_{\text{b}}}{m}\frac{\partial^2 f}{\partial v^2}=J_{\text{E}}^{\text{H}}[f,f],
\eeq
where $\theta\equiv T/T_\text{b}$ is the reduced temperature. In addition, for homogeneous states and after performing the angular integrals in Eq.\ \eqref{1.25}, the cooling rate $\zeta$ is
\begin{widetext}
\beq
\label{2.5}
\zeta(t)=-\frac{\pi^{(d-1)/2}}{d n T}m\sigma^{d-1} \chi \int \dd \mathbf{v}_1
\int \dd \mathbf{v}_2 \left[\frac{\Delta^2g}{\Gamma\left(\frac{d+1}{2}\right)}+\frac{\sqrt{\pi}}{d \Gamma\left(\frac{d}{2}\right)}\al \Delta g^2-\frac{g^3}{4 \Gamma\left(\frac{d+3}{2}\right)}(1-\al^2)\right]
f(\mathbf{v}_1,t)\;f(\mathbf{v}_2,t).
\eeq
\end{widetext}
Upon obtaining Eq.\ \eqref{2.5} use has been made of the result \cite{NE98}
\begin{equation}
\label{2.5.1}
B_k\equiv \int \dd\widehat{\boldsymbol{\sigma}}\; \Theta (\widehat{{ \boldsymbol{\sigma }}} \cdot \mathbf{g})\,
(\widehat{\boldsymbol{\sigma}} \cdot {\widehat{\mathbf{g}}})^k=\pi^{(d-1)/2} \frac{ \Gamma\left(\frac{k+1}{2
}\right)}{\Gamma\left(\frac{k+d}{2}\right)},
\end{equation}
for positive integers. Here, $\widehat{\mathbf{g}}=\mathbf{g}/g$.

For elastic collisions ($\al=1$) and $\Delta=0$, we have $\zeta=0$, so that  Eq.\ \eqref{2.4} admits the Maxwellian solution
\beq
\label{2.6}
f_0(v,t)=n \left(\frac{m}{2\pi T(t)}\right)^{d/2} \exp\left(-\frac{m v^2}{2T(t)}\right)
\eeq
where the temperature obeys the time--dependent equation
\beq
\label{2.7}
\partial_t T =2\gamma\left(T_{\text{b}}-T\right).
\eeq
The system therefore is in a time--dependent ``equilibrium state'' before reaching the asymptotic steady state where $T=T_{\text{b}}$.

For elastic collisions but $\Delta \neq 0$, $\zeta \neq 0$ and the solution to Eq.\
\eqref{2.4} is not known. Of course, in the most general case ($\al \neq 1$ and $\Delta \neq 0$), the solution is also not known to date. However, in the hydrodynamic regime, dimensional analysis suggests that the solution to equation \eqref{2.4} can be written in the scaled form \cite{BGMB13,BMGB14,GBS18}
\beq
\label{2.8}
f(\mathbf{v},\gamma, T_\text{b},\Delta)=n v_\text{th}^{-d}\varphi(\mathbf{c},\gamma^*, \theta)\equiv n v_\text{th}^{-d}\varphi(\mathbf{c},\lambda,\theta),
\eeq
where $\mathbf{c}\equiv \mathbf{v}/v_\text{th}$ is a dimensionless velocity and
$v_\text{th}=\sqrt{2T/m}$ is the thermal speed. In Eq.\ \eqref{2.8}, we have introduced the (reduced) friction coefficient $\gamma^*$ as
\beq
\label{2.9}
\gamma^*(\lambda,\theta)=\frac{\ell \gamma}{v_\text{th}}=\lambda \theta^{-1/2},
\eeq
where 
\beq
\label{2.9.0}
\lambda(\phi)=\frac{\gamma_\text{St} R(\phi)\ell}{\sqrt{2T_{\text{b}}/m}}=\frac{\sqrt{2}\pi^{d/2}}
{2^d d \Gamma\left(\frac{d}{2}\right)}\frac{R(\phi)}{\phi \sqrt{T_\text{b}^*}}.
\end{equation}
Here, $\ell=1/(n \sigma^{d-1})$ is proportional to the mean free path of hard spheres and we recall that $T_\text{b}^*\equiv T_\text{b}/(m \sigma^2 \gamma_\text{St}^2)$ is the (reduced) background gas temperature. For the sake of convenience, note that here the dependence of the scaled distribution $\varphi$ on the parameter $\Delta$ is through the dimensionless time--independent parameter $\Delta_\text{b}^*=\Delta/\sqrt{2T_\text{b}/m}$ instead of $\Delta^*=\Delta/\sqrt{2T/m}$ as in the \emph{dry} (no gas--phase) $\Delta$--model \cite{BGMB13,BMGB14,GBS18}. The use of $\Delta_\text{b}^*$ or $\Delta^*$ is only a matter of choice since $\Delta^*(t)=\theta(t)^{-1/2}\Delta_\text{b}^*$, where $\Delta_\text{b}^*$ does not depend on time.

In the absence of gas--phase ($\gamma=0$), it is important to remark that the consistency of the scaled solution \eqref{2.8} has been confirmed by computer simulations performed for the $\Delta$--model for monocomponent \cite{BGMB13,BMGB14} and multicomponent \cite{BSG20} granular fluids. We expect that this consistency is also kept in the presence of the surrounding gas. According to Eq.\ \eqref{2.8},
\beq
\label{2.10}
T \frac{\partial f}{\partial T}=-n v_\text{th}^{-d}\left(\frac{1}{2}\frac{\partial}{\partial\mathbf{c}}\cdot\left(\mathbf{c}\varphi\right)-\theta \frac{\partial \varphi}{\partial \theta}\right),
\eeq
and hence in dimensionless form Eq.\ \eqref{2.4} can be rewritten as
\begin{equation}
\label{2.11}
-\Lambda\left(\frac{1}{2}\frac{\partial }{\partial\mathbf{c}}\cdot(\mathbf{c}\varphi)-\theta \frac{\partial \varphi}{\partial \theta}\right)
-\gamma^*\frac{\partial}{\partial\mathbf{c}}\cdot(\mathbf{c}\varphi)-
\frac{\gamma^*}{2\theta}
\frac{\partial^2 \varphi}{\partial c^2}=J_{\text{E}}^*[\varphi,\varphi],
\end{equation}
where
\beq
\label{2.11.1}
\Lambda\equiv 2\gamma^*\left(\theta^{-1}-1\right)-\zeta^*,
\eeq
with $\zeta^*=\ell \zeta/v_\text{th}$, and $J_{\text{E}}^*=\ell v_\text{th}^{d-1}J_{\text{E}}^{\text{H}}/n$.

Since in the homogeneous state the distribution function $f(\mathbf{v};t)$ is isotropic in velocity space, then the heat flux vanishes ($\mathbf{q}=\mathbf{0}$) and the pressure tensor is diagonal, i.e., $P_{ij}=p\delta_{ij}$. According to Eqs.\ \eqref{1.21} and \eqref{1.22.1}, the (reduced) hydrostatic pressure $p^*=p/(nT)$ is given by
\begin{align}
\label{2.11.2}
p^* = 1+2^{d-2}\chi \phi \Bigg[1+\al+\frac{2}{\sqrt{\pi}}\frac{\Gamma\left(\frac{d}{2}\right)}{\Gamma\left(\frac{d+1}{2}\right)}
\theta^{-1/2}\Delta_\text{b}^{*}\nonumber\\
\times \int \dd\mathbf{c}_1\int \dd\mathbf{c}_2 \; g^* \varphi(\mathbf{c}_1)\varphi(\mathbf{c}_2)\Bigg],
\end{align}
where $\mathbf{g}^*=\mathbf{g}/v_\text{th}$.

Although the exact form of $\varphi$ is not known, indirect information on it can be obtained from the kurtosis or fourth cumulant
\begin{equation}
\label{2.12}
a_{2}=\frac{4}{d(d+2)}\int\; \dd{\bf c}\; c^4 \varphi(c)-1.
\end{equation}
The cumulant $a_{2}(\lambda,\theta)$ measures the deviation of $\varphi$ from its Maxwellian form $\varphi_\text{M}(c)=\pi^{-d/2}e^{-c^2}$. The coefficient $a_2$ can be obtained by multiplying Eq.\ \eqref{2.11} by $c^4$ and integrating over velocity. The result is
\beq
\label{2.13}
\frac{1}{2}\left(1+a_2+\frac{1}{2}\theta \frac{\partial a_2}{\partial \theta}\right)\Lambda+\gamma^*\left(1+a_2-\theta^{-1}\right)=\frac{\mu_4}{d(d+2)},
\eeq
where
\beq
\label{2.14}
\mu_4=\int \dd\mathbf{c}\; c^4\; J_{\text{E}}^*[\varphi,\varphi].
\eeq

To solve the differential equation \eqref{2.13} one needs to know $\zeta^*$ and $\mu_4$. Since these quantities are defined in terms of the unknown scaled distribution $\varphi(\mathbf{c})$, one has to estimate them by using an approximate form of $\varphi$. Here, as in Ref.\ \cite{GGG19a}, we replace the true distribution $\varphi(\mathbf{c})$ by its first Sonine approximation \cite{BP04}:   
\beq
\label{2.15}
\varphi(\mathbf{c}) \approx \varphi_\text{M}(\mathbf{c})\left[1+\frac{a_2}{2}\left(c^4-(d+2)c^2+\frac{d(d+2)}{4}\right)\right].
\eeq 
The evaluation of $\zeta^*$ and $\mu_4$ by using the approximation \eqref{2.15} has been carried out in Ref.\ \cite{BMGB14}. The expressions of $\zeta^*$ and $\mu_4$ by neglecting quadratic terms in $a_2$ can be written as \cite{BMGB14}
\beq
\label{2.16}
\zeta^*\to \zeta_0^{(0)}+\zeta_0^{(1)}a_2, 
\eeq
\beq
\label{2.16.0}
\mu_4\to \mu_4^{(0)}+\mu_4^{(1)}a_2,
\eeq
where
\beq
\label{2.17}
\zeta_0^{(0)}=\frac{2\sqrt{2}\pi^{\frac{d-1}{2}}}{d\Gamma\left(\frac{d}{2}\right)}\chi \left(\frac{1-\al^2}{2}-\sqrt{\frac{\pi}{2}}
\al \theta^{-1/2}\Delta_\text{b}^*-\theta^{-1}\Delta_\text{b}^{*2}\right), 
\eeq
\beq
\label{2.17.0}
\zeta_0^{(1)}=\frac{\sqrt{2}\pi^{\frac{d-1}{2}}}{8d\Gamma\left(\frac{d}{2}\right)}\chi \left(\frac{3}{2}(1-\al^2)+
\theta^{-1}\Delta_\text{b}^{*2}\right),
\eeq
\beq
\label{2.18}
\mu_4^{(0)}=\frac{d(d+2)}{8}\frac{C_0}{A_0}\zeta_0^{(0)}, 
\eeq
\beq
\label{2.18.0}
\mu_4^{(1)}=\frac{d(d+2)}{8}\left(
\frac{4 A_1+C_1}{A_0}\zeta_0^{(0)}-4\zeta_0^{(1)}\right).
\eeq
For the sake of completeness, the forms of the quantities $A_0$, $A_1$, $C_0$, and $C_1$ are displayed in the Appendix \ref{appA}.
In the limiting case $\Delta_\text{b}^{*}=0$, Eqs.\ \eqref{2.17}--\eqref{2.18.0} reduce to
\beq
\label{2.23}
\zeta_0^{(0)}=\frac{\sqrt{2}\pi^{\frac{d-1}{2}}}{d\Gamma\left(\frac{d}{2}\right)}\chi (1-\al^2), \quad
\zeta_0^{(1)}=\frac{3}{16}\zeta_0^{(0)},
\eeq
\beq
\label{2.24}
\mu_4^{(0)}=-\frac{d}{2}\left(d+\frac{3}{2}+\al^2\right)\zeta_0^{(0)}, 
\eeq
\beq
\label{2.24.0}
\mu_4^{(1)}=-\frac{d}{2}\Bigg[\frac{3}{32}\left(10d+39+10\al^2\right)+\frac{d-1}{1-\al}\Bigg]
\zeta_0^{(0)}.
\eeq
Equations \eqref{2.23}--\eqref{2.24.0} agree with previous results \cite{GGG19a} obtained in the conventional IHS model.

\vicente{If one neglects the non--Gaussian corrections to the scaled distribution $\varphi$ ($a_2=0$), then $\zeta^*\to \zeta_0^{(0)}$. In this approximation, when the influence of the surrounding interstitial gas is also neglected ($\gamma=0$), the steady--state condition $\zeta^*=0$ yields a quadratic equation in $\Delta^*$, whose physical solution (i.e., $\Delta^*=0$ if $\al=1$) provides the $\al$-dependence of the dimensionless quantity $\Delta^*$: 
\beq
\label{2.25}
\Delta^*(\al)=\frac{1}{2}\sqrt{\frac{\pi}{2}}\al \Bigg[\sqrt{1+\frac{4(1-\al^2)}{\pi \al^2}}-1\Bigg].
\eeq
Thus, in the usual $\Delta$--model \cite{BRS13}, given values of $\al$ and $\Delta$, Eq.\ \eqref{2.25} yields the steady temperature. It should be noted that molecular dynamics simulations have been compared to the theoretical predictions of Eq.\ \eqref{2.25}, showing in general a very good agreement with deviations smaller than 2 \%. More significant discrepancies between theory and simulation are found for small values of the coefficient of restitution and/or high densities \cite{BRS13}. However, in the presence of the interstitial gas ($\gamma\neq 0$), the parameters $\al$ and $\Delta^*$ must be considered as independent, even in the steady state.}


\subsection{Direct Simulation Monte Carlo (DSMC) method}

\rgg{To assess the reliability of the theoretical predictions in the homogeneous state, the Enskog kinetic equation \eqref{2.1} is solved numerically by means of the DSMC method \cite{B70b,B94}. We provide in this subsection some details on these simulations. Since the system is spatially homogeneous, the simulation only involves the evolution of particle velocities. Each time step is therefore divided into two stages: binary collisions and the action of the thermostat (stochastic bath with friction). To account for finite--density effects, the collision stage is modified according to the (homogeneous) Enskog description by introducing the pair correlation function at contact, $\chi$, which corrects the collision frequency through the number of candidate collision pairs \cite{MS97}. The rest of the algorithm proceeds as follows.

First, binary collisions between granular particles are simulated. The velocity distribution is represented by a finite ensemble of simulated particles according to
\beq
f^{(N)}(\mathbf{v},t)=\frac{n}{N}\sum_{k=1}^{N}\delta \left(\mathbf{v}-\mathbf{v}_k(t)\right),
\eeq
where $N$ is the number of simulated particles and $n$ is the number density.

At each time step, the number of collision attempts is determined according to the No Time Counter (NTC) algorithm, generalized to the Enskog equation through the inclusion of the pair correlation function $\chi$. The maximum relative velocity is estimated as $g^{\mathrm{max}}=Cv_{\mathrm{th}}$, where $v_{\mathrm{th}}=\sqrt{2T/m}$ is the thermal velocity and $C\simeq5$ \cite{B94}. A collision direction $\widehat{\boldsymbol{\sigma}}$ is then sampled uniformly over the unit sphere (or the unit circle for $d=2$), and a randomly selected pair $(k,\ell)$ is accepted for collision whenever
\beq
\left|\widehat{\boldsymbol{\sigma}}\cdot\mathbf{g}_{k\ell}\right|
>\mathrm{R}(0,1)\,g^{\mathrm{max}},
\eeq 
where $\mathrm{R}(0,1)$ is a uniform random number in $[0,1]$.
If a collision is accepted, the particle velocities are updated according to the collision rules \eqref{1.1} and \eqref{1.1.0}.

After the collision stage, the interaction with the surrounding fluid is incorporated through a Langevin thermostat. The velocity of each particle is updated at every time step as
\beq
\mathbf{v}\rightarrow
e^{-\gamma\delta t}\mathbf{v}
+\left(\frac{6\gamma T_\text{b}\delta t}{m}\right)^{1/2}\mathbf{R},
\eeq
where $\mathbf{R}$ is a $d$--dimensional random vector whose components are independently and uniformly distributed in the interval $[-1,1]$. It is worth noting that this update represents a discrete--time implementation of the Ornstein--Uhlenbeck process associated with the Langevin operator. An exact finite--time Ornstein--Uhlenbeck step would involve a Gaussian random increment with variance proportional to $1-e^{-2\gamma\delta t}$. In the present DSMC implementation, the random vector is sampled from a uniform distribution and its amplitude is chosen so that the first two velocity moments coincide with those of the continuous Langevin process up to leading order in $\delta t$. Therefore, in the limit in which $\delta t$ is much smaller than the mean time between collisions, the algorithm reproduces the same Fokker--Planck dynamics \cite{KG14}.}

\subsection{Homogeneous steady state (HSS)}

To obtain analytical results, one typically considers the long--time limit where the reduced temperature $\theta(t)$ achieves an asymptotic constant value $\theta_\text{s}$ independent of time. In the steady state ($\Lambda=0$), in the linear order in $a_2$, the left--hand side of Eq.\ \eqref{2.13} becomes
\beq
\label{2.26}
-\frac{1}{2}\zeta_0^{(0)}+\left(\lambda \theta_\text{s}^{-1/2}-\frac{1}{2}\zeta_0^{(1)}\right)a_2,
\eeq
where the steady temperature $\theta_\text{s}$ is consistently determined from the steady state condition $\Lambda=0$, namely, from the condition
\beq
\label{2.26.1}
2\lambda \theta_\text{s}^{-1/2}\left(\theta_{\text{s}}^{-1}-1\right)-\zeta_0^{(0)}=0.
\eeq
Thus, taking into account Eq.\ \eqref{2.26}, the \emph{steady} solution to Eq.\ \eqref{2.13} when only linear terms in $a_2$ are retained is
\beq
\label{2.27}
a_{2,\text{s}}=\frac{\mu_4^{(0)}+\frac{d(d+2)}{2}\zeta_0^{(0)}}{d(d+2)\left(\lambda \theta_\text{s}^{-1/2}-\frac{1}{2}\zeta_0^{(1)}\right)-\mu_4^{(1)}}.
\eeq
Note that all the quantities appearing in Eq.\ \eqref{2.27} are evaluated in the steady state. In the case that $a_{2,\text{s}}$ is not neglected in the steady state condition ($\Lambda=0$), the quantities $\theta_\text{s}$ and $a_{2,\text{s}}$ must be obtained by numerically solving Eq.\ \eqref{2.27} and the equation
\beq
\label{2.27.1}
2\lambda \theta_\text{s}^{-1/2}\left(\theta_{\text{s}}^{-1}-1\right)-\left(\zeta_0^{(0)}+\zeta_0^{(1)}a_{2,\text{s}}\right)=0.
\eeq
In addition, an estimate of $p^*$ can be made by considering the Sonine approximation \eqref{2.15} in Eq.\ \eqref{2.11.2}. Neglecting nonlinear terms in $a_2$, the expression of $p^*$ is
\begin{equation}
\label{2.28}
p^*=1+2^{d-2}\chi\phi\left[1+\al + 2\sqrt{\frac{2}{\pi}} \left(1-\frac{1}{16}a_2\right)\theta^{-1/2}\Delta_{\text{b}}^* \right].
\end{equation}

\begin{figure}[t]
\begin{center}
\begin{tabular}{lr}
\resizebox{6.5cm}{!}{\includegraphics{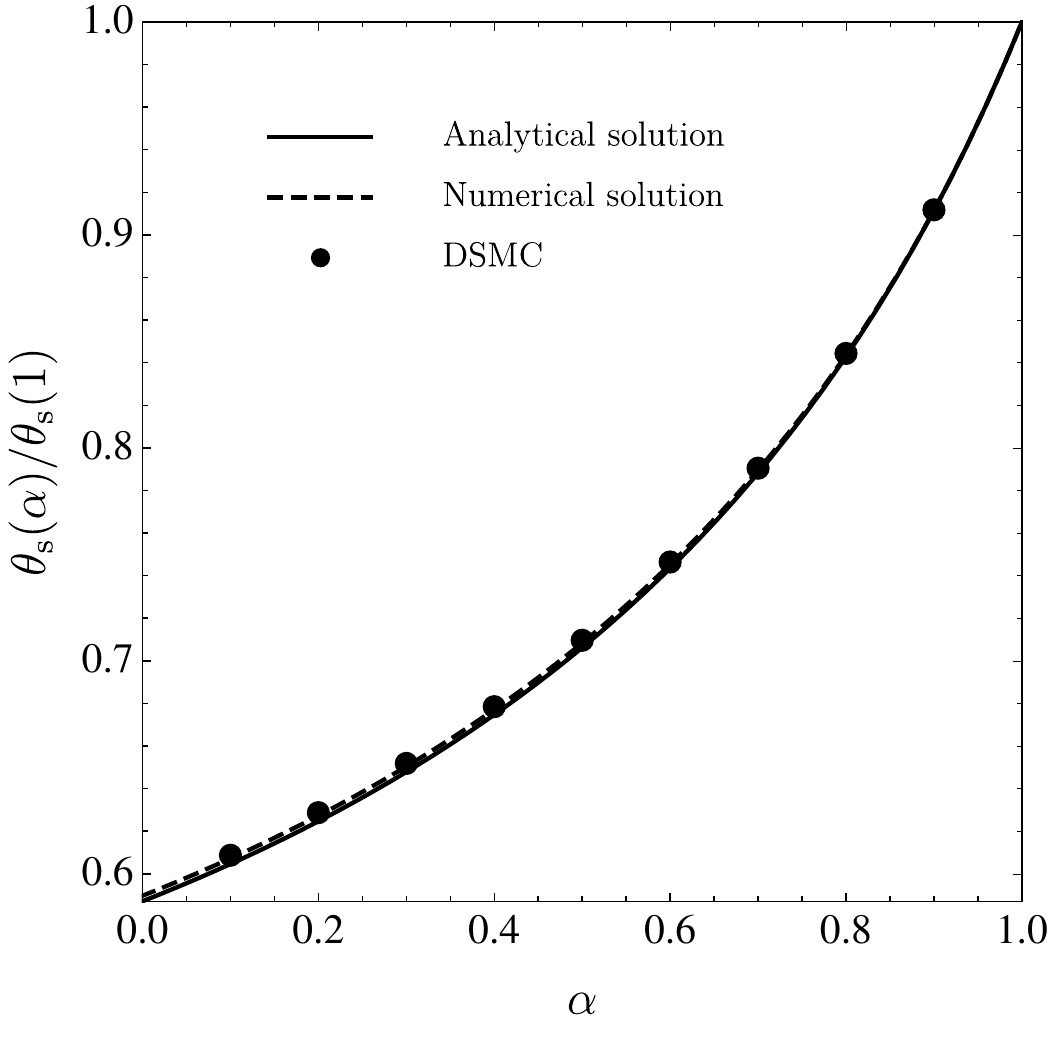}}
\end{tabular}
\end{center}
\caption{Plot of the (scaled) steady temperature $\theta_\text{s}(\alpha)/\theta_\text{s}(1)$ versus the coefficient of restitution $\al$
for $d=2$, $\phi=0.1$, $\Delta_{\text{b}}^*=1$, and $T_\text{b}^*=0.8$. Here, $\theta_\text{s}(1)$ refers to the value of the steady temperature for elastic collisions. The solid line is the theoretical result obtained from Eq.\ \eqref{2.26.1} while the dashed line corresponds to the theoretical result obtained by numerically solving Eqs.\ \eqref{2.27} and \eqref{2.27.1}. Symbols are the DSMC results. Here, we have assumed that $R(\phi)=1$.
\label{fig_temperatura}}
\end{figure}

\begin{figure}[!h]
\centering
\includegraphics[width=0.35\textwidth]{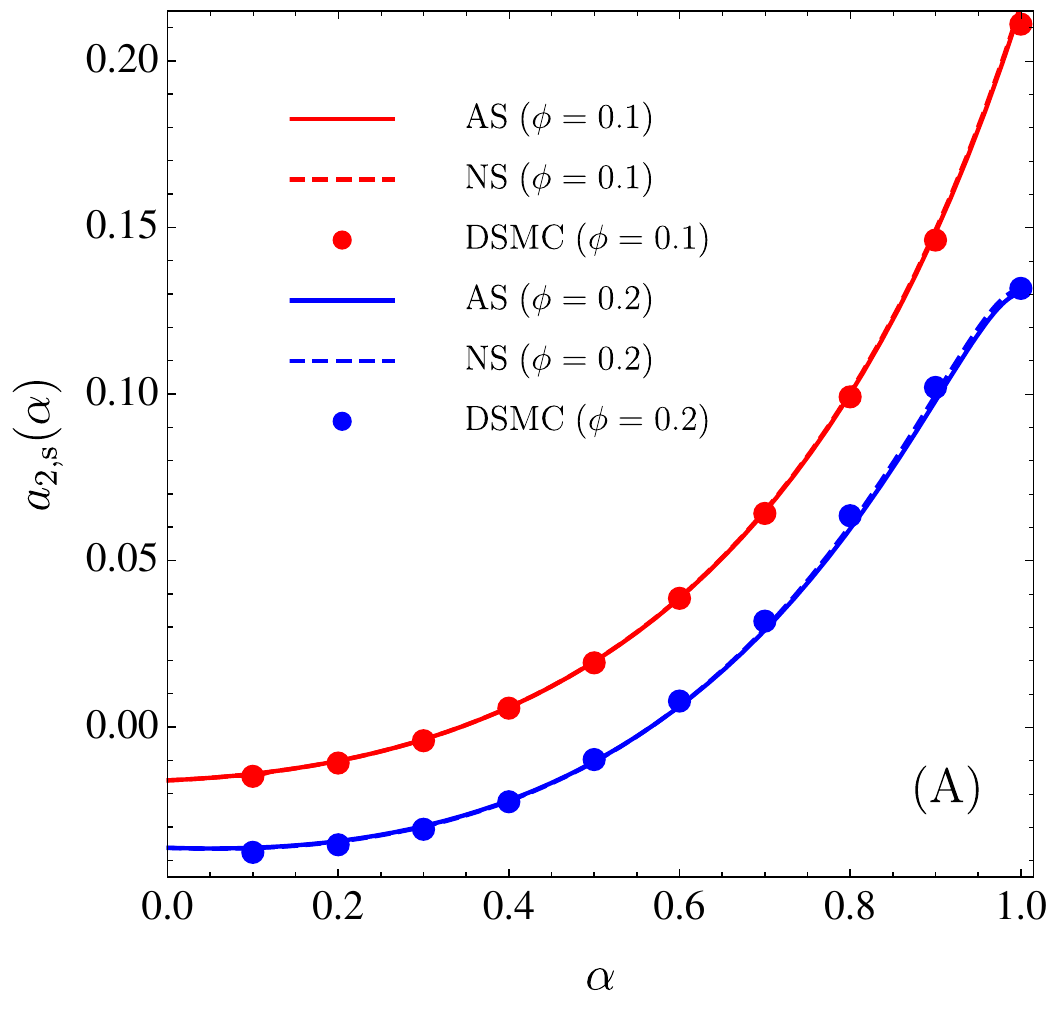}
\includegraphics[width=0.35\textwidth]{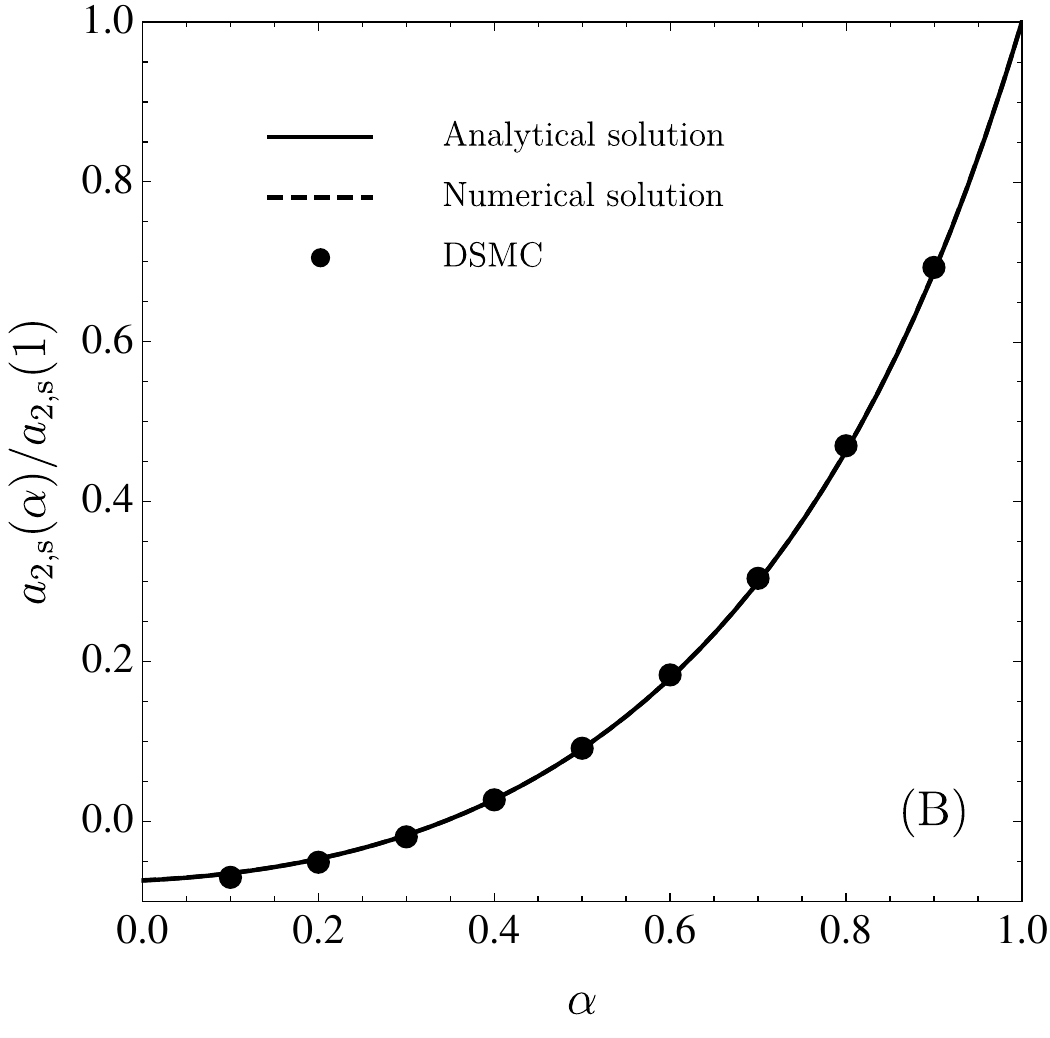}
\caption{(A) Plot of the steady fourth--cumulant $a_{2,\text{s}}(\alpha)$ versus the coefficient of restitution $\al$
for $d=2$, $\Delta_{\text{b}}^*=1$, $T_\text{b}^*=0.8$, and two different volume fractions: $\phi=0.1$ and $\phi=0.2$. The solid lines are the theoretical results obtained from Eq.\ \eqref{2.27} while the dashed lines correspond to the theoretical results obtained by numerically solving Eqs.\ \eqref{2.27} and \eqref{2.27.1}. Symbols are the DSMC results. Here, we have assumed that $R(\phi)=1$. (B) Plot of the (scaled) steady fourth--cumulant $a_{2,\text{s}}(\alpha)/a_{2,\text{s}}(1)$ versus the coefficient of restitution $\al$ for $d=2$, $\Delta_{\text{b}}^*=1$, $T_\text{b}^*=0.8$, and $\phi=0.1$. Here, $a_{2,\text{s}}(1)$ refers to the value of the steady fourth--cumulant for elastic collisions. The solid and dashed lines refer to the analytical and
numerical results, respectively, while symbols denote Monte Carlo simulations.}
\label{fig_a2}
\end{figure}

Figure \ref{fig_temperatura} shows the $\al$--dependence of the steady temperature $\theta_\text{s}(\alpha)/\theta_\text{s}(1)$ (scaled with respect to its elastic value $\theta_\text{s}(1)$) for hard disks ($d=2$) with the solid volume fraction $\phi=0.1$ and the parameters $\Delta_{\text{b}}^*=1$, and $T_\text{b}^*=0.8$. Since we are not aware in the granular literature of any prediction for the function $R(\phi)$ for a two--dimensional system, we have assumed here that $R(\phi)=1$. For hard disks, a good approximation for $\chi(\phi)$ is \cite{T95}
\beq
\label{2.29}
\chi(\phi)=\frac{1-\frac{7}{16}\phi}{(1-\phi)^2}.
\eeq
Figure \ref{fig_temperatura} provides two different theoretical estimates of the steady temperature $\theta_\text{s}$: (i) the analytical expression of $\theta_\text{s}$ obtained by neglecting $a_{2,\text{s}}$ and given by Eq.\ \eqref{2.26.1} and (ii) the numerical solution of the set of coupled equations \eqref{2.27} and \eqref{2.27.1} that provide $\theta_\text{s}$ and $a_{2,\text{s}}$. These theoretical results are compared with those obtained by numerically solving the Enskog equation from the DSMC method \cite{B94}. As expected, the steady temperature decreases with increasing inelasticity in collisions at a given value of $\Delta_{\text{b}}^*$ (parameter that accounts for the injection of energy). Additionally, the two theoretical predictions are indistinguishable and exhibit excellent agreement with Monte Carlo simulations across the full range of the coefficient of restitution values.

The dependence of the kurtosis $a_{2,\text{s}}(\alpha)$ on the coefficient of restitution $\al$ is plotted in panel (A) of Fig.\ \ref{fig_a2} for $d=2$, $\Delta_{\text{b}}^*=1$, $T_\text{b}^*=0.8$, and two different volume fractions ($\phi=0.1$ and $\phi=0.2$). For both densities \david{and the parameters considered here}, $a_{2,\text{s}}$ increases monotonically with $\alpha$, approaching larger positive values as the system approaches $\al=1$. In contrast, in the strongly inelastic regime ($\alpha \lesssim 0.3$), the kurtosis becomes small and can even take slightly negative values.
\david{It is quite apparent from Fig. \ref{fig_a2}A that}, at fixed $\alpha$, increasing the volume fraction systematically reduces the magnitude of the kurtosis, showing that denser systems are closer to Gaussian statistics \david{in the region where $a_{2,\text{s}}$ is positive}. 
\david{Moreover, the dependence of $a_{2,\text{s}}$ on the density $\phi$ in the $\Delta$--model appears to be weaker than the one observed in unconfined granular suspensions \cite{GGG19a}}. As with the case of $\theta_\text{s}$, there is excellent agreement between the two theoretical results and the simulations, even for extreme values of inelasticity and density. To complement panel (A), panel (B) shows the ratio $a_{2,\text{s}}(\alpha)/a_{2,\text{s}}(1)$ versus $\al$. As expected, this ratio decreases with increasing inelasticity.

\david{However, the behavior of the kurtosis $a_{2,\text{s}}$, as illustrated in Fig.\ \ref{fig_a2}, should be interpreted carefully. Although not shown here, an extensive analysis of the dependence of $a_{2,\text{s}}$ on the space of parameters reveals that the sign and magnitude of $a_{2,\text{s}}$ do not follow a general trend with either $\alpha$ or $\phi$, and may depend sensitively on the value of $\Delta_\text{b}^*$. This conclusion contrasts with the results obtained for the conventional IHS model \cite{BP04, G19}, where increasing inelasticity usually enhances the non--Gaussian corrections measured by the kurtosis $a_{2,\text{s}}$. In fact, the present $\Delta$--model involves two competing collisional mechanisms: the energy loss controlled by the coefficient of restitution $\alpha$ and the collisional energy injection controlled by the parameter $\Delta$.

Strictly speaking, a decrease in the absolute value of $a_{2,\text{s}}$ should be understood only as a reduction of the fourth--order deviation from the Maxwellian distribution. In other words, a lower absolute value of $a_{2,\text{s}}$ does not necessarily imply that the full velocity distribution is closer to a Gaussian distribution. Higher--order cumulants or high--energy tails may still exhibit significant non--Gaussian features \cite{BP06a, HOB00}. Within the first Sonine approximation used here, $a_{2,\text{s}}$ quantifies the leading correction to Maxwellian statistics; however, it should not be regarded as a complete measure of Gaussianity.
}

\section{Chapman--Enskog expansion. First order approximation}
\label{sec4}

Once the homogeneous state is well characterized, the next step is to solve the Enskog equation \eqref{1.14} by assuming that the time--dependent homogeneous state is slightly perturbed by small spatial gradients. These perturbations give rise to nonzero contributions to the pressure tensor and the heat flux. The knowledge of the fluxes enables us to identify the corresponding Navier--Stokes transport coefficients. As in previous studies \cite{GGG19a,GBS18}, to obtain those transport coefficients the Enskog equation is solved to first order in spatial gradients by means of the Chapman--Enskog method \cite{CC70} conveniently adapted to dissipative dynamics. The Chapman--Enskog method assumes the existence of a normal solution where all the space and time dependence of the distribution function only occurs through a functional dependence on the hydrodynamic fields $n$, $\mathbf{U}$, and $T$. For the sake of simplicity \cite{CC70}, this functional dependence can be made explicit by assuming small spatial gradients and hence, $f(\mathbf{r}, \mathbf{v}; t)$ is written as a series expansion in powers of $\Omega\equiv (\nabla n, \nabla_i U_j, \nabla T)$:
\beq
\label{3.0}
f=f^{(0)}+f^{(1)}+\cdots,
\eeq
where the distribution $f^{(k)}$ is of order $k$ in $\Omega$. The expansion \eqref{3.0} leads to similar expansions for the fluxes, the cooling rate and the Enskog collision operator:
\beq
\label{3.0.1}
\mathsf{P}=\mathsf{P}^{(0)}+\mathsf{P}^{(1)}+\cdots, \quad \mathbf{q}=\mathbf{q}^{(0)}+\mathbf{q}^{(1)}+\cdots,
\eeq
\beq
\label{3.0.2}
\zeta=\zeta^{(0)}+\zeta^{(1)}+\cdots, \quad J_\text{E}=J_\text{E}^{(0)}+J_\text{E}^{(1)}+\cdots.
\eeq
Similarly, the time derivatives $\partial_t$ must be also expanded as
\beq
\label{3.0.3}
\partial_t=\partial_t^{(0)}+\partial_t^{(1)}+\cdots.
\eeq
The action of the operators $\partial_t^{(k)}$ on the hydrodynamic fields $n$, $\mathbf{U}$, and $T$ is obtained by substituting the expansions \eqref{3.0.1}--\eqref{3.0.3} in the balance equations \eqref{1.18}--\eqref{1.20} and collecting terms of the same order in the spatial gradients. Moreover, in ordering the different levels of approximations in the Enskog kinetic equation, we have to characterize the magnitude of $\gamma$ and $\Delta \mathbf{U}$ relative to the gradients. As in previous studies \cite{GGG19a}, since $\gamma$ does not create any flux in the system it is considered to be of zeroth--order in gradients. With respect to the difference $\Delta \mathbf{U}$, it is expected to be at least of first order in gradients because $\mathbf{U}$ tends to $\mathbf{U}_\text{g}$ in the homogeneous state. In this paper, we will restrict the calculations to the first order in spatial gradients.

\subsection{Zeroth--order approximation}

In the absence of spatial gradients, $f^{(0)}$ verifies the Enskog equation
\beq
\label{3.1}
\frac{\partial f^{(0)}}{\partial t}-\gamma\frac{\partial}{\partial \mathbf{v}}\cdot \mathbf{v} f^{(0)}-\gamma \frac{T_{\text{b}}}{m}\frac{\partial^2 f^{(0)}}{\partial v^2}=J_\text{E}^{(0)}[\mathbf{V}|f^{(0)},f^{(0)}],
\eeq
where $J_\text{E}^{(0)}[\mathbf{V}|f^{(0)},f^{(0)}]$ is given by Eq.\ \eqref{2.2} with the replacement $f(\mathbf{v}; t)\to f^{(0)}(\mathbf{r}, \mathbf{v}; t)$. At this order, $\partial_t^{(0)}n=0$, $\partial_t^{(0)}\mathbf{U}=\mathbf{0}$, and
\beq
\label{3.2}
\partial_t^{(0)} T=2\gamma\left(T_{\text{b}}-T\right)-\zeta^{(0)} T.
\eeq
Here, $\zeta^{(0)}$ is determined from Eq.\ \eqref{2.5} to zeroth--order. A good approximation to $\zeta_0^*=\ell \zeta^{(0)}/v_\text{th}$ is given by Eqs.\ \eqref{2.16} and \eqref{2.17}.

Equation \eqref{3.1} has the same form as the corresponding Enskog equation \eqref{2.4} for a homogeneous state when one rewrites \eqref{3.1} in terms of the derivative $\partial_T f^{(0)}$. This means that $f^{(0)}(\mathbf{r}, \mathbf{v}; t)$ is in fact the \emph{local} version of the time--dependent homogeneous distribution function, as expected. Thus, the zeroth--order solution can be written as
\beq
\label{3.3}
f^{(0)}(\mathbf{r}, \mathbf{v},\gamma, T_\text{b},\Delta)=n(\mathbf{r};t) v_\text{th}(\mathbf{r};t)^{-d}\varphi(\mathbf{c},\gamma^*, \theta),
\eeq
where the scaled distribution $\varphi(\mathbf{c},\gamma^*, \theta)$ verifies the \emph{unsteady} equation
\begin{align}
\label{3.4}
\left[2\gamma^*\left(\theta^{-1}-1\right)-\zeta_0^*\right]\theta\frac{\partial\varphi}{\partial\theta}
+
\left(\frac{\zeta_0^*}{2}-\gamma^*\theta^{-1}\right)\frac{\partial}{\partial \mathbf{c}}\cdot \left(\mathbf{c}\varphi \right) \nonumber\\
-\frac{\gamma^*}{2\theta}\frac{\partial^2 \varphi}{\partial c^2}=J_{\text{E}}^*[\mathbf{c}|\varphi,\varphi].
\end{align}
Here, $J_{\text{E}}^*[\mathbf{c}|\varphi,\varphi]=\ell v_\text{th}^{d-1}J_{\text{E}}[\mathbf{v}|f,f]/n$.

Although Eq.~\eqref{3.4} formally determines the unsteady scaled distribution $\varphi$, its exact solution is not generally available. Instead, the leading deviations from Maxwellian behavior can be characterized through the kurtosis $a_2(\theta)$. Deriving its evolution equation is therefore useful both to describe the relaxation towards the steady state and to determine the transport coefficients later on. The evolution equation for $a_2(\theta)$ follows from Eq.~\eqref{3.4} as
\beq
\label{3.5}
\theta \frac{\partial a_2}{\partial \theta}=\frac{\frac{4}{d(d+2)}\mu_4
-2\Lambda(1+a_2)
-4\gamma^*\left(1+ a_2-\theta^{-1}\right)}{\Lambda},
\eeq
where the quantity $\Lambda$ is defined by Eq.\ \eqref{2.11.1} with the  replacement $\zeta^*\to \zeta_0^*$.
Equation \eqref{3.5} is equivalent to Eq.\ \eqref{2.13}. In the absence of gas phase, Eq.\ \eqref{3.5} has been numerically solved by using different initial conditions \cite{BMGB14}. The theoretical predictions have been also compared to numerical results obtained from the DSMC method \cite{B94} and a good agreement has been found. In the steady state ($\Lambda=0$), $a_2\equiv a_{2,\text{s}}$ is given by Eq.\ \eqref{2.27}. Moreover, as we will show later, to determine the transport coefficients in the steady state we need to know the derivatives $\Upsilon_\theta\equiv (\partial a_2/\partial \theta)_\text{s}$, $\Upsilon_\lambda\equiv (\partial a_2/\partial \lambda)_\text{s}$, and $\Upsilon_\chi\equiv (\partial a_2/\partial \chi)_\text{s}$. These derivatives must be evaluated in the  HSS and they give indirect information on the departure of the time--dependent zeroth--order solution $f^{(0)}$ from its stationary form. As expected, according to Eq.\ \eqref{3.5}, the derivative $\Upsilon_\theta$ becomes indeterminate since the numerator and denominator of the above equation vanish. To obtain $\Upsilon_\theta$ we expand first the numerator and denominator  of Eq.\ \eqref{3.5} to first order in $a_2$:
\begin{widetext}
\beq
\label{3.6}
\theta \frac{\partial a_2}{\partial \theta}=\frac{\frac{4}{d(d+2)}\mu_4^{(0)}+2\zeta_0^{(0)}+2\left(\frac{2}{d(d+2)}
\mu_4^{(1)}-2 \gamma^*\theta^{-1}+\zeta_0^{(0)}+\zeta_0^{(1)}\right)a_2}{2\gamma^* \left(\theta^{-1}-1\right)-\zeta_0^{(0)}-\zeta_0^{(1)}a_2}\equiv \frac{X_1+X_2 a_2}{Y_1+Y_2 a_2},
\eeq
\end{widetext}
where the expressions of $X_1$, $X_2$, $Y_1$ and $Y_2$ can be identified from Eq.\ \eqref{3.6}. The derivative $\Upsilon_\theta\equiv \partial_\theta a_2$ can be determined by applying l'H\^opital's rule to Eq.\ \eqref{3.6}. The result leads to the following quadratic equation for $\Upsilon_\theta$:
\beq
\label{3.7}
\theta Y_2 \Upsilon_\theta^2+\left[\theta\left(Y_{1,\theta}+Y_{2,\theta}a_{2,\text{s}}\right)-X_2\right]\Upsilon_\theta-\left(X_{1,\theta}+X_{2,\theta} a_{2,\text{s}}\right)=0,
\eeq
where $X_{i,\theta}\equiv \partial_\theta X_i$ and $Y_{i,\theta}\equiv \partial_\theta Y_i$. However, since the magnitude of $\Upsilon_\theta$ is in general quite small, the quadratic term in $\Upsilon_\theta$ in \eqref{3.7} can be neglected and hence, one obtains the simple expression
\beq
\label{3.8}
\Upsilon_\theta=\frac{X_{1,\theta}+X_{2,\theta} a_{2,\text{s}}}{\theta\left(Y_{1,\theta}+Y_{2,\theta}a_{2,\text{s}}\right)-X_2}.
\eeq

\begin{figure}[t]
\begin{center}
\begin{tabular}{lr}
\resizebox{6.55cm}{!}{\includegraphics{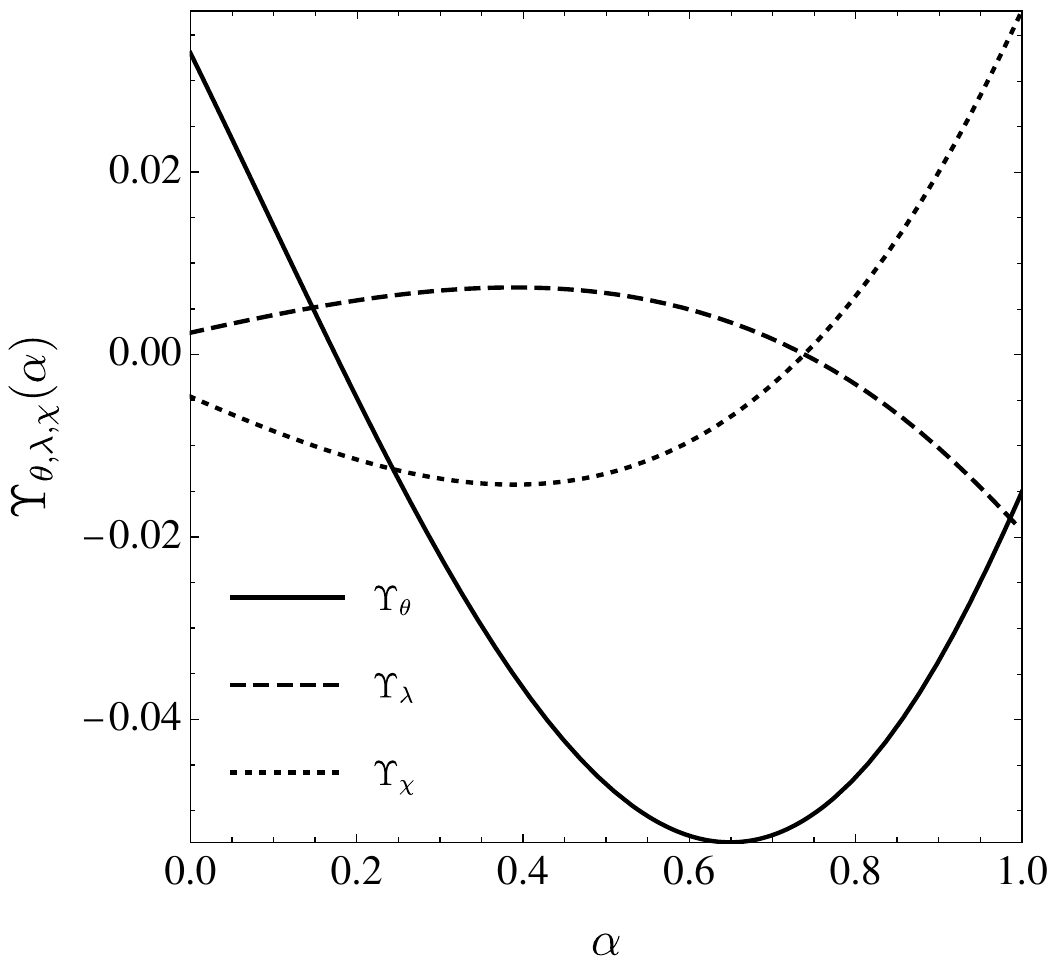}}
\end{tabular}
\end{center}
\caption{Plot of the derivatives $\Upsilon_\theta$, $\Upsilon_\lambda$, and $\Upsilon_\chi$ as a function of the coefficient of restitution $\al$
for $d=2$, $\phi=0.2$, $\Delta_{\text{b}}^*=0.5$, and $T_\text{b}^*=1$.
\label{fig_derivatives}}
\end{figure}

The derivatives $\Upsilon_\lambda\equiv \partial_\lambda a_2$ and $\Upsilon_\chi\equiv \partial_\chi a_2$ can be easily derived from Eq.\ \eqref{3.6} with the results
\beq
\label{3.9}
\Upsilon_\lambda=\frac{\left(Y_{1,\lambda}+Y_{2,\lambda}a_{2,\text{s}}\right)\theta\Upsilon_\theta-
X_{1,\lambda}-X_{2,\lambda}a_{2,\text{s}}}
{X_2-Y_2 \theta \Upsilon_\theta},
\eeq
\begin{equation}
\label{3.9.1}
\Upsilon_\chi=\frac{\left(Y_{1,\chi}+Y_{2,\chi}a_{2,\text{s}}\right)\theta\Upsilon_\theta-X_{1,\chi}-
X_{2,\chi}a_{2,\text{s}}}
{X_2-Y_2 \theta \Upsilon_\theta}.
\end{equation}

The $\al$--dependence of the above derivatives is illustrated in Fig.\ \ref{fig_derivatives} for $d=2$, $\phi=0.2$, $\Delta_{\text{b}}^*=0.5$, and $T_\text{b}^*=1$. We consider here the expression \eqref{3.8} for $\Upsilon_\theta$ for the sake of simplicity. We observe that, for the parameter values displayed in Fig.\ \ref{fig_derivatives}, the magnitude of these derivatives is in general quite small. The same conclusion has been found after exploring a broad region of the parameter space relevant to the present study. Nevertheless, since Eqs.\ \eqref{3.8}--\eqref{3.9.1} involve denominators that may become small for certain combinations of the parameters, this behavior should not be regarded as a general property over the whole parameter space. In what follows, the derivatives are evaluated at the HSS for each set of parameters considered. Finally, we also observe that these derivatives exhibit a complex dependence on the coefficient of restitution.

\subsection{First--order approximation}

Since the application of the Chapman--Enskog method to first order in spatial gradients follows similar mathematical steps to those used in the conventional IHS model (with gas--phase) \cite{GGG19a} and in the $\Delta$--model (without gas--phase) \cite{GBS18,GBS20,GBS26}, the complete derivation of the kinetic equation for the first--order distribution $f^{(1)}(\mathbf{V})$ is omitted in this paper. We refer the interested reader to these articles \cite{GGG19a,GBS18} for specific details.

The first--order velocity distribution function $f^{(1)}$ is given by
\beqa
\label{3.10}
f^{(1)}(\mathbf{V})&=&\boldsymbol{\mathcal{A}}(\mathbf{V})\cdot\nabla\ln T+\boldsymbol{\mathcal{B}}(\mathbf{V})\cdot\nabla\ln n \nonumber\\
& & + \,\mathcal{C}_{ij}(\mathbf{V})\frac{1}{2}\left(\frac{\partial U_i}{\partial r_j}+\frac{\partial U_j}{\partial r_i}-\frac{2}{d}\delta_{ij}\nabla\cdot\mathbf{U}\right)\nonumber\\
& & +\,\mathcal{D}(\mathbf{V})\nabla\cdot\mathbf{U}.
\eeqa
In the steady state ($\Lambda=0$), the quantities $\boldsymbol{\mathcal{A}}$, $\boldsymbol{\mathcal{B}}$, $\mathcal{C}_{ij}$, and $\mathcal{D}$ verify the following set of coupled linear integral equations:
\begin{align}
\label{3.11}
-\left(2\gamma\theta^{-1}
+\frac{1}{2}\zeta^{(0)}+\zeta^{(0)}\theta \frac{\partial\ln \zeta_0^*}{\partial\theta}\right)
\boldsymbol{\mathcal{A}}-\gamma\frac{\partial}{\partial\mathbf{v}}\cdot\mathbf{V}\boldsymbol{\mathcal{A}}\nonumber\\
-\frac{\gamma T_{\text{b}}}{m}\frac{\partial^2}{\partial v^2}\boldsymbol{\mathcal{A}}+\mathcal{L}\boldsymbol{\mathcal{A}}=\mathbf{A},
\end{align}
\beqa
\label{3.12}
& & -\gamma\frac{\partial}{\partial\mathbf{v}}\cdot\mathbf{V}\boldsymbol{\mathcal{B}}-\frac{\gamma T_{\text{b}}}{m}\frac{\partial^2}{\partial v^2}\boldsymbol{\mathcal{B}}+\mathcal{L}\boldsymbol{\mathcal{B}}=\mathbf{B}+
\nonumber\\
& & 
\left[\zeta^{(0)}\left(1+\phi
\frac{\partial\ln\chi}{\partial\phi}\right)
+\chi\phi\frac{\partial\chi}{\partial\phi}\frac{\partial}{\partial\chi}\left(\frac{\zeta^{(0)}}{\chi}\right)\right.\nonumber\\
& &\left.-\lambda
\left(1-\phi\frac{\partial\ln R}{\partial \phi}\right)\frac{\partial \zeta^{(0)}}{\partial \lambda}
-2\gamma\left(\theta^{-1}-1\right)\phi \frac{\partial\ln R}{\partial \phi}\right]\boldsymbol{\mathcal{A}},
\nonumber\\
\eeqa
\beq
\label{3.13}
-\gamma\frac{\partial}{\partial\mathbf{v}}\cdot\mathbf{V}\mathcal{C}_{ij}-\frac{\gamma T_{\text{b}}}{m}\frac{\partial^2}{\partial v^2}\mathcal{C}_{ij}+\mathcal{L}\mathcal{C}_{ij}=C_{ij},
\eeq
\beq
\label{3.14}
-\gamma\frac{\partial}{\partial\mathbf{v}}\cdot\mathbf{V}\mathcal{D}-\frac{\gamma T_{\text{b}}}{m}\frac{\partial^2}{\partial v^2}\mathcal{D}-\zeta^{(1,1)} T\frac{\partial f^{(0)}}{\partial T}+\mathcal{L}\mathcal{D}=D.
\eeq
In Eqs.\ \eqref{3.11}--\eqref{3.14},
\beq
\label{3.14.1}
\mathcal{L} X=-\left(J_\text{E}^{(0)}[f^{(0)},X]+J_\text{E}^{(0)}[X,f^{(0)}]\right)
\eeq
is the linearized Boltzmann--Enskog collision operator and
the functions $\mathbf{A}(\mathbf{V})$, $\mathbf{B}(\mathbf{V})$, $C_{ij}(\mathbf{V})$, and $D(\mathbf{V})$ are
\begin{align}
\label{3.15}
\mathbf{A}(\mathbf{V})=-\mathbf{V}T\frac{\partial f^{(0)}}{\partial T}-\frac{p}{\rho}\left(1+\theta\frac{\partial \ln p^*}{\partial \theta}\right)\frac{\partial f^{(0)}}{\partial\mathbf{V}}\nonumber\\
-\boldsymbol{\mathcal{K}}\left[T\frac{\partial f^{(0)}}{\partial T}\right],
\end{align}
\beqa
\label{3.16}
\mathbf{B}(\mathbf{V})&=&-\mathbf{V}n\frac{\partial f^{(0)}}{\partial n}-\frac{p}{\rho}\left(1+\phi\frac{\partial \ln p^*}{\partial\phi}\right)\frac{\partial f^{(0)}}{\partial\mathbf{V}}
\nonumber\\
& &
-\boldsymbol{\mathcal{K}}\left[n
\frac{\partial f^{(0)}}{\partial n}\right]-\frac{1}{2}\phi\left(\frac{\partial \ln\chi}{\partial\phi}\right)\boldsymbol{\mathcal{K}}\left[f^{(0)}\right],
\nonumber\\
\eeqa
\beq
\label{3.17}
C_{ij}(\mathbf{V})=V_i\frac{\partial f^{(0)}}{\partial V_j}+\mathcal{K}_i\left[\frac{\partial f^{(0)}}{\partial V_j}\right],
\eeq
\beqa
\label{3.18}
D(\mathbf{V})&=&\frac{1}{d}\frac{\partial}{\partial\mathbf{V}}\cdot\left(\mathbf{V}f^{(0)}\right)+
\left(\zeta^{(1,0)}+\frac{2}{d}p^*\right)
T\frac{\partial f^{(0)}}{\partial T}\nonumber\\
& & -f^{(0)}+n\frac{\partial f^{(0)}}{\partial n}+\frac{1}{d}\mathcal{K}_i\left[\frac{\partial f^{(0)}}{\partial V_i}\right].
\eeqa
In Eqs.\ \eqref{3.15}--\eqref{3.18}, the operator $\boldsymbol{\mathcal{K}}[X]$ is defined as \cite{GBS18}:
\beqa
\label{3.19}
& & \boldsymbol{\mathcal{K}}[X]= -\sigma^{d}\chi\int \dd \mathbf{v}_{2}\int \dd\widehat{\boldsymbol {\sigma
}}\;\Theta (-\widehat{\boldsymbol {\sigma}} \cdot
\mathbf{g}-2\Delta)\nonumber\\
& & \times (-\widehat{\boldsymbol {\sigma }}\cdot
\mathbf{g}-2\Delta)
\widehat{\boldsymbol {\sigma }} \,\alpha^{-2}f^{(0)}(\mathbf{v}_{1}'')X(\mathbf{v}_{2}'')+\sigma^{d}\chi
\nonumber\\
& &  \times \int \dd \mathbf{v}_{2}\int \dd\widehat{\boldsymbol {\sigma
}}\;\Theta (\widehat{\boldsymbol {\sigma}} \cdot
\mathbf{g}) (\widehat{\boldsymbol {\sigma }}\cdot
\mathbf{g})
\widehat{\boldsymbol {\sigma }} f^{(0)}(\mathbf{v}_{1})X(\mathbf{v}_{2}).\nonumber\\
\eeqa

In addition, upon writing Eqs.\ \eqref{3.14} and \eqref{3.18} we have accounted for the fact that the first--order contribution to the cooling rate $\zeta^{(1)}=\zeta_U \nabla \cdot \mathbf{U}$, where
\beq
\label{3.20}
\zeta_U=\zeta^{(1,0)}+\zeta^{(1,1)},
\eeq
\beqa
\label{3.21}
\zeta^{(1,0)}&=&-\frac{2^{d-1}}{d}\phi \chi \left[\frac{3}{2}(1-\al^2)-2 \theta^{-1}\Delta_\text{b}^{*2}\right.
\nonumber\\
& & \left.
-4\sqrt{\frac{2}{\pi}}\left(1-\frac{1}{16}a_2\right)\alpha \theta^{-1/2}\Delta_\text{b}^{*}\right],
\eeqa
\begin{align}
\label{3.22}
\zeta^{(1,1)}=-\frac{2m\pi^{\frac{d-1}{2}}}{dnT}\chi \sigma^{d-1}\int \dd\mathbf{v}_1 \int \dd\mathbf{v}_2
\Bigg[\frac{\Delta^2}{\Gamma\left(\frac{d+1}{2}\right)}g \nonumber \\+\frac{\sqrt{\pi}}{d\Gamma\left(\frac{d}{2}\right)}\al \Delta g^2 -\frac{1-\al^2}{4\Gamma\left(\frac{d+3}{2}\right)}g^3\Bigg]f^{(0)}(\mathbf{V}_1)\mathcal{D}(\mathbf{V}_2).
\end{align}
Upon obtaining the expression \eqref{3.21} for $\zeta^{(1,0)}$ use has been made of the Sonine approximation \eqref{2.15} and only linear terms in $a_2$ have been retained.

When $\Delta_\text{b}^*=0$, the expressions obtained in this subsection agree with those derived in the IHS model for granular suspensions \cite{GGG19a}. Moreover, when $\gamma=0$, the results obtained within the $\Delta$--model for a dry granular fluid are also recovered, as expected \cite{GBS18,GBS26}.

\section{Navier--Stokes transport coefficients and first--order contribution to the cooling rate}
\label{sec5}

\subsection{Pressure tensor}

The constitutive equation for the pressure tensor $P_{ij}^{(1)}$  to first order in the spatial gradients is given by
\beq
\label{4.1}
P_{ij}^{(1)}=-\eta \left(\frac{\partial U_i}{\partial r_j}+\frac{\partial U_j}{\partial r_i}-\frac{2}{d}\delta_{ij}\nabla\cdot\mathbf{U}\right)
-\eta_\text{b}\delta_{ij}\nabla \cdot \mathbf{U},
\eeq
where $\eta$ is the shear viscosity and $\eta_\text{b}$ is the bulk viscosity. While $\eta$ has kinetic and collisional contributions ($\eta=\eta_\text{k}+\eta_\text{c}$), $\eta_\text{b}$ has only collisional contributions and hence, it vanishes for dilute gases. The forms of the collisional contributions to both transport coefficients are exactly the same as those obtained in the dry granular case (namely, in the absence of the interstitial gas). These expressions were previously derived in preceding works \cite{GBS18,GBS26} by neglecting both non--Gaussian corrections ($a_{2,\text{s}}=0$) to the zeroth--order solution $f^{(0)}$ and one contribution to the bulk viscosity. Here, we extend these expressions by considering these corrections  \footnote{While the present study was being conducted, some typos were found in the forms obtained in Ref.\ \cite{GBS18}. The expressions displayed here are the corrected results.}. The expression of the collisional contribution $\eta_\text{c}$ to the shear viscosity $\eta$ is
\begin{align}
\label{4.2}
\eta_\text{c}=2^{d-1}\chi \phi \Bigg[\frac{1+\al}{(d+2)} +\frac{d}{\sqrt{\pi}(d+1)}  
\frac{\Gamma\left(\frac{d}{2}\right)}{\Gamma\left(\frac{d+1}{2}\right)}
 \nonumber\\
\times \theta^{-1/2}\Delta_\text{b}^*I_\eta
\Bigg]\eta_\text{k}+\frac{d}{d+2}\eta_\text{b}',
\end{align}
\begin{align}
\label{4.3}
\eta_\text{b}'=\frac{2^{2d+\frac{1}{2}}}{(d+2)\sqrt{\pi}}\chi \phi^2 \Bigg[\frac{(d+1)}{2\sqrt{\pi}}\frac{\Gamma\left(\frac{d}{2}\right)}{\Gamma\left(\frac{d+3}{2}\right)}
(1+\al)
I_{\eta_\text{b}}'\nonumber\\
+\theta^{-1/2}\Delta_\text{b}^*\Bigg]\eta_0,
\end{align}
where
\beq
\label{4.4}
\eta_0(T)=\frac{d+2}{8}\Gamma\left(\frac{d}{2}\right)
\pi^{-\frac{d-1}{2}}\sigma^{1-d}\sqrt{mT}
\end{equation}
is the low density value of the shear viscosity in the elastic limit. In Eqs.\ \eqref{4.2} and \eqref{4.3}, we have introduced the dimensionless integrals
\beqa
\label{4.5}
I_\eta &=& 4\int \dd\mathbf{c}_1 \int \dd\mathbf{c}_2 \; g^{*-1}g_x^{*}g_y^{*} c_{1,x}c_{1,y}\Bigg[ 1 + \frac{a_2}{2}
\nonumber\\
& & 
\times \left( c_2^4 - (d+2)c_2^2 + \frac{d(d+2)}{4} \right) \Bigg]\varphi_\text{M}(\mathbf{c}_1)
\varphi_\text{M}(\mathbf{c}_2),
\nonumber\\
\eeqa
\beq
\label{4.6}
I_{\eta_\text{b}}'=\int \dd \mathbf{c}_1
\int \dd\mathbf{c}_2\; g^*\; \varphi(\mathbf{c}_1)\varphi(\mathbf{c}_2).
\eeq
It must be remarked that upon obtaining Eq.\ \eqref{4.2} we have considered the leading Sonine expression for the unknown $\mathcal{C}_{ij}(\mathbf{V})$, namely,
\beq
\label{4.7}
\mathcal{C}_{ij}(\mathbf{V})\to -\frac{\eta_\text{k}}{n T^2}D_{ij}(\mathbf{V})f_\text{M}(\mathbf{V}),
\eeq
where $f_\text{M}(\mathbf{V})=n v_\text{th}^{-d}\varphi_\text{M}(\mathbf{c})$ is the Maxwellian distribution function and $D_{ij}(\mathbf{V})$ is the traceless tensor
\beq
\label{4.7.1}
D_{ij}(\mathbf{V})=m \left(V_iV_j-\frac{1}{d}V^2 \delta_{ij}\right).
\eeq

The bulk viscosity $\eta_\text{b}$ is
\beq
\label{4.8}
\eta_\text{b}=\eta_\text{b}'+\eta_\text{b}'',
\eeq
where $\eta_\text{b}'$ is defined by Eq.\ \eqref{4.3} and $\eta_\text{b}''$ is given in terms of the coefficient $e_D$ which is involved in the evaluation of the first--order contribution $\zeta^{(1,1)}$ to the cooling rate. The expression of $\eta_\text{b}''$ is
\beq
\label{4.9}
\eta_\text{b}''=-\frac{2^d\Gamma\left(\frac{d}{2}\right)}{\sqrt{\pi}\Gamma\left(\frac{d+1}{2}\right)}\chi \phi I_{\eta_\text{b}}'' \theta^{-1/2}\Delta_\text{b}^* n T e_D,
\eeq
where
\begin{align}
\label{4.10}
I_{\eta_\text{b}}''=\int \dd \mathbf{c}_1
\int \dd\mathbf{c}_2\; g^*\;  &\left[c_1^4-(d+2)c_1^2+\frac{d(d+2)}{4}\right]\nonumber\\
& \qquad \qquad \times \varphi_\text{M}(\mathbf{c}_1)\varphi(\mathbf{c}_2),
\end{align}
and the coefficient $e_D$ will be evaluated later. As in the case of $\eta_\text{c}$, to obtain Eq.\ \eqref{4.9} use has been made of the approximation
\beq
\label{4.10.1}
\mathcal{D}(\mathbf{V})\to e_D E(\mathbf{V})f_\text{M}(\mathbf{V}),
\eeq
where
\beq
\label{4.10.2}
E(\mathbf{V})=\left(\frac{m V^2}{2T}\right)^2-(d+2)\left(\frac{m V^2}{2T}\right)+\frac{d(d+2)}{4}.
\eeq
The contribution $\eta_\text{b}''$ to $\eta_\text{b}$ was neglected in the derivation performed within the $\Delta$--model for a dry (no gas--phase) granular gas \cite{GBS18}.

The kinetic shear viscosity $\eta_\text{k}$ is defined as
\beq
\label{4.11}
\eta_\text{k}=- \frac{1}{(d-1)(d+2)}\int\; \dd \mathbf{v}\; D_{ij}(\mathbf{V})\; \mathcal{C}_{ij}(\mathbf{V}).
\eeq
To determine $\eta_\text{k}$ one multiplies by $D_{ij}(\mathbf{V})$ both sides of Eq.\ \eqref{3.13} and integrates over velocity.
After some algebra, one gets the result
\beqa
\label{4.12}
& & \eta_\text{k}=\frac{n T}{2\gamma+\nu_\eta}\Bigg[1-\frac{2^{d-2}}{d+2}\chi \phi (1+\al)(1-3\al)\nonumber\\
& & -\frac{2^d}{d+2}\chi \phi 
 \theta^{-1/2}\Delta_\text{b}^*\left(\frac{2\Gamma\left(\frac{d}{2}\right)}{\sqrt{\pi}
\Gamma\left(\frac{d+1}{2}\right)} {I}_\eta' - \theta^{-1/2}\Delta_\text{b}^*\right)\Bigg],
\nonumber\\
\eeqa
where we have introduced the collision frequency
\beq
\label{4.13}
\nu_\eta=\frac{\displaystyle \int \dd\mathbf{v}\,D_{ij}(\mathbf{V}) \mathcal{L}\mathcal{C}_{ij}(\mathbf{V})}
{\displaystyle \int \dd\mathbf{v} \,D_{ij}(\mathbf{V})\mathcal{C}_{ij}(\mathbf{V})},
\eeq
and the (dimensionless) integral
\beq
\label{4.14}
{I}_\eta'=\int \dd{\bf c}_1\int \dd{\bf c}_2\;
\varphi({\bf c}_1)\varphi({\bf c}_2)
\left[g^{*-1} (\mathbf{g}^*\cdot \mathbf{c}_1)-(1+\al)g^*\right].
\eeq
In addition, to obtain Eq.\ \eqref{4.12} we have accounted for the result
\beqa
\label{4.15}
& & \int\; \dd \mathbf{v}
\,D_{ij}(\mathbf{V}) {\cal K}_i\left[\frac{\partial f^{(0)}}{\partial V_j}\right]
\nonumber\\
& & 
=2^{d-2}(d-1)\chi \phi (1+\al)(1-3\al)n T+2^d (d-1)\chi \phi  \nonumber\\
& & \times \theta^{-1/2}
\Delta_\text{b}^*
\left[\frac{2\Gamma\left(\frac{d}{2}\right)}{\sqrt{\pi}
\Gamma\left(\frac{d+1}{2}\right)} {I}_\eta' - \theta^{-1/2}\Delta_\text{b}^*\right] n T.
\eeqa

\subsection{Heat flux}

The heat flux to first order in spatial gradients is
\beq
\label{4.16}
\mathbf{q}^{(1)}=-\kappa \nabla T-\mu \nabla n,
\eeq
where $\kappa$ is the thermal conductivity and $\mu$ is the diffusive heat conductivity. This latter coefficient vanishes for ordinary gases ($\al=1$) in the absence of the external bath ($\gamma=0$). However, $\mu\neq 0$ even for elastic collisions when the gas is driven by a nonconservative external force \cite{PG14}.

The heat flux transport coefficients have kinetic and collisional contributions. As in the case of the shear viscosity, to get the collisional contributions $\kappa_\text{c}$ and $\mu_\text{c}$ to $\kappa$ and $\mu$, respectively, one has to take the leading Sonine approximations to the unknowns $\boldsymbol{\mathcal{A}}(\mathbf{V})$ and $\boldsymbol{\mathcal{B}}(\mathbf{V})$:
\beq
\label{4.17}
\boldsymbol{\mathcal{A}}(\mathbf{V})\to -\frac{2}{d+2}\frac{m}{n T^2}\kappa_\text{k} f_\text{M}(\mathbf{V}) \mathbf{S}(\mathbf{V}), 
\eeq
\beq
\label{4.17.0}
\boldsymbol{\mathcal{B}}(\mathbf{V})\to -\frac{2}{d+2}\frac{m}{T^3}\mu_\text{k} f_\text{M}(\mathbf{V}) \mathbf{S}(\mathbf{V}),
\eeq
where $\kappa_\text{k}$ and $\mu_\text{k}$ are the kinetic contributions to the coefficients $\kappa$ and $\mu$, respectively, and
\beq
\label{4.18}
\mathbf{S}(\mathbf{V})=\left(\frac{m}{2}V^2-\frac{d+2}{2}T\right)\mathbf{V}.
\eeq
The collisional contributions $\kappa_\text{c}$ and $\mu_\text{c}$ can be obtained by expanding to first order in gradients in Eq.\ \eqref{1.23} and by using the approximations \eqref{4.17}. After some algebra, one gets
\beqa
\label{4.19}
\kappa_\text{c}&=&3\frac{2^{d-2}}{d+2}\chi \phi
\Bigg[1+\al+\frac{16}{3(d+1)}\frac{\Gamma\left(\frac{d}{2}\right)}
{\sqrt{\pi}\Gamma\left(\frac{d+1}{2}\right)}\nonumber\\
& & \times 
\theta^{-1/2}\Delta_\text{b}^* I_\kappa'\Bigg]\kappa_\text{k}
+\frac{2^{d+\frac{1}{2}}}{d}\frac{(d-1)}{\pi(d+2)}\chi \phi^2 \nonumber\\
& & \times 
\left[\frac{d\Gamma\left(\frac{d}{2}\right)}
{\Gamma\left(\frac{d+3}{2}\right)}
(1+\al) I_\kappa+2\sqrt{\pi}\theta^{-1/2}\Delta_\text{b}^* I_\kappa''\right]\kappa_0,
\nonumber\\
\eeqa
\begin{align}
\label{4.20}
\mu_\text{c}= 3\frac{2^{d-2}}{d+2}\chi \phi
\Bigg[1+\al+\frac{16}{3(d+1)}\frac{\Gamma\left(\frac{d}{2}\right)}
{\sqrt{\pi}\Gamma\left(\frac{d+1}{2}\right)}\nonumber\\
 \times 
\theta^{-1/2}\Delta_\text{b}^* I_\kappa'\Bigg]\mu_\text{k},
\end{align}
where
\beq
\label{4.21}
\kappa_0=\frac{d(d+2)}{2(d-1)}\frac{\eta_0}{m}
\eeq
is the low density value of the thermal conductivity of an elastic gas. The dimensionless
integrals $I_\kappa$, $I_\kappa'$, and $I_\kappa''$ appearing in Eqs.\ \eqref{4.19} and \eqref{4.20} are given by
\begin{align}
\label{4.22}
I_\kappa=\int\ \dd\mathbf{c}_1 \int\ \dd\mathbf{c}_2\; &\varphi(\mathbf{c}_1)
\varphi(\mathbf{c}_2)\left[g^{*-1}({\bf g}^*\cdot {\bf G}^*)^{2}+g^*G^{*2}\right.\nonumber\\
&\quad \left.
+\frac{3}{2}g^*({\bf g}^*\cdot {\bf G}^*)+\frac{1}{4}g^{*3}\right],
\end{align}
\begin{align}
\label{4.23}
I_\kappa'=\int\ \dd\mathbf{c}_1 \int\dd\mathbf{c}_2 \; \varphi_\text{M}(\mathbf{c}_1)&\varphi(\mathbf{c}_2)g^{*-1}\left[
(\mathbf{g}^*\cdot \mathbf{S}^*)(\mathbf{g}^*\cdot \mathbf{G}^*)\right.
\nonumber\\
& \qquad \left.
+\;g^{*2}(\mathbf{G}^*\cdot \mathbf{S}^*)\right],
\end{align}
\beq
\label{4.24}
I_\kappa''=\frac{d}{2}-2\theta
\int\ \dd\mathbf{c}_1 \int\ \dd\mathbf{c}_2 \; \varphi(\mathbf{c}_1)\;\frac{\partial \varphi(\mathbf{c}_2)}{\partial \theta}\;(\mathbf{g}^*\cdot \mathbf{G}^*).
\eeq
Here, $\mathbf{G}^*\equiv \mathbf{G}/v_\text{th}$ and $\mathbf{S}^*(\mathbf{c})=\left(c^2-\frac{d+2}{2}\right){\bf c}$.

The kinetic contributions to the thermal conductivity $\kappa_\text{k}$ and the diffusive heat conductivity $\mu_\text{k}$ are defined as
\beq
\label{4.25}
\kappa_\text{k}=-\frac{1}{dT}\int \dd{\bf v}\; {\bf S}({\bf V})\cdot {\boldsymbol {\mathcal A}}({\bf V}), 
\eeq
\beq
\label{4.25.0}
\mu_\text{k}=-\frac{1}{dn}\int \dd{\bf v} \;{\bf S}({\bf V})\cdot {\boldsymbol {\mathcal B}}({\bf V}).
\eeq

The kinetic coefficient $\kappa_\text{k}$ can be determined by multiplying both sides of Eq.\ \eqref{3.11} by ${\bf S}({\bf V})$ and integrating over $\mathbf{v}$. The result is
\beqa
\label{4.26}
& & \left(\nu_\kappa+\gamma -\frac{3}{2}\zeta^{(0)}-\zeta^{(0)}\theta \frac{\partial \ln \zeta_0^*}{\partial \theta}\right)\kappa_\text{k}=\frac{d+2}{2m}n T
\nonumber\\
& & \times 
\left(1+2a_2+\theta \Upsilon_\theta\right)
+\frac{nv_\text{th}^{-d}}{d T}
\int \dd\mathbf{V}\;\mathbf{S}(\mathbf{V})\cdot\boldsymbol{\mathcal{K}}\left[\theta \frac{\partial \varphi}
{\partial \theta}\right]
\nonumber\\
& & 
-\frac{1}{2dT}\int\; \dd\mathbf{V}\,
\mathbf{S}(\mathbf{V})\cdot \boldsymbol{\mathcal{K}}\left[\frac{\partial}{\partial \mathbf{V}}\cdot \mathbf{V}f^{(0)}\right],\nonumber\\
\eeqa
where
\begin{equation}
\label{4.27}
\nu_\kappa=\frac{\displaystyle\int \dd{\bf v} \;{\bf S}({\bf V})\cdot {\cal L}\boldsymbol{\mathcal{A}}({\bf V})}
{\displaystyle\int \dd{\bf v}\;{\bf S}({\bf V})\cdot \boldsymbol{\mathcal{A}}({\bf V})}.
\end{equation}
The last two integrals of the right hand side of Eq.\ \eqref{4.26} involving the collision operator $\boldsymbol{\mathcal{K}}$ have been evaluated in the Appendix \ref{appA} by considering the Sonine approximation \eqref{2.15} and neglecting as usual nonlinear terms in $a_2$. As in the case of $\nu_\eta$, the collision frequency $\nu_\kappa$ was also estimated in Ref.\ \cite{BBMG15} by considering the leading Sonine approximation \eqref{4.17} for $\boldsymbol{\mathcal{A}}(\mathbf{V})$ and by replacing the true $\varphi$ by its Sonine approximation \eqref{2.15}. The explicit form of $\nu_\kappa$ is also displayed in the Appendix \ref{appA}.

As for $\kappa_\text{k}$, to determine the kinetic coefficient $\mu_\text{k}$, Eq.\ \eqref{3.12} is multiplied by ${\bf S}({\bf V})$ and then integrated over the velocity:
\beqa
\label{4.28}
& & \left(\nu_\mu-3\gamma\right)\mu_\text{k}=-\frac{1}{dn}\int \dd\mathbf{v}\;\mathbf{S}\cdot \mathbf{B}
+\Bigg[\zeta^{(0)}\left(1+\phi
\frac{\partial\ln\chi}{\partial\phi}\right)
\nonumber\\
& & 
+\chi\phi\frac{\partial\chi}{\partial\phi}\frac{\partial}{\partial\chi}\left(\frac{\zeta^{(0)}}{\chi}\right)-\lambda
\left(1-\phi\frac{\partial\ln R}{\partial \phi}\right)\frac{\partial \zeta^{(0)}}{\partial \lambda}\nonumber\\
& & 
-2\gamma\left(\theta^{-1}-1\right)\phi \frac{\partial\ln R}{\partial \phi}\Bigg]\frac{T}{n}\kappa_\text{k},
\eeqa
where
\begin{equation}
\label{4.29}
\nu_\mu=\frac{\displaystyle \int \dd{\bf v} \;{\bf S}({\bf V})\cdot {\cal L}\boldsymbol{\mathcal{B}}({\bf V})}
{\displaystyle\int \dd{\bf v}\;{\bf S}({\bf V})\cdot \boldsymbol{\mathcal{B}}({\bf V})}.
\end{equation}
The first term on the right--hand side of Eq.\ \eqref{4.28} is given by
\beqa
\label{4.30}
& & -\frac{1}{dn}\int \dd\mathbf{v}\;\mathbf{S}\cdot \mathbf{B}
=\frac{d+2}{2}\frac{T^2}{m}\left(a_2+\phi\frac{\partial a_2}{\partial \phi}\right)
+ \frac{1}{dn}
\nonumber\\
& & 
\times \left(1+\frac{1}{2}\phi\frac{\partial \ln\chi}{\partial\phi}\right)
\int \dd\mathbf{v}\;\mathbf{S}\cdot \boldsymbol{\mathcal{K}}\left[f^{(0)}\right]\nonumber\\
& &
+\frac{v_\text{th}^{-d}}{d}
\int \dd\mathbf{v}\; \mathbf{S}\cdot \boldsymbol{\mathcal{K}}\left[\phi
\frac{\partial \varphi}{\partial \phi}\right].
\eeqa
Here, the integrals involving the collision operator $\boldsymbol{\mathcal{K}}$ have been also estimated in the Appendix \ref{appA} by using the Sonine approximation \eqref{2.15}. The collision frequency $\nu_\mu=\nu_\kappa$ when the leading Sonine approximation \eqref{4.17} for $\boldsymbol{\mathcal{B}}(\mathbf{V})$ is considered.

\subsection{First--order contribution to the cooling rate}

The first--order contribution $\zeta_U$ is given by Eq.\ \eqref{3.20} where $\zeta^{(1,0)}$ is defined by Eq.\ \eqref{3.21} and $\zeta^{(1,1)}$ is given by Eq.\ \eqref{3.22} in terms of the Sonine coefficient $e_D$. To get this coefficient, one multiplies both sides of Eq.\ \eqref{3.14} by $E(\mathbf{V})$ and integrates over velocity. After some algebra, one obtains the result
\begin{align}
\label{4.31}
\left[\nu_\zeta+4\gamma-\overline{\zeta}^{(1,1)}\left(a_2+\frac{1}{2}\theta \Upsilon_\theta\right)\right]e_D=\frac{2}{d(d+2)n}\nonumber\\
\times \int \dd\mathbf{v}\, E(\mathbf{V}) D(\mathbf{V}),
\end{align}
where
\beq
\label{4.32}
\nu_\zeta=\frac{\displaystyle \int \dd\mathbf{v}\, E(\mathbf{V}) \mathcal{L}\mathcal{D}(\mathbf{V})}
{\displaystyle\int \dd\mathbf{v}\, E(\mathbf{V})\mathcal{D}(\mathbf{V})}.
\eeq
The coefficient $\overline{\zeta}^{(1,1)}=\zeta^{(1,1)}/e_D$, where $\zeta^{(1,1)}$ is defined by Eq.\ \eqref{3.22}. An estimate of $\overline{\zeta}^{(1,1)}$ is obtained when one takes the Sonine approximation \eqref{4.10.1} for $\mathcal{D}(\mathbf{V})$. In the linear order in $a_2$,    
the coefficient $\overline{\zeta}^{(1,1)}$ is
\beqa
\label{4.33}
\overline{\zeta}^{(1,1)}&=&
\frac{\sqrt{2}\,\pi^{\frac{d-1}{2}}}{256\,d\,\Gamma\!\left(\frac d2\right)}\;
\chi\,\frac{v_{\mathrm{th}}}{\ell}\;
\Big[
96\left(1-\alpha^2\right)+64\,\theta^{-1}\Delta_{b}^{*2}
\nonumber\\
& & 
+a_2\left(9\left(1-\alpha^2\right)+30\,\theta^{-1}\Delta_{b}^{*2}\right)
\Big].
\eeqa
Moreover, the right--hand side of Eq.\ \eqref{4.31} can be easily computed as
\beqa
\label{4.34}
& & \frac{2}{d(d+2)n}\int \dd\mathbf{v}\, E(\mathbf{V}) D(\mathbf{V})=-\frac{2}{d}a_2+\left(\zeta^{(1,0)}+\frac{2}{d}p^*\right)
\nonumber\\
& & \times 
\left(a_2+\frac{1}{2}\theta \Upsilon_\theta\right)+\frac{1}{2}\phi \frac{\partial a_2}{\partial \phi}+\frac{2}{d^2(d+2)n}
\nonumber\\
& & \times 
\int \dd\mathbf{v} E(\mathbf{V})\mathcal{K}_i\left[\frac{\partial f^{(0)}}{\partial V_i}\right].
\eeqa
The expressions of the collision frequency $\nu_\zeta$ as well as the last integral in Eq.\ \eqref{4.34} have been also displayed in the Appendix \ref{appA}.

\subsection{Low--density limit}

Let us consider the low--density limit. In this case ($\phi\to 0$), the collisional contributions to the transport coefficients vanish and there are only kinetic contributions. In addition, $\chi=1$ and the coefficient $a_2$ depends only on the reduced temperature $\theta$ but not on $\lambda$ and $\chi$. In the dilute regime, the expressions of the transport coefficients for a dilute confined granular suspension are
\beq
\label{4.35}
\eta=\frac{n T}{\nu_\eta+2\gamma},
\eeq
\beq
\label{4.36}
\kappa=\frac{d+2}{2}\frac{1+2a_2+\theta \Upsilon_\theta}{\nu_\kappa+\gamma -\frac{3}{2}\zeta^{(0)}-\zeta^{(0)}\theta \frac{\partial \ln \zeta_0^*}{\partial \theta}}\frac{nT}{m},
\eeq
\beq
\label{4.37}
\mu=\frac{\frac{T\zeta^{(0)}}{n}\kappa+\frac{d+2}{2}\frac{T^2}{m}a_2}{\nu_\kappa-3\gamma}.
\eeq
In addition, when $\phi\to 0$, $\zeta^{(1,0)}=0$ but $\zeta^{(1,1)}\neq 0$. This result differs from the one obtained in the IHS model where $\zeta_U=0$ in the low-density limit \cite{BDKS98}. From Eqs.\ \eqref{4.31} and \eqref{4.34}, the coefficient $e_D$ is given by
\beq
\label{4.38}
e_D=\frac{d^{-1}\theta \Upsilon_\theta}{\nu_\zeta+4\gamma-\overline{\zeta}^{(1,1)}\left(a_2+\frac{1}{2}\theta \Upsilon_\theta\right)}.
\eeq
The first--order contribution $\zeta_U=\overline{\zeta}^{(1,1)} e_D$, where $\overline{\zeta}^{(1,1)}$ is given by Eq.\ \eqref{4.33} with $\chi=1$.

\begin{table*}[p!]
    \caption{Explicit expressions of the scaled transport coefficients for a two--dimensional monocomponent granular suspension $(d = 2)$ at the stationary temperature. These forms have been written in terms of the parameter $\Delta^* = \theta^{-1/2}\Delta_\text{b}^*$.}
    \label{table1}
    \centering
    {\scriptsize
    \setlength{\jot}{1pt}
    \setlength{\tabcolsep}{0pt}
    \renewcommand{\arraystretch}{1.12}
    \begin{tabular}{@{}c@{}} 
        \Xhline{2\arrayrulewidth}\addlinespace[0.5mm]
        \Xhline{2\arrayrulewidth}\addlinespace[1mm]

        $\begin{aligned}
        \eta= \Bigg\{1+\chi\phi \left[ \frac{1+\al}{2}
        +\frac{1}{\sqrt{2\pi}} \Delta^*\left( 1+\frac{3}{32}a_2 \right) \right]\Bigg\}\eta_\text{k}
        +\frac{1}{2}\eta_\text{b}^{\prime}
        \end{aligned}$ 
        \\ \addlinespace[1mm]
        
        $\begin{aligned}
        \eta_\text{k} = \frac{\sqrt{2\pi}}{\nu_\eta^* + 2\gamma^*}
        \Bigg\{ 1-\frac{1}{4}\chi\phi(1+\alpha)(1-3\alpha)
        +\chi\phi\Delta^*\left[\sqrt{\frac{2}{\pi}}(1+2\alpha)
        \left(1-\frac{1}{16}a_2\right)+\Delta^*\right] \Bigg\}\eta_0
        \end{aligned}$
        \\ \addlinespace[1mm]
        
        $\begin{aligned}
        \nu_\eta^* = \frac{\sqrt{2\pi}}{8}\chi\left[
        (1+\al)(7-3\al)\Big(1-\frac{1}{32}a_2\Big)
        +2\sqrt{2\pi}(1-\al)\Delta^*
        -2 \Big(1+\frac{3}{32}a_2\Big)\Delta^{*2}\right]
        \end{aligned}$ 
        \\ \addlinespace[1mm]
        
        $\begin{aligned}
        \eta_\text{b}^{\prime} =
        \frac{32\sqrt{2\pi}}{\pi^2}\chi\phi^2
        \left[ \frac{1+\al}{4}\sqrt{\frac{\pi}{2}}
        \left( 1-\frac{1}{16}a_2 \right) + \frac{\pi}{8}\Delta^* \right]\eta_0
        \end{aligned}$ 
        \\ \addlinespace[1mm]
        
        $\begin{aligned}
        \kappa = \Bigg\{ 1+\frac{3}{4}\chi\phi
        \left[ 1+\alpha+\sqrt{\frac{2}{\pi}}\Delta^*
        \left(1-\frac{1}{32}a_2\right) \right] \Bigg\}\kappa_\text{k}
        + \frac{2}{\pi}\chi\phi^2
        \left[ (1+\alpha)\left(1+\frac{7}{16}a_2\right)
        + \sqrt{\frac{\pi}{2}}\Delta^* \right]\kappa_0
        \end{aligned}$
        \\ \addlinespace[1mm]

        $\begin{aligned}
        \kappa_{\mathrm{k}}={}&
        \left(\nu_\kappa^*+\gamma^*-\frac{3}{2}\zeta^{*}_0
        -\zeta^{*}_0\theta\frac{\partial\ln\zeta_0^*}{\partial\theta}\right)^{-1}
        \sqrt{\frac{\pi}{2}}
        \Bigg\{1+2a_2+\theta\Upsilon_\theta
        \\
        &-\chi\phi\Bigg[
        \frac{3}{8}(1+\alpha)^2\big(1-2\alpha-a_2(1+\alpha)\big)
        +\frac{1}{\sqrt{2\pi}}\Delta^*
        \left[ \left(\frac{3}{4}-\frac{97}{64}a_2\right)
        \right.
        \\
        &\left.
        +3(1+\alpha)\left(1-\frac{1}{2}\sqrt{2\pi}\,\Delta^*\right)
        -\frac{a_2}{8}\left(21+\frac{27}{4}\alpha\right)\alpha
        -\frac{9}{2}(1+\alpha)^2
        -\Delta^{*2}\left(1-\frac{1}{16}a_2\right)
        \right] \Bigg]
        \\
        &+\frac{3}{16}\chi\phi\,\theta\Upsilon_\theta
        \Bigg[(1+\alpha)^3+\frac{1}{2\sqrt{2\pi}}\Delta^*
        \left(13+2\alpha(8+3\alpha)-\frac{4}{3}\Delta^{*2}\right)\Bigg]
        \Bigg\}\kappa_0
        \end{aligned}$
        \\ \addlinespace[1mm]

        $\begin{aligned}
        \nu_\kappa^* =
        \frac{\sqrt{2\pi}}{16}\,\chi\left[
        (1+\alpha)(19-15\alpha)
        +\frac{(1+\alpha)(7-3\alpha)}{32}\,a_2
        -2\sqrt{\frac{\pi}{2}}\,(5\alpha-1)\,\Delta^*
        -\left(10-\frac{3}{16}a_2\right)\Delta^{*2}
        \right]
        \end{aligned}$
        \\ \addlinespace[1mm]

        $\begin{aligned}
        \mu = \Bigg\{ 1+\frac{3}{4}\chi\phi
        \left[ 1+ \alpha + \sqrt{\frac{2}{\pi}}\Delta^*
        \left(1-\frac{1}{32}a_2\right) \right] \Bigg\}\mu_\text{k}
        \end{aligned}$ 
        \\ \addlinespace[1mm]

        $\begin{aligned}
        \mu_\text{k} = {} &
        \left(\nu^*_\kappa-3\gamma^*\right)^{-1}
        \sqrt{\frac{\pi}{8}}\Bigg\{
        2 \left(a_2 + \phi \frac{\partial a_2}{\partial \phi}\right)
        \\ 
        & + 4\left(1+\frac{1}{2}\phi\frac{\partial \ln \chi}{\partial \phi}\right)
        \chi \phi \Bigg[
        \frac{3}{8}(1+\alpha)
        \left[\alpha(\alpha-1)
        +\frac{a_2}{6}(14-3\alpha + 3\alpha^2)\right]
        \\ 
        & +\frac{1}{\sqrt{2\pi}}\Delta^*
        \left[
        2\Delta^{*2}
        -3\left(\frac{1}{2}-\alpha^2\right)
        +\frac{3}{2}\sqrt{2\pi}\,\alpha\,\Delta^*
        + \frac{a_2}{32}(47 + 18\alpha^2 - 4\Delta^{*2})
        \right] \Bigg]  
        \\
        & + \frac{3}{8}\chi\phi^2 \frac{\partial a_2}{\partial \phi}
        \Bigg[
        \left(1+\alpha\right)^3
        +\frac{\Delta^*}{2\sqrt{2\pi}}
        \left[13+2\alpha(8+3\alpha)-\frac{4}{3}\Delta^{*2}\right]
        \Bigg]
        \\
        & + \sqrt{\frac{8}{\pi}}
        \Bigg[
        \zeta^{*}_0\left(1+\phi\frac{\partial\ln\chi}{\partial\phi}\right)
        +\chi\phi\frac{\partial\chi}{\partial\phi}
        \frac{\partial}{\partial\chi}\left(\frac{\zeta^*_0}{\chi}\right)
        -\lambda\left(1-\phi\frac{\partial\ln R}{\partial \phi}\right)
        \frac{\partial \zeta^{*}_0}{\partial \lambda}
        \\
        &\hspace{2.4cm}
        -2\gamma^*\left(\theta^{-1}-1\right)
        \phi \frac{\partial\ln R}{\partial \phi}
        \Bigg] \frac{\kappa_\text{k}}{\kappa_0}
        \Bigg\}\kappa_0 \frac{T}{n}
        \end{aligned}$ 
        \\ \addlinespace[1mm]

        $\begin{aligned}
        \eta_\text{b} = \eta_\text{b}' + \eta_\text{b}''
        \end{aligned}$ 
        \\ \addlinespace[1mm]
        
        $\begin{aligned}
        \eta_\text{b}'' =
        \frac{1}{2} \chi\phi\,\Delta^* e_D^*
        \left( 1 + \frac{15}{32}a_2 \right)\eta_0
        \end{aligned}$ 
        \\ \addlinespace[1mm]
        
        $\begin{aligned}
        e_D^* =
        \left[
        \nu_\zeta^* + 4\gamma^*
        -\overline{\zeta}^{(1,1)*}
        \left(a_2+\frac{1}{2}\theta\Upsilon_\theta\right)
        \right]^{-1}
        \left[
        -a_2+\left(\zeta^{(1,0)}+p^*\right)
        \left(a_2 + \frac{1}{2}\theta\Upsilon_\theta\right)
        +\frac{1}{2}\phi\frac{\partial a_2}{\partial \phi}
        +\frac{1}{8n}I_\zeta
        \right]
        \end{aligned}$ 
        \\ \addlinespace[1mm]

        $\begin{aligned}
        \nu_\zeta^* = {} &
        -\sqrt{\frac{\pi}{2}}\,
        \frac{(1+\alpha)}{96}\,\chi\,
        \left(30\alpha^3-30\alpha^2+153\alpha-185\right)
        \\
        &-\frac{\sqrt{2\pi}}{16384}\chi
        \Bigg\{
        -4 \Delta^{*2}\left(32+15 a_2\right)
        +36\al^2\Delta^{*2}\left(32+3a_2\right)
        +6 \Delta^{*2}\left(416-33 a_2\right)
        \\
        &\hspace{1.4cm}
        +512 \sqrt{\frac{\pi}{2}}\Delta^{*}
        \left[2+7\al+3\al^3-4(1-\al)\right]
        +2048\, \Delta^{*2}
        \left(1+\frac{15}{32}a_2\right)
        \Bigg\}
        \end{aligned}$
        \\ \addlinespace[1mm]
        
        $\begin{aligned}
        \zeta^{(1,0)} =
        \chi \phi
        \left[
        2 \Delta^{*2}
        +4\sqrt{\frac{2}{\pi}}
        \left(1-\frac{1}{16}a_2\right)\al \Delta^{*}
        -\frac{3}{2}(1-\al^2)
        \right]
        \end{aligned}$
        \\ \addlinespace[1mm]
        
        $\begin{aligned}
        \overline{\zeta}^{(1,1)*} =
        \frac{\sqrt{2\pi}}{512}\;
        \chi
        \left[
        96\left(1-\alpha^2\right)+64\,\Delta^{*2}
        +a_2\Big(9\left(1-\alpha^2\right)+30\,\Delta^{*2}\Big)
        \right]
        \end{aligned}$ 
        \\ \addlinespace[1mm]
        
        $\begin{aligned}
        I_\zeta =
        n\chi\phi\Bigg\{
        \frac{3}{4} \left(1+\alpha\right)
        \left[(1-\alpha^2)(5\alpha-1)
        -\frac{a_2}{6}
        \left(15\alpha^3-3\alpha^2+69\alpha-41\right)\right]
        - \frac{2}{\sqrt{2\pi}} \Delta^* (I_0 + a_2 I_1)
        \Bigg\}
        \end{aligned}$ 
        \\ \addlinespace[1mm]
        
        $\begin{aligned}
        I_0 =
        -6+4\alpha(-3+3\alpha+4\alpha^2)
        +3\sqrt{2\pi}(1+\alpha)(-1+3\alpha)\Delta^*
        +8(1+2\alpha)\Delta^{*2}
        +2\sqrt{2\pi}\,\Delta^{*3}
        \end{aligned}$ 
        \\ \addlinespace[1mm]
        
        $\begin{aligned}
        I_1 =
        \frac{1}{8}
        \left[
        31+2\alpha(47+9\alpha+12\alpha^2)
        -4(1+2\alpha)\Delta^{*2}
        \right]
        \end{aligned}$ 
        \\ \addlinespace[1mm]
        
        $\begin{aligned}
        \zeta_U = \zeta^{(1,0)} + \overline{\zeta}^{(1,1)*}e_D^*
        \end{aligned}$
        \\ \addlinespace[1mm]
        \Xhline{2\arrayrulewidth}\addlinespace[0.5mm]
        \Xhline{2\arrayrulewidth}
    \end{tabular}
    }
\end{table*}

\section{Some illustrative systems in the two--dimensional case}
\label{sec6}

The results obtained in Sec.\ \ref{sec5} provide the collisional and kinetic contributions to the Navier--Stokes transport coefficients, as well as the first--order contribution to the cooling rate. These expressions are given in terms of integrals involving the scaled distribution function $\varphi$. These integrals are estimated in Appendix \ref{appA} by approximating $\varphi$ by its leading Sonine form \eqref{2.15}.

\begin{figure}[t]
\begin{center}
\begin{tabular}{lr}
\resizebox{6.5cm}{!}{\includegraphics{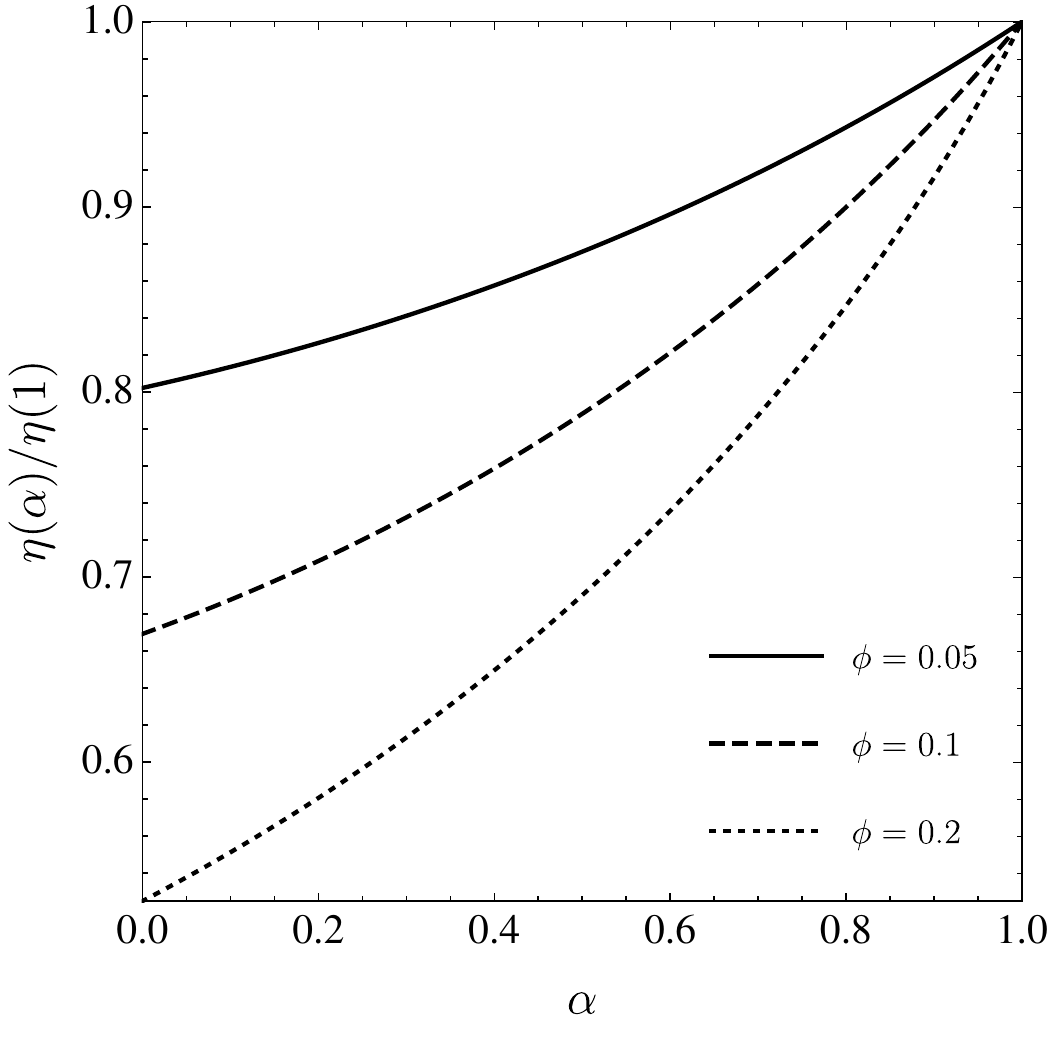}}
\end{tabular}
\end{center}
\caption{Plot of the (scaled) shear viscosity $\eta(\alpha)/\eta(1)$ versus the coefficient of restitution $\al$
for $d=2$, $\Delta_{\text{b}}^*=0.5$, and $T_\text{b}^*=1$. Three different values of $\phi$ have been considered: $\phi = 0.05$ (solid), $\phi = 0.1$ (dashed) and $\phi = 0.2$ (dotted). Here, $\eta(1)$ refers to the shear viscosity coefficient for elastic collisions.
\label{fig_eta}}
\end{figure}
\begin{figure}[t]
\begin{center}
\begin{tabular}{lr}
&\resizebox{6.5cm}{!}{\includegraphics{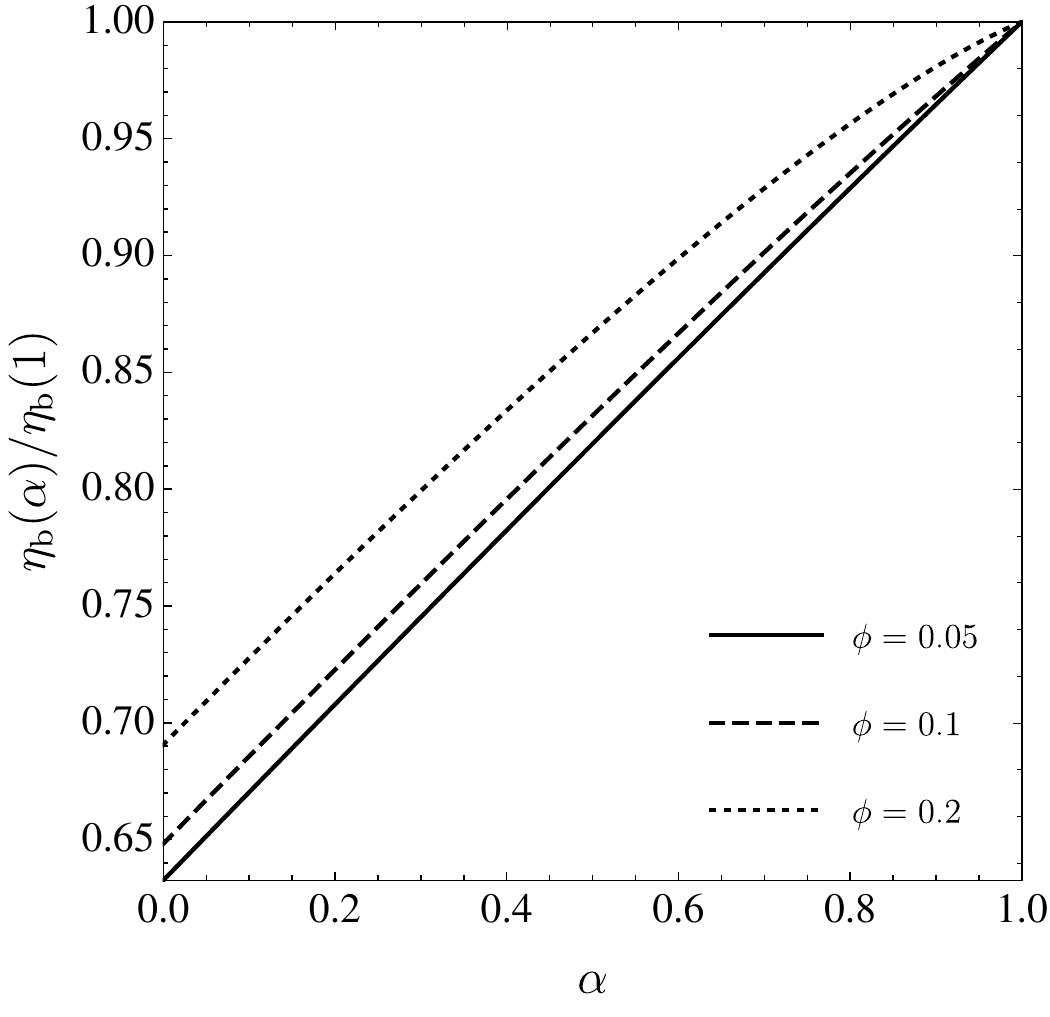}}
\end{tabular}
\end{center}
\caption{Plot of the (scaled) bulk viscosity $\eta_\text{b}(\alpha)/\eta_\text{b}(1)$ versus the coefficient
of restitution $\al$ for $d=2$, $\Delta_{\text{b}}^*=0.5$, and $T_\text{b}^*=1$. Three different values of $\phi$ have been considered: $\phi = 0.05$ (solid), $\phi = 0.1$ (dashed) and $\phi = 0.2$ (dotted). Here, $\eta_\text{b}(1)$ refers to the bulk viscosity coefficient for elastic collisions.
\label{fig_etab}}
\end{figure}

Although the above analytical results apply to a $d$--dimensional system ($d=2$ for hard disks and $d=3$ for hard spheres), the $\Delta$--model was primarily proposed \cite{BRS13} to effectively account for the effect of confinement in a quasi--two--dimensional setup. For this reason, we are mainly interested in the case $d=2$ here. Table \ref{table1} displays for hard disks ($d=2$) the explicit forms of the transport coefficients and the cooling rate as functions of the coefficient of restitution $\al$, the volume fraction $\phi$, the (dimensionless) bath temperature $T_\text{b}^*$, and the (dimensionless) parameter $\Delta_ \text{b}^*$. Additionally, these coefficients are defined in terms of the (reduced) temperature $\theta$, the kurtosis $a_2$, and the derivatives $\Upsilon_\theta$, $\Upsilon_\lambda$, and $\Upsilon_\chi$.

As expected, the transport coefficients and cooling rate depend intricately on the system's parameters. Since we are primarily interested in the influence of inelasticity on transport, we scale the transport coefficients with respect to their values for elastic collisions to illustrate the differences between granular and molecular suspensions. Furthermore, we consider a two--dimensional system with $T_\text{b}^*=1$, $\Delta_ \text{b}^*=0.5$, and three different values of the solid volume fraction: $\phi=0.05$ (very dilute system), $\phi=0.1$, and $\phi=0.2$ (moderately dense system). This choice of parameters is also useful for comparison with the granular suspension studied in Ref.~\cite{GGG19a}, thereby allowing us to assess more clearly the impact of the additional $\Delta$-collisional energy injection on the transport coefficients.

Figure \ref{fig_eta} shows the $\al$--dependence of the (scaled) shear viscosity $\eta(\alpha)/\eta(1)$, where $
\eta(1)$ corresponds to the shear viscosity of a granular suspension for elastic collisions. For a given density, it is quite apparent that the impact of inelasticity in collisions on the shear viscosity is in general quite important since the ratio $\eta(\alpha)/\eta(1)$ differs significantly from 1. Figure \ref{fig_eta} also highlights that the shear viscosity decreases with increasing inelasticity (decreasing $\al$) regardless of density. Regarding the influence of density on $\eta$, we observe that the shear viscosity (scaled with respect to its elastic value) decreases significantly with density at a given value of $\al$. This behavior contrasts with the results derived in the $\Delta$--model for a dry granular gas \cite{GBS18}, where the scaled shear viscosity exhibits  weak dependence on density. A plausible explanation is that, in the present case, the interstitial gas tends to thermalize the suspension, enhancing the frequency of particle collisions and thereby making excluded--volume and collisional transfer effects associated with finite density more pronounced. For this reason, the dependence of $\eta(\alpha)/\eta(1)$ on both density and the coefficient of restitution, as found in the conventional IHS model for granular suspensions \cite{GGG19a}, agrees qualitatively well with the results obtained here in the $\Delta$--model.

The dependence of the (scaled) bulk viscosity $\eta_\text{b}(\alpha)/\eta_\text{b}(1)$ versus the coefficient of restitution $\al$ is shown in Fig.\ \ref{fig_etab}. Note that in the context of the $\Delta$--model, the expression of $\eta_\text{b}$ obtained here for granular suspensions differs from that derived for a dry granular gas
\cite{GBS18} because the contribution $\eta_\text{b}''$ to $\eta_\text{b}$ was neglected in the latter study \cite{GBS18}. This contrasts with the IHS model \cite{GGG19a}, where the expressions of the bulk viscosity with and without the gas phase are the same. As in the case of the shear viscosity, Fig.\ \ref{fig_etab} shows that the ratio $\eta_\text{b}(\alpha)/\eta_\text{b}(1)$ decreases with decreasing $\al$. Moreover, since $\eta_\text{b}=0$ for dilute gases ($\phi=0$), the bulk viscosity (scaled with its elastic value) increases with density at a given value of $\al$, as expected.

\begin{figure}[h]
\begin{center}
\begin{tabular}{lr}
\resizebox{6.5cm}{!}{\includegraphics{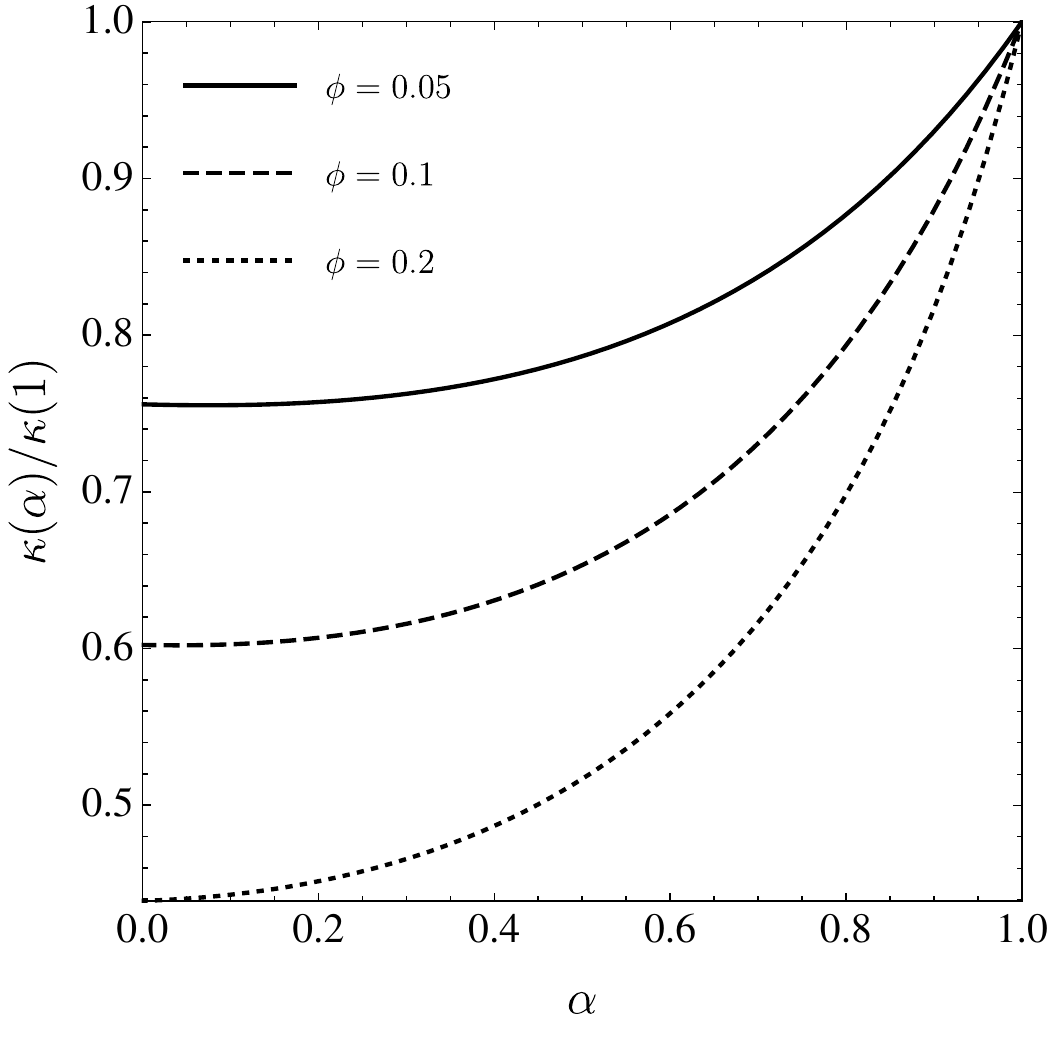}}
\end{tabular}
\end{center}
\caption{Plot of the (reduced) thermal conductivity $\kappa(\alpha)/\kappa(1)$ versus the coefficient of restitution $\al$ for $d=2$, $\Delta_{\text{b}}^*=0.5$, and $T_\text{b}^*=1$. Three different values of $\phi$ have been considered: $\phi = 0.05$ (solid), $\phi = 0.1$ (dashed) and $\phi = 0.2$ (dotted).
Here, $\kappa(1)$ refers to the thermal conductivity coefficient for elastic collisions.
\label{fig_kappa}}
\end{figure}

\begin{figure}[h]
\begin{center}
\begin{tabular}{lr}
\resizebox{6.55cm}{!}{\includegraphics{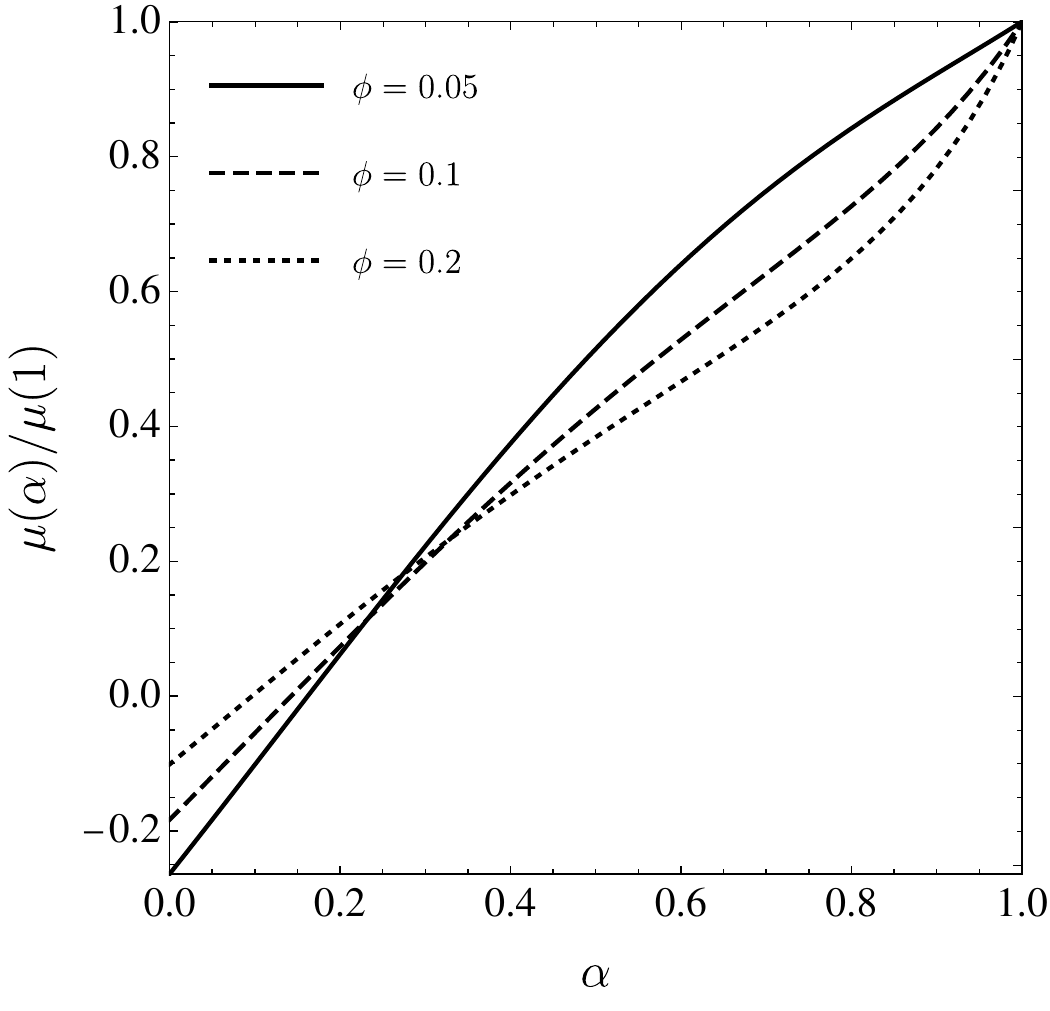}}
\end{tabular}
\end{center}
\caption{Plot of the (reduced) heat diffusive coefficient $\mu(\alpha)/\mu(1)$
versus the coefficient of restitution $\al$ for $d=2$, $\Delta_{\text{b}}^*=0.5$, and $T_\text{b}^*=1$.
Three different values of $\phi$ have been considered: $\phi = 0.05$ (solid), $\phi = 0.1$ (dashed) and $\phi = 0.2$ (dotted).
Here, $\mu(1)$ refers to the heat diffusive coefficient for elastic collisions.
\label{fig_mu}}
\end{figure}

Now, we will consider the transport coefficients associated with the heat flux. Figure \ref{fig_kappa} shows the $\al$--dependence of the (scaled) thermal conductivity $\kappa(\alpha)/\kappa(1)$. As can be seen, the thermal conductivity (scaled by its elastic value) depends on both $\al$ and $\phi$
in a manner similar to that found for the shear viscosity. This behavior qualitatively agrees well with that reported in Ref.\ \cite{GGG19a} for the conventional IHS model of granular suspensions. Regarding the coefficient $\mu$, unlike the IHS model \cite{GGG19a}, we see that $\mu\neq 0$ for elastic collisions. The contribution to the heat flux from the density gradient is also present in relativistic gases \cite{GLW80,CK02} as well as in ordinary (elastic) gases subjected to a force proportional to the particle velocity \cite{PG14}. Additionally, in contrast to the dry (no gas phase) $\Delta$--model \cite{GBS18}, $\mu\neq 0$ for a dilute granular gas even if the kurtosis $a_2=0$ [see Eq.\ \eqref{4.37}]. The dependence of the (scaled) heat diffusive coefficient $\mu(\alpha)/\mu(1)$ on $\al$ is shown in Fig.\ \ref{fig_mu} for different densities. It is apparent that this coefficient decreases with increasing inelasticity. However, unlike with the shear viscosity and the thermal conductivity, we observe that the ratio $\mu(\alpha)/\mu(1)$ decreases with density for not quite small values of the coefficient of restitution (let's say, $\al \gtrsim 0.5$), while the opposite occurs for quite extreme values of dissipation (let's say, $\al \lesssim 0.2$). In fact, it can be negative for very small values of $\al$ regardless of density. Comparing these results with those derived from the $\Delta$--model in the absence of the gas phase
\cite{GBS18,GBS26} reveals significant qualitative differences. This indicates that the gas phase significantly impacts heat transport.





\begin{figure}[h]
\begin{center}
\begin{tabular}{lr}
\resizebox{6.55cm}{!}{\includegraphics{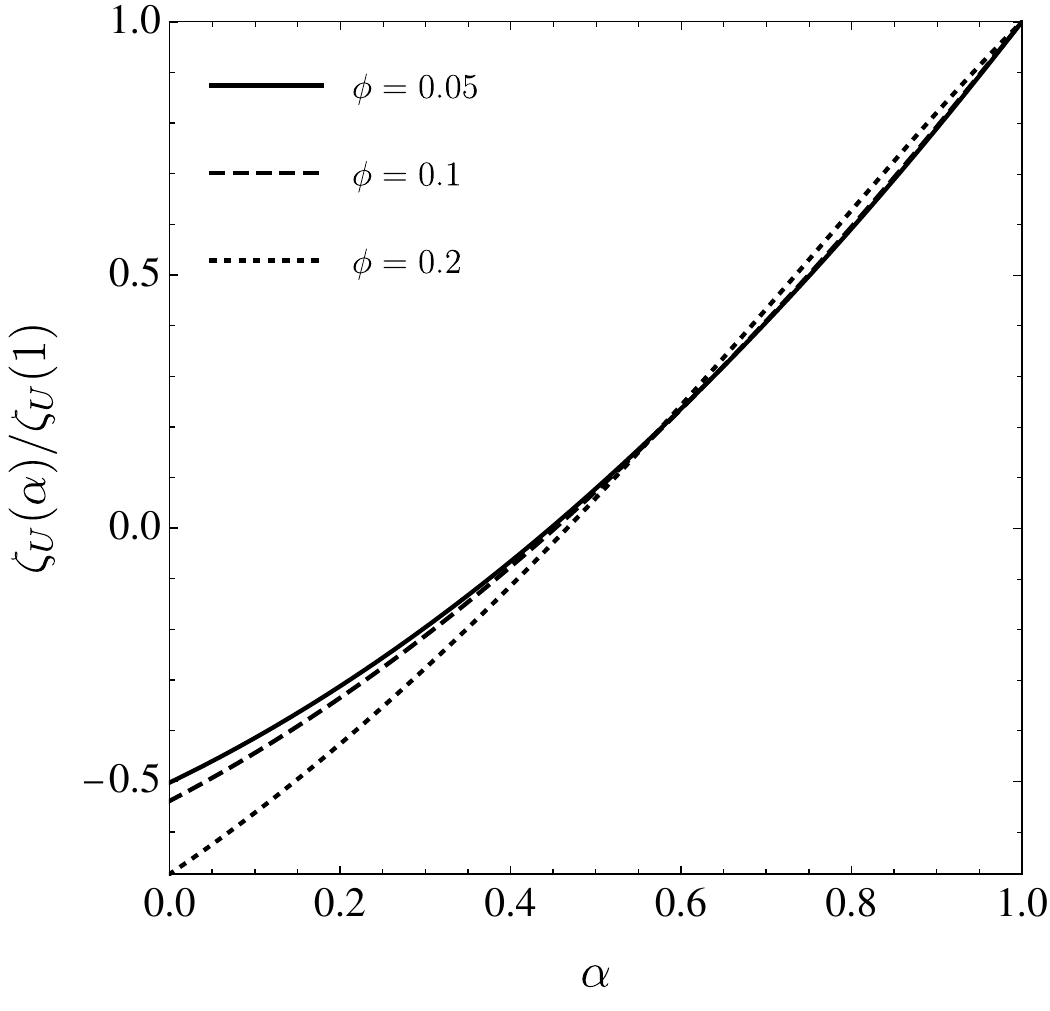}}
\end{tabular}
\end{center}
\caption{Plot of the (reduced) first--order contribution to the cooling rate $\zeta_U(\alpha)/\zeta_U(1)$ as a function of the coefficient of restitution $\alpha$ for $d=2$, $\Delta_{\text{b}}^*=0.5$, and $T_\text{b}^*=1$. Three different values of $\phi$ have been considered: $\phi = 0.05$ (solid), $\phi = 0.1$ (dashed) and $\phi = 0.2$ (dotted). Here, $\zeta_U(1)$ refers to the first--order contribution to the cooling rate for elastic
collisions.
\label{fig_zetaU_red}}
\end{figure}

Finally, the $\al$--dependence of the (scaled) first--order contribution $\zeta_U(\al)/\zeta_U(1)$ to the cooling rate is plotted in Fig.\ \ref{fig_zetaU_red} for the same parameters employed in Figs.\ \ref{fig_eta}--\ref{fig_mu}. Here, in contrast to the conventional $\Delta$--model \cite{GBS18}, $\zeta_U$ is different from zero for elastic collisions ($\zeta_U(1)\neq 0$). We observe that the ratio $\zeta_U(\al)/\zeta_U(1)$ decreases with increasing inelasticity regardless of the density considered. Moreover, the influence of density on $\zeta_U$ (reduced with respect to its elastic value) is very weak.

In summary, the study performed in this Section for confined granular suspensions has clearly shown that the dependence of the transport coefficients on the coefficient of normal restitution $\al$ differs significantly from the one previously found \cite{GBS18,BSG26} in the absence of the gas phase. Thus, the combined effect of both parameters $\Delta$ (which accounts for the injection of energy into the system) and $\gamma$ (which accounts for the effect of the interstitial fluid) on momentum and heat transport cannot be neglected for finite values of $\al$. This conclusion contrasts with the results derived in Ref.\ \cite{MPSF25} where it was assumed that the transport coefficients do not depend on $\Delta$ and/or $\gamma$.      

\section{Stability of the homogeneous steady state}
\label{sec7}

The knowledge of the explicit forms of the (steady) Navier--Stokes transport coefficients and the cooling rate allows us to solve the hydrodynamic equations for the fields $n$, $\mathbf{U}$, and $T$ for states near the HSS. The nonlinear Navier--Stokes hydrodynamic equations can be easily written when one substitutes the equation of state \eqref{2.28} for the hydrostatic pressure, the constitutive equations \eqref{4.1} and \eqref{4.16} for the pressure tensor and the heat flux, respectively, and the equation $\zeta=\zeta^{(0)}+\zeta_U \nabla\cdot \mathbf{U}$ (where $\zeta^{(0)}$ and $\zeta_U$ are given by Eqs.\ \eqref{2.16} and \eqref{3.20}, respectively) for the cooling rate into the exact balance equations \eqref{1.18}--\eqref{1.20}. The corresponding Navier--Stokes hydrodynamic equations are
\begin{equation}
\label{7.1}
\text{D}_t n+n\nabla\cdot\mathbf{U}=0,
\end{equation}
\begin{align}
\label{7.2}
\text{D}_t U_i+\rho^{-1}\nabla_i p=\rho^{-1}\nabla_j&\Bigg[\eta\left(\nabla_iU_j+\nabla_jU_i-\frac{2}{d}\delta_{ij}
\nabla\cdot\mathbf{U}\right)
\nonumber\\
& 
+\eta_\text{b}\delta_{ij}\nabla\cdot\mathbf{U}\Bigg]-\gamma \Delta U_i,
\end{align}
\beqa
\label{7.3}
& & \Big(\text{D}_t+2\gamma\left(1-\theta^{-1}\right)+\zeta^{(0)}\Big)T = \frac{2}{dn}\nabla\cdot\left(\kappa\nabla T+\mu\nabla n\right)\nonumber\\
& &
+\frac{2}{dn}\bigg[\eta\left(\nabla_iU_j+\nabla_jU_i-\frac{2}{d}\delta_{ij}
\nabla\cdot\mathbf{U}\right)+
\eta_\text{b}\delta_{ij}\nabla\cdot\mathbf{U}\bigg]
\nonumber\\
& & \qquad \qquad \qquad \times \,\nabla_iU_j
-T\zeta_U\nabla\cdot\mathbf{U}-\frac{2}{dn}p\nabla\cdot\mathbf{U},
\nonumber\\
\eeqa
where $\nabla_i\equiv \partial/\partial r_i$. 

As it has been already discussed in several previous papers \cite{G05,GMD06,GGG19a},  at Navier--Stokes order the general form of the cooling rate $\zeta$ should include second--order gradient contributions of the form $\zeta_n \nabla^2 n$ and $\zeta_T \nabla^2 T$ in Eq.\ \eqref{7.3}. Nevertheless, as shown for a dilute (dry) granular gas \cite{BDKS98}, given that the ratios $\zeta_n/\mu$ and $\zeta_T/\kappa$ were shown to be very small for not very inelastic particles, the terms $\zeta_n \nabla^2 n$ and $\zeta_T \nabla^2 T$ were neglected in the stability analysis performed in Refs.\ \cite{G05,GMD06,GGG19a}. We assume that the same happens here in the $\Delta$--model so that the second--order contributions to $\zeta$ can be neglected for practical purposes.

In order to analyze the stability of the homogeneous solution, Eqs.\ \eqref{1.18}--\eqref{1.20} must be linearized around the homogeneous steady state. In this state, the hydrodynamic fields take the homogeneous steady values $n_{\text{s}} \equiv \text{const}.$, $T_{\text{s}} \equiv \text{const}.$, and $\mathbf{U}_{\text{g},s} = \mathbf{U}_s \equiv \mathbf{0}$.
For small spatial gradients, we assume that the deviations $\delta y_{\beta}(\mathbf{r}, t) = y_\beta(\mathbf{r}, t) - y_{\beta,s}$ are small, where $\delta y_{\beta}(\mathbf{r}, t)$ denotes the deviations of the hydrodynamic fields $\{y_\beta ; \beta = 1,..., d + 2\} = \{n, \mathbf{U}, T \}$ with respect to their values in the HSS. Moreover, as usual in granular suspensions \cite{GGG19a}, we assume that the gas phase is not disturbed and hence, $\mathbf{U}_\text{g} = \mathbf{U}_{\text{g},s} \equiv \mathbf{0}$.

The linearized equations for the perturbations $\delta n$, $\delta \mathbf{U}$, and $\delta T$ can be easily derived from Eqs.\ \eqref{7.1}--\eqref{7.3} as
\begin{widetext}
\begin{equation}
\label{7.4}
    \frac{\partial}{\partial t} \delta n + n\nabla \cdot \delta \textbf{U} = 0,
\end{equation}
\begin{equation}
\label{7.5}
    \frac{\partial}{\partial t} \delta U_i + \rho^{-1}\left(\frac{\partial p}{\partial T}\right)_n\nabla_{i}\delta T + \rho^{-1}\left(\frac{\partial p}{\partial n}\right)_T\nabla_{i}\delta n = \rho^{-1}\eta\nabla^2\delta U_i + \rho^{-1} \left( \frac{d-2}{d}\eta + \eta_\text{b} \right)\nabla_i (\nabla \cdot \delta \textbf{U}) - \gamma \delta U_i,
\end{equation}
\begin{align}
\label{7.6}
    \frac{\partial}{\partial t} \delta T + \left[ 2(1-\theta^{-1})\left(\frac{\partial \gamma}{\partial n}\right) + \left(\frac{\partial \zeta^{(0)}}{\partial n}\right)\right]T\delta n + \left[ 2\gamma + \zeta^{(0)} + T\left(\frac{\partial\zeta^{(0)}}{\partial T}\right) \right]\delta T = \frac{2}{dn}\left(\kappa \nabla^2\delta T + \mu\nabla^2\delta n\right) \nonumber \\ - \zeta_U T\;\nabla \cdot \delta \textbf{U} - \frac{2}{dn}p\; \nabla \cdot \delta \textbf{U}.
\end{align}
\end{widetext}
It must be recalled that all the quantities involved in Eqs.\ \eqref{7.4}--\eqref{7.6} are evaluated in the HSS. Henceforth, the subscript $\text{s}$ will be removed for the sake of simplicity.

\noindent In order to compare our results with those previously obtained for granular fluids \cite{G05} and non--confined granular suspensions \cite{GGG19a}, we will introduce the following space and time (dimensionless) variables:
\begin{equation}
\label{7.7}
    \tau = \frac{1}{2\sqrt{2}}\frac{v_{\text{th}}}{\ell}t = \frac{1}{2}n\sigma^{d-1}\sqrt{\frac{T}{m}}t, \qquad \textbf{r}' = \frac{1}{2\ell} \textbf{r} = \frac{1}{2}n\sigma^{d-1}\textbf{r}.
\end{equation}
The dimensionless time scale $\tau$ is a measure of the average number of collisions per particle in the time interval between $0$ and $t$ while $\textbf{r}'$ is proportional to the mean free path of solid particles. As usual in the linear stability analysis \cite{G05,GMD06,GGG19a}, we introduce the following set of Fourier transformed dimensionless variables:
\begin{equation}
\label{7.8}
    \rho_{\mathbf{k}}(\tau) = \frac{\delta n_{\mathbf{k}}(\tau)}{n}, \quad
    \mathbf{w}_{\mathbf{k}}(\tau) = \frac{\delta \mathbf{U}_{\mathbf{k}}(\tau)}{\sqrt{T/m}}, \quad
    \theta_{\mathbf{k}}(\tau) = \frac{\delta T_{\mathbf{k}}(\tau)}{T},
\end{equation}
where $\delta y_{\mathbf{k}\beta}(\tau)\equiv \left\{\rho_{\mathbf{k}}(\tau), \mathbf{w}_{\mathbf{k}}(\tau), \theta_{\mathbf{k}}(\tau)\right\}$ is defined as
\begin{equation}
\label{7.9}
    \delta y_{\mathbf{k}\beta}(\tau) = \int \mathrm{d}\mathbf{r}' e^{-\mathrm{i}\mathbf{k} \cdot \mathbf{r}'} \delta y_{\beta}(\mathbf{r}', \tau).
\end{equation}
Note that here the wave vector $\mathbf{k}$ is dimensionless.

\begin{widetext}
In the Fourier space, the set of (coupled) differential equations \eqref{7.4}--\eqref{7.7} can be rewritten as
\begin{equation}
\label{7.10}
    \frac{\partial}{\partial \tau} \rho_{\textbf{k}} + i\textbf{k}\cdot \textbf{w}_{\textbf{k}} = 0,
\end{equation}
\begin{equation}
\label{7.11}
    \frac{\partial}{\partial \tau} w_{\textbf{k},i} + ik_ip^*\left( C_{p,T} \theta_{\textbf{k}}+C_{p,n}\rho_{\textbf{k}} \right) + 2\sqrt{2}\gamma^*w_{\textbf{k},i} + \frac{1}{2}\eta^* k^2w_{\textbf{k},i} + \frac{1}{2}\left( \frac{d-2}{d}\eta^* + \eta_\text{b}^* \right)k_i(\textbf{k}\cdot\textbf{w}_{\textbf{k}}) = 0,
\end{equation}
\begin{align}
\label{7.12}
    \frac{\partial}{\partial \tau}\theta_\textbf{k} + \left[2\sqrt{2}\left( \zeta_0^*C_\chi + \zeta_0^{(1)}C_n + C_\gamma\right) + \mu^*k^2 \right]\rho_\textbf{k} + \left[2\sqrt{2}\left(2\gamma^*\theta^{-1} + \frac{1}{2}\zeta_0^* + \theta \frac{\partial \zeta_0^*}{\partial\theta}\right) + D_T^*k^2\right] \theta_\textbf{k} \nonumber \\
    + \frac{2}{d}i\left(p^* + \frac{d}{2}\zeta_U\right)(\textbf{k}\cdot \textbf{w}_\textbf{k}) = 0.
\end{align}
In Eqs.\ \eqref{7.10}--\eqref{7.12}, we have introduced the reduced quantities
\begin{equation}
\label{7.13}
    \eta^* = \frac{v_{\text{th}}}{\sqrt{2}nT\ell}\eta, \qquad  \eta^*_\text{b} = \frac{v_{\text{th}}}{\sqrt{2}nT\ell}\eta_\text{b}, \qquad D^*_T = \frac{1}{d\sqrt{2}}\frac{m v_{\text{th}}}{nT \ell}\kappa, \qquad \mu^* = \frac{1}{d\sqrt{2}} \frac{m v_{\text{th}}}{T^2\ell}\mu,
\end{equation}
\begin{equation}
\label{7.14}
    C_{p,T} = \frac{1}{p^*n}\left(\frac{\partial p}{\partial T}\right)_n = 1 + \theta\left(\frac{\partial \ln p^*}{\partial \theta}\right)_\phi, \qquad C_{p,n} = \frac{1}{p^*T}\left(\frac{\partial p}{\partial n}\right)_\theta = 1+\phi\left(\frac{\partial \ln p^*}{\partial \phi}\right)_\theta,
\end{equation}
and
\begin{equation}
\label{7.15}
    C_\chi = 1+\phi\frac{\partial \ln \chi}{\partial \phi}, \qquad C_n = \phi\frac{\partial \chi}{\partial \phi}\Upsilon_\chi + \phi\frac{\partial \lambda}{\partial \phi}\Upsilon_\lambda, \qquad C_\gamma = 2(1-\theta^{-1})\gamma^* \phi \frac{\partial \ln R}{\partial \phi}.
\end{equation}
\end{widetext}

As expected, in Eq.\ \eqref{7.11} the $d-1$ transverse
velocity components $\textbf{w}_{\textbf{k}\perp} = \textbf{w}_{\textbf{k}} -(\textbf{w}_{\textbf{k}} \cdot \hat{\textbf{k}})\hat{\textbf{k}}$ (orthogonal to the wave vector $\textbf{k}$) decouple from the other three modes. Their evolution equation is
\begin{equation}
\label{7.16}
    \frac{\partial \textbf{w}_{\textbf{k}\perp}}{\partial \tau} + \left( 2\sqrt{2}\gamma^* + \frac{1}{2}\eta^*k^2 \right)\textbf{w}_{\textbf{k}\perp} = \textbf{0}.
\end{equation}
\noindent The solution to Eq.\ \eqref{7.16} is
\begin{equation}
\label{7.17}
    \textbf{w}_{k\perp}(\tau) = \textbf{w}_{k\perp}(0) \exp \left[ -\left( \frac{1}{2}\eta^* k^2 +
    2\sqrt{2}\gamma^* \right) \tau \right].
\end{equation}
\noindent Since both the (reduced) friction coefficient $\gamma^*$ and the (reduced) shear viscosity coefficient $\eta^*$ are positive, then the transversal shear modes of the confined granular suspension are linearly stable.

The remaining (longitudinal) modes correspond to $\rho_\textbf{k}$, $\theta_\textbf{k}$ and the longitudinal velocity component of the velocity field, $w_{\textbf{k}\parallel} = \textbf{w}_\textbf{k} \cdot \hat{\textbf{k}}$ (parallel to \textbf{k}). These modes are coupled and obey the equation
\begin{equation}
\label{7.18}
    \frac{\partial \delta y_{\mathbf{k}\beta}(\tau)}{\partial \tau} + \mathsf{M}_{\beta\mu} \delta y_{\mathbf{k}\mu}(\tau) = 0,
\end{equation}
where $\delta y_{\mathbf{k}\beta}(\tau)$ denotes now the set $\{\rho_{\mathbf{k}}, w_{\mathbf{k}\|}, \theta_{\mathbf{k}}\}$ and $\mathsf{M}$ is the square matrix
\begin{widetext}
\begin{equation}
\label{7.19}
\mathsf{M} = \begin{pmatrix}
0 & ik & 0 \\
ikp^* C_{p,n} & 2\sqrt{2}\gamma^* + \nu_{\ell}^* k^2 & ikp^* C_{p,T} \\
2\sqrt{2} \left( \zeta_0^* C_{\chi} + \zeta_0^{(1)} C_n + C_{\gamma} \right) + \mu^* k^2 & \frac{2}{d} ik \left( p^* + \frac{d}{2} \zeta_U \right) & 2\sqrt{2} \left( 2\gamma^* \theta^{-1} + \frac{1}{2} \zeta_0^* + \theta \partial_\theta \zeta_0^* \right) + D_{\mathrm{T}}^* k^2
\end{pmatrix}.
\end{equation}
\end{widetext}
Here,
\begin{equation}
    \nu_\ell^* = \frac{1}{2}\left( 2\frac{d-1}{d}\eta^* + \eta^*_\text{b}\right).
\end{equation}

The three longitudinal modes have the form $\exp\left[-\delta_\beta(k)\tau\right]$ for $\beta = 1,2,3 $, where $\delta_\beta(k)$ are the eigenvalues of the matrix M, namely, the roots of the characteristic polynomial
\begin{equation}
\label{7.20}
    \delta^3 - X(k)\delta^2 + Y(k)\delta - Z(k) = 0,
\end{equation}
where
\begin{equation}
\label{7.21}
    X(k) = \sqrt{2}\left( \zeta_0^* + 2\theta\frac{\partial \zeta_0^*}{\partial \theta} + 4\gamma^*\theta^{-1} + 2\gamma^*\right) + k^2(D_T^* + \nu_\ell^*),
\end{equation}
\beqa
\label{7.22}
    Y(k) &=& (2\sqrt{2}\gamma^* + \nu_\ell^* k^2)\Bigg[ k^2D_T^* + \sqrt{2} \left( \zeta_0^* + 2\theta\frac{\partial \zeta_0^*}{\partial \theta}\right. 
\nonumber\\
& & \left.+ 4\gamma^*\theta^{-1} \right) \Bigg] 
+ k^2p^*\left[C_{p,n} + C_{p,T}\left(\zeta_U + \frac{2}{d}p^*\right)\right],
\nonumber\\
\eeqa
\beqa
\label{7.23}
    Z(k) &=& p^*k^2\Bigg[ k^2\left( C_{p,n}D_T^* - C_{p,T}\mu^* \right) + \sqrt{2}C_{p,n}\nonumber\\
& & \left(\zeta_0^* + 2\theta\frac{\partial \zeta_0^*}{\partial \theta} 
 + 4\gamma^*\theta^{-1}\right)-2\sqrt{2}C_{p,T}\left(\zeta_0^*C_\chi
 \right.
 \nonumber\\
& & \left.
+\zeta_0^{(1)}C_n + C_\gamma\right)\Bigg].
\eeqa
In the case $\Delta_\text{b}^* = 0$, we have that $C_{p,T} = 1$, $\partial_\theta\zeta_0^* = \zeta_0^{(1)}\Upsilon_\theta$ and hence, Eqs.\ \eqref{7.20}--\eqref{7.23} agree with the stability analysis performed in Ref.\ \cite{GGG19a}.

\noindent The linear stability of the HSS is determined by the sign of the real part of the eigenvalues $\delta_\beta(k)$. In our convention, the modes decay as $\exp[-\delta_\beta(k)\tau]$, so the HSS is linearly stable provided $\Re[\delta_\beta(k)]>0$ for $\beta=1,2,3$ and for all $k\geq0$. It is quite apparent that the determination of the dependence of the real part of the eigenvalues $\delta_\beta(k)$ on the (dimensionless) wave vector $k$ and the parameters of the system is not really a simple problem. Therefore, to gain some insight into the general problem, it is convenient to study first the solution to the cubic equation \eqref{7.20} in the extreme long wavelength limit, $k=0$.

\begin{figure}[!h]
\centering
\includegraphics[width=0.35\textwidth]{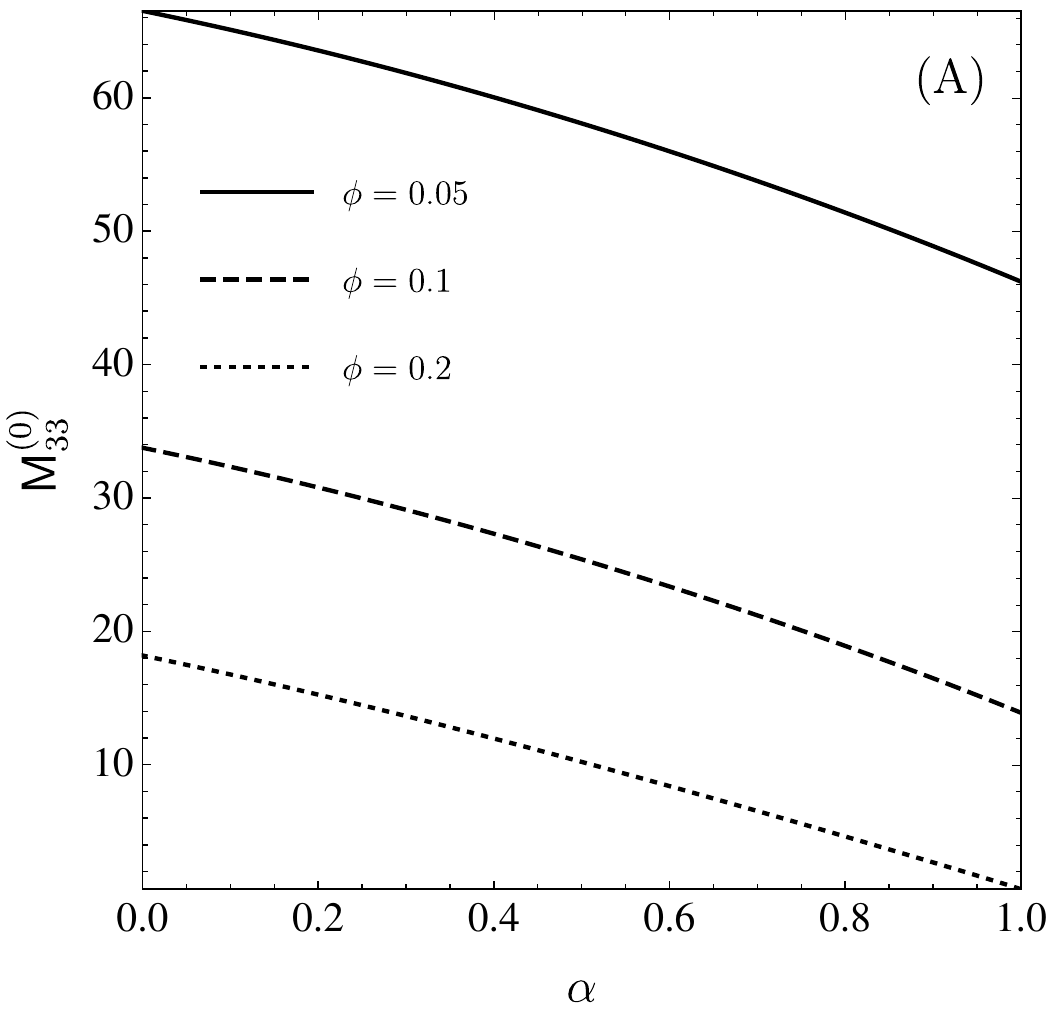}
\includegraphics[width=0.35\textwidth]{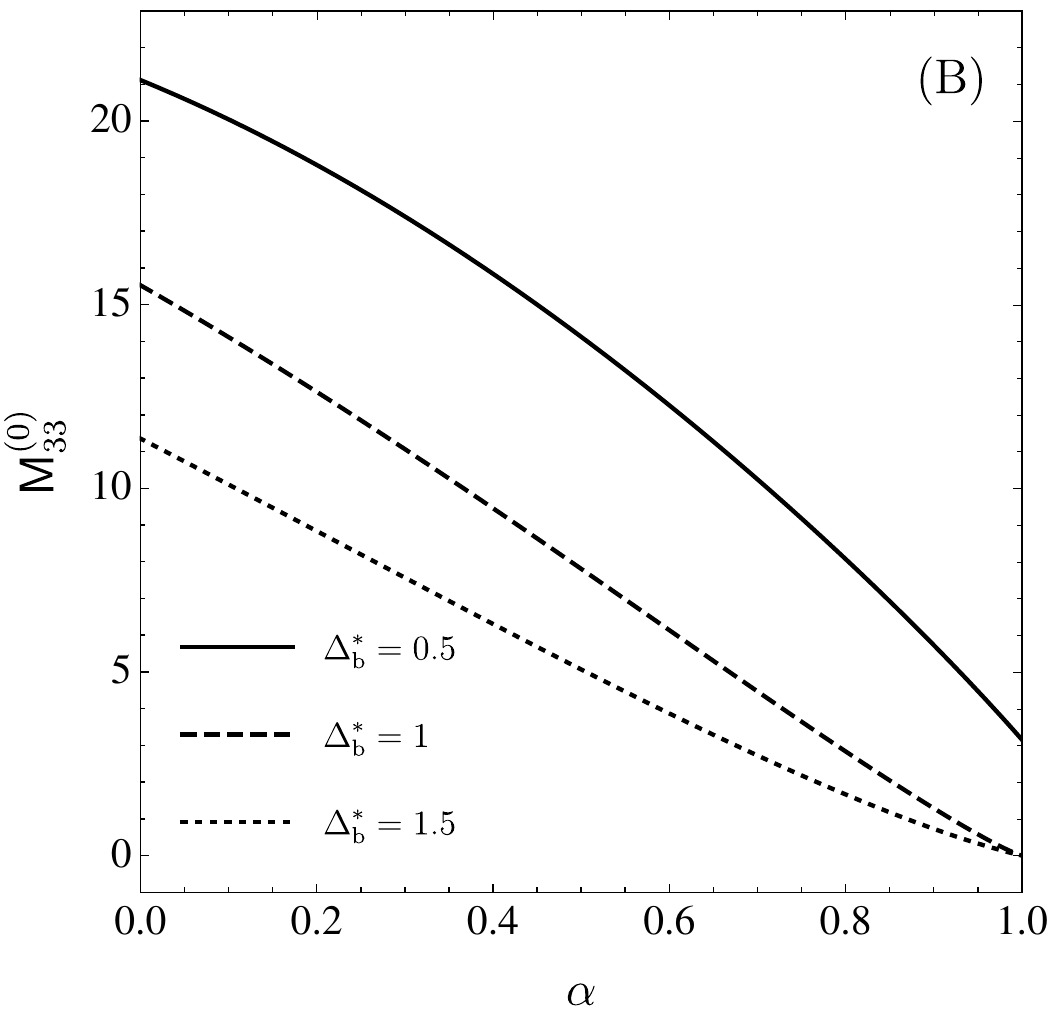}
\caption{(A): Plot of the eigenvalue $\mathsf{M}_{33}^{(0)}$ versus the coefficient of restitution $\al$ for $d=2$, $\Delta_{\text{b}}^*=1$, $T_\text{b}^*=0.9$, and three different values of $\phi$: $\phi = 0.05$ (solid), $\phi = 0.1$ (dashed) and $\phi = 0.2$ (dotted). (B): Plot of the eigenvalue $\mathsf{M}_{33}^{(0)}$ versus the coefficient of restitution $\al$ for $d=2$, $\phi=0.25$, $T_\text{b}^*=0.9$, and three different values of $\Delta_\text{b}^{*}$: $\Delta_\text{b}^{*}=0.5$ (solid), $\Delta_\text{b}^{*}=1$ (dashed) and $\Delta_\text{b}^{*}=1.5$ (dotted).
\label{fig_m33}}
\end{figure}

\subsection{Extreme long wavelength limit ($k=0$)}

In the long wavelength limit ($k = 0$), the square matrix $\mathsf{M}$ reduces to the matrix $\mathsf{M}^{(0)}$, given by
\begin{equation}
\label{7.24}
\mathsf{M}^{(0)} = \begin{pmatrix}
0 & 0 & 0 \\
0 & 2\sqrt{2}\gamma^* & 0 \\
\mathsf{M}_{31}^{(0)} & 0 & \mathsf{M}_{33}^{(0)}
\end{pmatrix},
\end{equation}
where
\beq
\mathsf{M}_{31}^{(0)}\equiv 2\sqrt{2}\left( \zeta_0^* C_{\chi} + \zeta_0^{(1)} C_n + C_{\gamma} \right),
\eeq
\beq
\mathsf{M}_{33}^{(0)}\equiv 2\sqrt{2}\left( 2\gamma^* \theta^{-1} + \frac{1}{2} \zeta_0^* + \theta \partial_\theta \zeta_0^* \right). 
\eeq

\noindent In this simpler case, it is quite straightforward to get closed solutions for the fields $\rho_\textbf{0}$ and $\textbf{w}_\textbf{0}$, with the results
\begin{equation}
\label{7.25}
    \rho_\textbf{0}(\tau) = \rho_\textbf{0}(0), \qquad \textbf{w}_\textbf{0}(\tau) = \textbf{w}_\textbf{0}(0)
    e^{- 2\sqrt{2}\gamma^* \tau}.
\end{equation}

\noindent According to Eq.\ \eqref{7.25}, the mode associated with the density is (marginally) stable (which is consistent with mass density conservation), whereas all the modes associated with the velocity field are (asymptotically) stable (since $\gamma^*$ is always positive). On the other hand, $\theta_\textbf{0}$ obeys the (linear) inhomogeneous ordinary differential equation
\begin{equation}
\label{7.26}
    \partial_\tau\theta_\textbf{0} + \mathsf{M}_{33}^{(0)}\theta_\textbf{0} + \mathsf{M}_{31}^{(0)}\rho_{\textbf{0}} = 0.
\end{equation}
The solution to Eq.\ \eqref{7.26} is
\begin{equation}
\label{7.27}
    \theta_\textbf{0}(\tau) = \left( \theta_\textbf{0}(0) + \frac{\mathsf{M}_{31}^{(0)}}{\mathsf{M}_{33}^{(0)}}\rho_\textbf{0}(0) \right)e^{-\mathsf{M}_{33}^{(0)}\tau} - \frac{\mathsf{M}_{31}^{(0)}}{\mathsf{M}_{33}^{(0)}}\rho_\textbf{0}(0).
\end{equation}

\noindent An exhaustive exploration of the space of parameters reveals that $\mathsf{M}_{33}^{(0)}$ is always positive, so
\begin{equation}
    \theta_\textbf{0}(\infty) =  - \frac{\mathsf{M}_{31}^{(0)}}{\mathsf{M}_{33}^{(0)}}\rho_\textbf{0}(0),
\end{equation}
and we conclude that the mode associated with $\theta_\textbf{0}$ relaxes exponentially to its asymptotic value $\theta_\textbf{0}(\infty)$.

Figure \ref{fig_m33} illustrates the dependence of the eigenvalue  $\mathsf{M}_{33}^{(0)}$ on the coefficient of restitution $\al$ for different systems. We observe that in all the cases $\mathsf{M}_{33}^{(0)}$ increases with increasing inelasticity.

\subsection{General case}

A necessary condition for the loss of stability is that one eigenvalue reaches the imaginary axis for some $k_\text{c}$, i.e.\ $\Re[\delta(k_\text{c})]=0$. If the critical eigenvalue is real, the instability is non--oscillatory and occurs through the condition  $\delta(k_\text{c})=0$ \cite{N20}. In this case, the term $Z(k)$ of the characteristic polynomial equation \eqref{7.20} must vanish ($Z(k_\text{c})=0$). Therefore, for $k\neq 0$, one obtains a candidate threshold wave number $k_\text{h}$ from the condition $Z(k_\text{h})=0$. This yields
\begin{widetext}
\begin{equation}
\label{7.28}
k_\text{h}^2=
\frac{
2\sqrt{2}\,C_{p,T}\left(\zeta_0^*C_\chi+\zeta_0^{(1)}C_n+C_\gamma\right)
-\sqrt{2}\,C_{p,n}\left(\zeta_0^*+2\theta\,\partial_\theta\zeta_0^*+4\gamma^*\theta^{-1}\right)
}{
C_{p,n}D_T^*-C_{p,T}\mu^*
}.
\end{equation}
\end{widetext}
In the regions of the parameter space where the three eigenvalues are real, the absence of physical solutions of Eq.\ \eqref{7.28} (namely, $k_\text{h}^2<0$) rules out non--oscillatory instabilities. Our numerical exploration indicates that $k_\text{h}^2$ is negative for all parameters explored, so that no non--oscillatory instability is found.

However, the characteristic polynomial may admit a complex conjugate pair of eigenvalues, $\delta_{2,3}(k)=\delta_\text{R}(k)\pm i\delta_\text{I}(k)$, as we have explicitly observed for some parameters. In that situation, the stability threshold does \emph{not} necessarily occur at $\delta=0$: an oscillatory (Hopf--type) instability may take place when $\delta_\text{R}(k_\text{c})=0$ with $\delta_\text{I}(k_\text{c})\neq 0$ \cite{N20}. Therefore, the absence of physical solutions of $Z(k) = 0$ for $k>0$ discards non--oscillatory instabilities but, in principle, it is not sufficient to guarantee stability. A convenient sufficient--and--necessary test for linear stability can be obtained from the Routh--Hurwitz criterion \cite{N20}. For the case considered here, the Routh--Hurwitz conditions reduce to
\begin{equation}
\label{7.29}
X(k)>0,\quad Y(k)>0,\quad Z(k)>0,\quad X(k)Y(k)>Z(k),
\end{equation}
for all $k>0$ (the case $k = 0$ has been studied separately). In particular, the equality $X(k)Y(k)=Z(k)$ (with $X,Y,Z>0$) corresponds to the onset of an oscillatory instability. Our numerical exploration of the parameter space shows that the inequalities \eqref{7.29} are satisfied for all $k$ in the hydrodynamic range considered, and hence $\Re[\delta_\beta(k)]>0$ for all $\beta$. As a consequence, we conclude that the HSS is linearly stable.

\section{Discussion}
\label{sec8}

In this paper, we have developed an Enskog--kinetic theory for a confined quasi--two--dimensional granular suspension. The study has been carried out within the context of the so--called $\Delta$--model \cite{BRS13}: a coarse-grained collisional model that attempts to incorporate collisional energy injection into the dynamics of granular particles via the positive parameter $\Delta$. Additionally, the influence of the surrounding interstitial gas on solid particles has been modeled by a viscous drag force and a stochastic bath. By means of the generalization of the conventional Chapman--Enskog method \cite{CC70} to dissipative dynamics, we have obtained the Navier--Stokes transport coefficients and the first--order contribution to the cooling rate under steady--state conditions. Since granular gases (modeled as hard spheres with inelastic collisions) lack equilibrium states, one relevant point in solving the kinetic equation via the Chapman--Enskog expansion is choosing the base reference state in the perturbation method. However, the zeroth--order approximation to the kinetic equation (the solution of the Enskog equation in the absence of spatial gradients) cannot be chosen a priori, as it must be obtained consistently \cite{G19}. Here, the local version of the time--dependent homogeneous solution to the Enskog equation has been considered as the reference state in the Chapman--Enskog expansion. Using this reference state introduces new conceptual  difficulties not present in the conventional Chapman--Enskog method
 \cite{CC70}.

Before considering inhomogeneous situations, we have characterized the HSS. As said before, the time--dependent version of this state is considered as the zeroth--order distribution in the Chapman--Enskog perturbation solution. As expected, for times longer than the mean free time, the distribution function of the HSS adopts the scaled form \eqref{2.8}. However, since the exact form of the scaled distribution $\varphi(\mathbf{c},\lambda,\theta)$ is not known to date, indirect information on this distribution is provided by the knowledge of its first velocity moments. In particular, in this paper we have estimated 
the steady temperature $\theta_\text{s}$ as well as the kurtosis or fourth cumulant $a_{2,\text{s}}$ (which measures the departure of $\varphi$ from its Maxwellian form) by using the first Sonine approximation \eqref{2.15} to the true distribution $\varphi$. The theoretical results obtained for $\theta_\text{s}$ and $a_{2,\text{s}}$ from that Sonine approximation \eqref{2.15} have been compared against computer simulations for several systems. 
The excellent agreement found between theory and DSMC results for both quantities supports the consistency of the reference distribution used in the present work and provides confidence in the (approximate) expressions of the transport coefficients derived from it.

The analysis carried out here shows that the surrounding gas has, in general, a significant impact on momentum and heat transport of granular particles. In particular, the dependence of the shear viscosity $\eta$, bulk viscosity $\eta_\text{b}$, thermal conductivity $\kappa$, diffusive heat conductivity $\mu$, and the first--order contribution $\zeta_U$ to the cooling rate on the coefficient of restitution $\al$ is markedly different from that obtained for the \textit{dry} (in the absence of gas phase) granular gas of the $\Delta$--model. These differences become especially apparent as dissipation increases and, for some coefficients, also as density increases. Therefore, the combined action of the collisional injection parameter $\Delta$ and the friction coefficient $\gamma$ cannot be neglected when describing confined granular suspensions at finite inelasticity. In this sense, the present results indicate that the use of transport coefficients corresponding either to elastic hard disks or to dry confined systems may miss important quantitative, and in some cases qualitative, features of the dynamics of a confined granular suspension. 

Another relevant outcome of the paper is that, within the Navier--Stokes description considered here, the HSS is linearly stable in the region of parameter space explored. Both the transverse and longitudinal perturbation modes decay in time, and no evidence of either stationary or oscillatory instabilities has been found in the hydrodynamic range analyzed. This result indicates that, in the Navier--Stokes hydrodynamic order and under the assumptions of the model, the HSS  constitutes a robust reference state for the study of weakly inhomogeneous confined granular suspensions. The present theory is, however, subject to the simplifications inherent in these assumptions. First, the gas phase has been incorporated into the kinetic equation of the solid particles through  an effective way by means of a drag force plus stochastic forcing, while its influence on the binary collision process itself is neglected (in contrast to the more sophisticated kinetic description recently carried out in Refs.\ \cite{GG22,GChG24,GG25}). 
Thus, the theory developed in this paper is expected to be reliable when the effect of the interstitial gas on granular particles during a collision is weak, but not in situations such as liquid--solid flows, where fluid effects during collisions may become important. Second, the expressions of the transport coefficients have been obtained by retaining only the leading terms in a Sonine polynomial expansion (the so--called first Sonine correction).
Although approximations of this sort have been shown to be quite accurate for not too strong dissipation (see for instance, Fig.\ 3 of the review article \cite{BSG26}), one expects that the accuracy of the theory increases as higher--order corrections in the Sonine polynomial expansion (such as the so--called second Sonine approximation) are retained \cite{GM04,GV09,GMV13a,GABG23,GGBS24}.  
Third, in the two--dimensional illustrations we have adopted the simple choice $R(\phi)=1$, since a more accurate expression for the density dependence of the drag in confined hard disks is not currently available. 

A natural continuation of the present work is to particularize the theory developed here to the synchronization--dependent $\Delta$--model recently considered by Maire \emph{et al.} \cite{MPSF25}. In that framework, synchronization is incorporated through an effective collision injection amplitude $\overline{\Delta}\equiv \Delta(\phi,T,\tau_s)$, which reflects the competition between the collision frequency and the synchronization time $\tau_\text{s}$. Up to now, the hydrodynamic treatment of that active branch has relied on using the Navier--Stokes transport coefficients of elastic hard disks. However, the results derived in the present paper clearly show that the combined action of collisional injection and gas--phase dissipation may substantially modify the transport coefficients. It is therefore of clear interest to suppress the stochastic bath when appropriate and derive the \emph{proper} Navier--Stokes transport coefficients and cooling--rate corrections for the synchronized $\Delta$--model itself, instead of employing the elastic forms of the transport coefficients. Such a program would put the hydrodynamic description of the synchronized model on a firmer footing and would allow one to reassess, with quantitatively consistent inputs, its stability properties, hydrodynamic modes, and long--wavelength structure factors. Work along this line is in progress.

\acknowledgments

D. Gonz\'alez M\'endez and V. Garz\'o acknowledge financial support from grant no. PID2024--156352NB--I00 funded by MCIU/AEI/10.13039/501100011033 and by ERDF/EU and from grant no. GR24022 funded by Junta de Extremadura (Spain) and by European Regional Development Fund (ERDF) ``A way of making Europe''. The research of D. Gonz\'alez M\'endez has been supported by the predoctoral fellowship FPU24/01056 from the Spanish Government.


\begin{widetext}

\appendix

\section{Expressions of several auxiliary quantities}
\label{appA}

In this Appendix we provide the expressions of the different auxiliary quantities appearing in the calculations for determining the transport coefficients and the first--order contribution to the cooling rate. Except for the integral $I_\eta$, these expressions are provided for a $d$--dimensional gas.

First, the quantities $A_0$, $A_1$, $C_0$, and $C_1$ defining the collisional moment $\mu_4$  are given by \cite{BMGB14}:
\beq
\label{a1}
A_0(\al, \theta, \Delta_\text{b}^*)=(d+2)\left(\frac{1-\al^2}{2}-\sqrt{\frac{\pi}{2}}
\al \theta^{-1/2}\Delta_\text{b}^*-\theta^{-1}\Delta_\text{b}^{*2}\right),
\eeq
\beq
\label{a2}
A_1(\al, \theta, \Delta_\text{b}^*)=\frac{(d+2)}{16}\left(\frac{3}{2}(1-\al^2)+
\theta^{-1}\Delta_\text{b}^{*2}\right),
\eeq
\beqa
\label{a3}
C_0(\al, \theta, \Delta_\text{b}^*)&=&\sqrt{2\pi}(1+2d+3\al^2+4 \theta^{-1}\Delta_\text{b}^{*2})\al \theta^{-1/2}\Delta_\text{b}^{*}-3+4 \theta^{-2}\Delta_\text{b}^{*4}+\al^2+2\al^4-2d(1-\al^2-2 \theta^{-1}\Delta_\text{b}^{*2})
\nonumber\\
& &
+2 \theta^{-1}\Delta_\text{b}^{*2}(1+6\al^2),
\eeqa
\beqa
\label{a4}
C_1(\al, \theta, \Delta_\text{b}^*)&=&\sqrt{\frac{\pi}{2}}\Big[2-2d(1-\al)+7\al+3\al^3\Big]\theta^{-1/2}\Delta_\text{b}^{*}
-\frac{1}{16}\Big\{85+4 \theta^{-2}\Delta_\text{b}^{*4}-18\left(3+2\al^2\right)\theta^{-1}\Delta_\text{b}^{*2}\nonumber\\
& &
-\left(32+87\al+30\al^3\right)\al-2d\left[
6\theta^{-1}\Delta_\text{b}^{*2}-(1+\al)(31-15\al)\right]\Big\}.
\eeqa

We consider now the quantities appearing in the evaluation of $\eta$ and $\eta_\text{b}$. For a two--dimensional gas and by using the Sonine approximation \eqref{2.11.1}, the integral $I_\eta$ is
\beq
\label{a5}
I_\eta = \frac{3}{8}\sqrt{\frac{\pi}{2}}\left(1+\frac{3}{32}\,a_2\right),
\eeq
while the integrals $I_{\eta_\text{b}}'$, $I_{\eta_\text{b}}''$, and $I_{\eta}'$ for $d$ dimensions are given, respectively, by
\beq
\label{a6}
I_{\eta_\text{b}}'=\frac{\Gamma\left(\frac{d+1}{2}\right)}{\Gamma\left(\frac{d}{2}\right)}\sqrt{2}\left(1-\frac{1}{16}a_2\right),
\quad I_{\eta_\text{b}}''=-\frac{\sqrt{2}}{16}\frac{\Gamma\left(\frac{d+1}{2}\right)}{\Gamma\left(\frac{d}{2}\right)}
\left(1+\frac{15}{32}a_2\right), \quad I_\eta'=-\frac{\Gamma\left(\frac{d+1}{2}\right)}{\sqrt{2}\Gamma\left(\frac{d}{2}\right)}(1+2\alpha)
\left(1-\frac{1}{16}a_2\right).
\eeq
Moreover, the collision frequency $\nu_\eta$ has been estimated in Ref.\ \cite{BBMG15} by considering the leading Sonine approximation \eqref{4.7} for $\mathcal{C}_{ij}(\mathbf{V})$ and by replacing the true $\varphi$ by the approximation \eqref{2.15}. Neglecting nonlinear terms in $a_2$, the expression of $\nu_\eta$ is
\beq
\label{a7}
\nu_\eta=\frac{\sqrt{2}\pi^{\frac{d-1}{2}}n\sigma^{d-1}v_\text{th}\chi}{d(d+2)\Gamma\left(\frac{d}{2}\right)}\left[
(1+\al)(2d+3-3\al)\left(1-\frac{a_2}{32}\right)+\sqrt{2\pi}(d-2\al)\theta^{-1/2}\Delta_\text{b}^*-2 \Big(1+\frac{3a_2}{32}\Big)
\theta^{-1}\Delta_\text{b}^{*2}\right].
\eeq

The integrals $I_\kappa$, $I_\kappa'$, and $I_\kappa''$ appearing in the collisional contributions to the heat flux transport coefficients are
\beq
\label{a8}
I_\kappa=\sqrt{2}\frac{\Gamma\!\left(\frac{d+3}{2}\right)}{\Gamma\!\left(\frac{d}{2}\right)}
\left(1+\frac{7}{16}a_2\right), \quad I_\kappa'=\frac{3\sqrt{2}}{16}(d+1)\frac{\Gamma\left(\frac{d+1}{2}\right)}{\Gamma\left(\frac{d}{2}\right)}
\left(1-\frac{1}{32}a_2\right), \quad I_\kappa''=\frac{d}{2}.
\eeq
The collisional integrals involving the operator $\boldsymbol{\mathcal{K}}$ appearing in the evaluation of the kinetic coefficients $\kappa_\text{k}$ and $\mu_\text{k}$ are given by
\beqa
\label{a9}
\int \dd\mathbf{V} \;\mathbf{S}(\mathbf{V})\cdot \boldsymbol{\mathcal{K}}\left[\frac{\partial}{\partial \mathbf{V}}\cdot \mathbf{V}f^{(0)}\right] &=& 2^d d \frac{n T^2}{m}\phi \chi \Bigg\{\frac{3}{8}(1+\al)^2\left[1-2\al-a_2(1+\alpha)\right] +\frac{1}{\sqrt{2\pi}}\theta^{-1/2}\Delta_\text{b}^*\nonumber \\
& & \times \Bigg[ \left[ \frac{3}{4} + \frac{a_2}{8}\left( d - \frac{113}{8} \right) \right] + 3(1+\alpha)\left(1-\frac{1}{2}\sqrt{2\pi}\theta^{-1/2}\Delta_\text{b}^*\right) \nonumber \\
& & -\frac{a_2}{8}\left(21 + \frac{27}{4}\alpha\right)\alpha - \frac{9}{2}(1+\alpha)^2 - \theta^{-1}\Delta_{\text{b}}^{*2}\left(1- \frac{1}{16}a_2\right)\Bigg]\Bigg\},
\eeqa
\begin{equation}
\label{a10}
\frac{nv_\text{th}^{-d}}{d T}
\int \dd\mathbf{V}\;\mathbf{S}(\mathbf{V})\cdot\boldsymbol{\mathcal{K}}\left[\theta \frac{\partial \varphi}
{\partial \theta}\right]=\frac{3}{32}\;2^{d} \chi\phi \frac{n T}{m}\theta \Upsilon_\theta\Bigg\{\left(1+\alpha\right)^3 + \frac{\theta^{-1/2}\Delta_{\text{b}}^*}{2\sqrt{2\pi}}\left[13 + 2\alpha(8+3\alpha)-\frac{4}{3}\theta^{-1}\Delta^{*2}_{\text{b}}\right]\Bigg\},
\end{equation}
\begin{align}
\label{a11}
\int \dd\mathbf{V}\;\mathbf{S}(\mathbf{V})\cdot\boldsymbol{\mathcal{K}}\left[f^{(0)}\right]=2^d d\frac{nT^2}{m}\chi\phi\Bigg\{
\frac{3}{8}\left(1+\alpha\right)\left[\alpha\left(\alpha-1\right)+
\frac{a_2}{6}\left(10+2d-3\alpha+3\alpha^2\right)\right] \nonumber \\
+ \frac{1}{\sqrt{2\pi}} \theta^{-1/2}\Delta^*_{\text{b}}\Bigg[2\theta^{-1}\Delta_{\text{b}}^{*2}-3\left(\frac12-\alpha^{2}\right)
+\frac{3}{2}\sqrt{2\pi}\,\alpha\,\theta^{-1/2}\Delta_{\text{b}}^{*} \nonumber \\ + \frac{a_2}{32}\left(31+8d+18\alpha^{2}-4\theta^{-1}\Delta_{\mathrm b}^{*2}\right)\Bigg]\Bigg\},
\end{align}
\begin{equation}
\label{a12}
\frac{v_{\text{th}}^{-d}}{d}
\int \dd\mathbf{V}\;\mathbf{S}(\mathbf{V})\cdot\boldsymbol{\mathcal{K}}\left[\phi \frac{\partial \varphi}
{\partial \phi}\right]
=
\frac{3}{32}\,2^{d}\,\chi\phi^2\,\frac{T^2}{m} \frac{\partial a_2}{\partial \phi}
\Bigg\{\left(1+\alpha\right)^3
+\theta^{-1/2}\Delta_{\text{b}}^*\,\frac{1}{2\sqrt{2\pi}}
\left[13+2\alpha(8+3\alpha)-\frac{4}{3}\theta^{-1}\Delta_{\text{b}}^{*2}\right]\Bigg\}.
\end{equation}

Finally, we consider the quantities involved in the coefficient $e_D$. This coefficient provides the explicit form of the first--order contribution $\zeta^{(1,1)}$ to the cooling rate. The expression of the collision integral appearing in Eq.\ \eqref{4.34} can be written as
\beqa
\label{a13}
\int \dd\mathbf{v} E(\mathbf{V})\mathcal{K}_i\left[\frac{\partial f^{(0)}}{\partial V_i}\right]&=&
3d\; 2^{d-5}\;n \phi \chi \left(1+\alpha\right)\Bigg\{(1-\alpha^2)(5\alpha-1)-
\frac{a_2}{6}
\left[15\alpha^3-3\alpha^2+3\left(4d+15\right)\alpha-\left(20d+1\right)\right]\Bigg\}
\nonumber\\
& &-\frac{2^{d-2}d}{\sqrt{2\pi}}n\chi\phi\;\theta^{-1/2}\Delta^*_{\text{b}}\Bigg\{-6+4\alpha(-3+3\alpha+4\alpha^2) + 3\sqrt{2\pi}(1+\alpha)(-1+3\alpha)\theta^{-1/2}\Delta_{\text{b}}^* \nonumber\\
& & + 8(1+2\alpha)\theta^{-1}\Delta_{\text{b}}^{*2} + 2\sqrt{2\pi}\,\theta^{-3/2}\Delta_\text{b}^{*3}+\frac{a_2}{8}\Bigg[47-8d+2\alpha(31+8d+9\alpha+12\alpha^2)
\nonumber\\
& & -4(1+2\alpha)\theta^{-1}\Delta_\text{b}^{*2}\Bigg]\Bigg\}.
\eeqa
The collision frequency $\nu_\zeta$ is given by
\beqa
\label{a14}
\nu_{\zeta}&=&-\frac{\pi^{\frac{d-1}{2}}}{\Gamma\left(\frac{d}{2}\right)}\frac{1}{24\sqrt{2}(d+2)}\chi \frac{v_\text{th}}{\ell}\Bigg\{
(1+\alpha)\left[30\alpha^3-30\alpha^2+\left(105+24d\right)\alpha-56d-73\right]
\nonumber\\
& & +\frac{3}{128d}\Bigg[-4 \theta^{-1}\Delta_\text{b}^{*2}\left(32+15 a_2\right)+36\al^2\theta^{-1}\Delta_\text{b}^{*2}\left(32+3a_2\right)+6 \theta^{-1}\Delta_\text{b}^{*2}\left(288-45 a_2\right.\nonumber\\
& & \left.
+64d+6d a_2\right)+512 \sqrt{\frac{\pi}{2}}\theta^{-1/2}\Delta_\text{b}^{*}\left[2+7\al+3\al^3-2d(1-\al)\right]\Bigg]
-12 \frac{(d+2)}{d}\theta^{-1}\Delta_\text{b}^{*2}\left(1+\frac{15}{32}a_2\right)\Bigg\}.
\nonumber\\
\eeqa
\end{widetext}

\bibliography{difdelta}
\end{document}